\documentclass[]{JHEP3}
\usepackage{epsfig}

\DeclareMathAlphabet{\scr}{U}{rsfs}{m}{n}
\usepackage{latexsym}
\usepackage[mathscr]{eucal}
\usepackage{amsfonts}
\usepackage{amscd}
\usepackage{cite}
\usepackage{array}   
\usepackage{amssymb}
\usepackage{colordvi}
\usepackage[centertags]{amsmath}
\usepackage{enumerate}
\usepackage{graphicx}
\usepackage{booktabs}
\usepackage{theorem}

\def\be{\begin{equation}}
\def\ee{\end{equation}}
\def\beq{\begin{equation}}
\def\eeq{\end{equation}}
\def\bea{\begin{eqnarray}}
\def\eea{\end{eqnarray}}

\newcommand{\newc}{\newcommand}
\newc{\ol}{\overline}
\newc{\wt}{\widetilde}
\newc{\bs}{\boldsymbol}
\newc{\ma}{\mathcal}
\newc{\vl}{\langle}
\newc{\vr}{\rangle}
\def\vev#1{\langle #1 \rangle}
\newc{\sg}{S}
\newc{\ug}{U}
\newc{\tg}{T}

\title{Neutrino Mass and Mixing with Discrete Symmetry\,\footnote{Review article 
submitted for publication in Reports on Progress in Physics}}

\author{Stephen F. King\,\footnote{e-mail address: king@soton.ac.uk} \\
Department of Physics and Astronomy,
University of Southampton,\\
Southampton, SO17 1BJ, U.K.}

\author{Christoph Luhn\,\footnote{e-mail address: christoph.luhn@durham.ac.uk}\\
Institute for Particle Physics Phenomenology, University of Durham,\\
Durham, DH1 3LE, U.K.}


\abstract{This is a review article about neutrino mass and mixing and flavour model building strategies based on discrete family symmetry. After a pedagogical introduction and overview of the whole of neutrino physics,
we focus on the PMNS mixing matrix and the latest global fits following the Daya Bay and RENO experiments which measure the reactor angle. 
We then describe the simple bimaximal, tri-bimaximal and golden ratio
patterns of lepton mixing and the deviations required for a non-zero reactor angle,
with solar or atmospheric mixing sum rules resulting from charged lepton corrections or 
residual trimaximal mixing. 
The different types of see-saw mechanism are then reviewed as well as the sequential dominance mechanism. 
We then give a mini-review of finite group theory, which may be used as a discrete family symmetry
broken by flavons either completely, or with different subgroups preserved in the neutrino and charged lepton sectors.
These two approaches are then reviewed in detail in separate chapters including mechanisms for flavon vacuum alignment and different model building strategies that have been proposed to generate the reactor angle. We then briefly review grand unified theories (GUTs) and how they may be combined with discrete family symmetry to describe all quark and lepton masses and mixing.
Finally we discuss three model examples which combine an $SU(5)$ GUT with the discrete family symmetries $A_4$, $S_4$ and $\Delta(96)$.}

\preprint{IPPP-12-100 \\ DCPT-12-200}

\begin{document}

\section{\label{sec:intro}Introduction}

\subsection{Historical overview }

In 1930 the Austrian physicist Wolfgang Pauli proposed the existence
of particles called neutrinos, denoted as $\nu$,
as a ``desperate remedy'' to account for
the missing energy in a type of radioactivity called beta decay. At
the time physicists were puzzled because nuclear beta decay appeared
to violate energy conservation. In beta decay, a neutron in an
unstable nucleus transforms into a proton and emits an electron, where
the radiated electron was found to have a continuous energy
spectrum. This came as a great surprise to many physicists because
other types of radioactivity involved gamma rays and alpha particles
with discrete energies.  Pauli deduced that some of the energy must
have been taken away by a new particle emitted in the decay process,
the neutrino, which carries energy and has spin 1/2, but which is
massless, electrically neutral and very weakly interacting.  
Because neutrinos interact so
weakly with matter, Pauli bet a case of champagne that nobody would
ever detect one, and they became known as ``ghost particles''.
Indeed it was not until a quarter of a century later, in 
1956, that Pauli lost his bet and neutrinos were discovered when Clyde
Cowan and Fred Reines detected antineutrinos emitted from a nuclear
reactor at Savannah River in South Carolina, USA.  

Since then, after decades of painstaking experimental and theoretical work,
neutrinos have become enshrined as an essential part of the accepted
quantum description of fundamental particles and forces, the Standard
Model (SM) of particle physics.
This is a highly successful theory in which
elementary building blocks of matter are divided into three generations
of two kinds of particle - quarks and leptons. It also includes three
of the fundamental forces of Nature, the strong ($g$), electromagnetic 
($\gamma$) and weak ($W,Z$) forces carried by spin 1 force carrying
bosons (shown in parentheses) but does not include gravity. 
There are six flavours of quarks.
The leptons consist of the charged electron $e^-$, muon $\mu^-$
and tau $\tau^-$, together with
three electrically neutral particles - the electron neutrino $\nu_e$, muon
neutrino $\nu_{\mu}$ and tau neutrino $\nu_{\tau}$ which are our main
concern here.

The first clues that neutrinos have mass came from an experiment deep
underground, carried out by an American scientist Raymond Davis Jr., detecting
solar neutrinos~\cite{Cleveland:1998nv}. It revealed only about one-third of
the number predicted by theories of how the Sun works pioneered by John
Bahcall~\cite{Cleveland:1998nv}.  The result puzzled both solar and neutrino
physicists. 
Based on the original idea of neutrino oscillation, first introduced by
Pontecorvo in 1957~\cite{Pontecorvo:1957qd} and independently by Maki, Nakagawa
and Sakata in 1962~\cite{MNS}, some Russian researchers, Mikheyev and Smirnov,
developing ideas proposed previously by Wolfenstein in the U.S., suggested
that the solar neutrinos might be changing into something else. 
Only electron neutrinos are emitted by the Sun and they could be converting
into muon and tau neutrinos which were not being detected by Davis's experiment. 
The precise mechanism for ``solar neutrino oscillations'' proposed by
Mikheyev, Smirnov and Wolfenstein involved the resonant enhancement of
neutrino oscillations due to matter effects in the Sun, and is known as the MSW
effect~\cite{MSW}.

\begin{figure}[htb]
\centering
\includegraphics[width=0.86\textwidth]{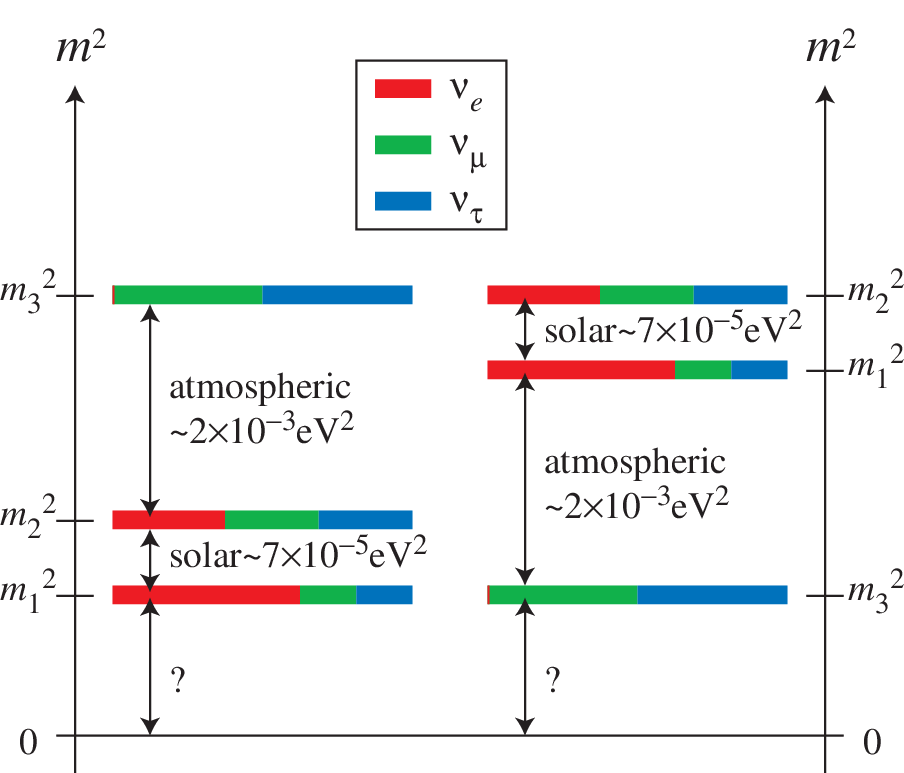}
\caption{\label{mass1}\small{The probability that a particular neutrino
mass state contains a particular SM state may be represented by colours as
shown in the key. Note that neutrino oscillation experiments only determine
the difference between the squared values of the masses. Also, while
$m_2^2>m_1^2$, it is presently unknown whether $m_3^2$ is heavier or lighter
than the other two, corresponding to the left and right panels of the figure,
referred to as normal or inverted mass squared ordering, respectively. Finally
the value of the lightest neutrino mass (sometimes referred to as the neutrino
mass scale) is presently unknown and is represented by a question mark in each
case.}}
\vspace*{-2mm}
\end{figure}

Neutrino oscillations are analogous to coupled pendulums, where
oscillations in one pendulum induce oscillations in another pendulum.
The coupling strength is defined in terms of something called the
``lepton mixing matrix'' $U$ 
which relates the basic Standard Model 
neutrino states, $\nu_e$, $\nu_{\mu}$, $\nu_{\tau}$,
associated with the electron, muon and tau,
to the neutrino mass states $\nu_1$, $\nu_2$, and $\nu_3$
with (real and positive) masses $m_1$, $m_2$, and $m_3$~\cite{MNS},
\begin{equation}
\left(\begin{array}{c} \nu_e \\ \nu_\mu \\ \nu_\tau \end{array} \\ \right)=
\left(\begin{array}{ccc}
U_{e1} & U_{e2} & U_{e3} \\
U_{\mu1} & U_{\mu2} & U_{\mu3} \\
U_{\tau1} & U_{\tau2} & U_{\tau3} \\
\end{array}\right)
\left(\begin{array}{c} \nu_1 \\ \nu_2 \\ \nu_3 \end{array} \\ \right)
\; .
\label{MNS0}
\end{equation}
According to quantum mechanics it is not necessary that the Standard
Model states $\nu_e$, $\nu_{\mu}$, $\nu_{\tau}$ be identified in a one-one
way with the mass eigenstates $\nu_1$, $\nu_2$, and $\nu_3$,
and the matrix elements of $U$ give the quantum amplitude that
a particular Standard Model state contains an admixture of a 
particular mass eigenstate. The probability that a particular neutrino mass
state contains a particular SM state may be represented by colours as
in Fig.~\ref{mass1}. Note that neutrino oscillations are only sensitive to the 
differences between the squares of the neutrino masses
$\Delta m_{ij}^2\equiv m_i^2-m_j^2$, and 
gives no information about the absolute value of the neutrino mass squared
eigenvalues $m_i^2$.
There are basically two
patterns of neutrino mass squared orderings
consistent with the atmospheric and solar data as shown in
Fig.~\ref{mass1}.

As with all quantum amplitudes,
the matrix elements of $U$ are expected to be complex numbers in
general. The lepton mixing matrix $U$ is also frequently
referred to as the Maki-Nakagawa-Sakata (MNS) matrix
$U_{\mathrm{MNS}}$~\cite{MNS}, and sometimes the name of Pontecorvo is added
at the beginning to give $U_{\mathrm{PMNS}}$. The standard parameterisation of
the PMNS matrix in terms of three angles and at least one complex phase, as
recommended by the Particle Data Group (PDG)~\cite{Nakamura:2010zzi}, will be
discussed later. 

Before getting into details, here is a quick executive summary of the implications of neutrino mass and mixing
following from Fig.~\ref{mass1}:
\begin{itemize}
\item Lepton flavour is not conserved, so the individual lepton numbers $L_e$, $L_{\mu}$, $L_{\tau}$ are separately broken
\item Neutrinos have tiny masses which are not very hierarchical
\item Neutrinos mix strongly unlike quarks
\item The SM parameter count is increased by at least 7 new parameters (3 neutrino masses, 3 mixing angles
and at least one complex phase)
\item It is the first (and so far only) new physics beyond the SM
\end{itemize}

The idea of neutrino oscillations was first confirmed in 1998 by the Japanese experiment
Super-Kamiokande (SK)~\cite{SK}
which showed that there was a deficit of muon
neutrinos reaching Earth when cosmic rays strike the upper atmosphere,
the so-called ``atmospheric neutrinos''. Since most neutrinos pass
through the Earth unhindered, Super-Kamiokande was able to detect
muon neutrinos coming from above and below, and found that while the
correct number of muon neutrinos came from above, only about a half
of the expected number came from below. The
results were interpreted as half the 
muon neutrinos from below oscillating into tau
neutrinos over an oscillation length $L$ of the diameter of the Earth,
with the muon neutrinos from above having a negligible oscillation
length, and so not having time to oscillate, yielding the 
expected number of muon neutrinos from above.

In 2002, the Sudbury Neutrino Observatory (SNO) in Canada
spectacularly confirmed the flavour conversion in ``solar
neutrinos''~\cite{Ahmed:2003kj}. The 
experiment measured both the flux of the electron neutrinos and the
total flux of all three types of neutrinos. The SNO data revealed that
physicists' theories of the Sun were correct after all, and the solar
neutrinos $\nu_e$ were produced at the standard rate but were
oscillating into $\nu_{\mu}$ and $\nu_{\tau}$, with only about a third
of the original $\nu_e$ flux arriving at the Earth.

Since then, neutrino oscillations consistent with solar neutrino observations
have been seen using man made neutrinos from nuclear
reactors at KamLAND in Japan~\cite{Eguchi:2002dm}
(which, for the first time, observed the periodic pattern characteristic for
neutrino oscillations), and neutrino oscillations consistent with atmospheric
neutrino observations have been seen using neutrino 
beams fired over hundreds of kilometres 
as in the K2K experiment in Japan~\cite{Ahn:2006zz}, 
the Fermilab-MINOS experiment in the U.S.~\cite{Michael:2006rx}
or the CERN-OPERA experiment in Europe.
Further long-baseline neutrino beam experiments are in the pipeline,
and neutrino oscillation physics is entering the precision era,
with superbeams and a neutrino factory on the horizon.

Following these results several research groups 
showed that the electron neutrino has a mixing matrix element of $|U_{e2}|\approx 1/\sqrt{3}$
which is the quantum amplitude for $\nu_e$ to contain 
an admixture of the mass eigenstate $\nu_2$ corresponding to 
a massive neutrino of mass 
$m_2\approx 0.008$ electronvolts (eV) or greater (where $\sqrt{m_2^2-m_1^2}\approx 0.008$ eV). 
By comparison the electron has a mass of about half a
megaelectronvolt (MeV). 
Put another way, the mass state $\nu_2$ contains roughly equal probabilities of 
$\nu_e$, $\nu_{\mu}$ and $\nu_{\tau}$ sometimes called
 trimaximal mixing, corresponding to the three equal red, green and blue colours associated with $m_2^2$
in Fig.~\ref{mass1}. The muon and tau neutrinos were observed to contain 
approximately equal amplitudes of the third neutrino 
$\nu_3$ of mass $m_3$, 
$|U_{\mu 3}|\approx |U_{\tau 3}| \approx 1/\sqrt{2}$,
where a normalised amplitude of $1/\sqrt{2}$ corresponds
to a 1/2 fraction of $\nu_3$ in each of $\nu_{\mu}$ and $\nu_{\tau}$,
leading to a maximal mixing and oscillation of 
$\nu_{\mu} \leftrightarrow \nu_{\tau}$. 
Put another way, the mass state $\nu_3$ contains roughly equal probabilities of 
$\nu_{\mu}$ and $\nu_{\tau}$ called
maximal mixing, corresponding to the two equal green and blue colours associated with $m_3^2$
in Fig.~\ref{mass1}.
Interestingly, the value of $m_3$ is not determined and it could be anywhere between
zero and 0.3 eV, depending on the mass scale and ordering.
Although at least one neutrino mass must be 0.05 eV or greater
(where $\sqrt{|m_3^2-m_2^2|}\approx 0.05$ eV),
this could be either $m_3$ or $m_2$, as shown in Fig.~\ref{mass1}.

According to the early results from the CHOOZ nuclear reactor
experiment~\cite{Apollonio:1999ae}, 
the electron neutrino $\nu_e$ could only contain a very small 
amount of the third neutrino mass
eigenstate $\nu_3$, $|U_{e3}|<0.2$.
Evidence for non-zero $U_{e3}$ was
first provided by T2K, MINOS and Double
Chooz~\cite{Abe:2011sj}. 
Recently the Daya Bay~\cite{DayaBay}, RENO~\cite{RENO}, and Double
Chooz~\cite{DCt13} collaborations have measured $|U_{e3}|\approx 0.15$.
Put another way, the mass state $\nu_3$ has a probability of 
containing $\nu_e$ of about $(0.15)^2$, corresponding to the small amount of
red colour associated with $m_3^2$ in Fig.~\ref{mass1}.
As we shall see, this element being non-zero excludes a number of simple
mixing patterns and models which were previously proposed, and has led
to a number of new developments.

\subsection{\label{milestones}Where we stand}
The main experimental milestones from 1998-2012 may be summarised as follows:
\begin{itemize}
\item 1998 - SK confirms that atmospheric $\nu_{\mu}$ are converted to another neutrino type, probably $\nu_{\tau}$ consistent with near maximal mixing $|U_{\mu 3}|\approx |U_{\tau 3}| \approx 1/\sqrt{2}$
\item 2002 - SK, SNO and the older neutrino experiments such as Homestake and
  the Gallium experiments results are combined in a global fit pointing
  towards the large (but non-maximal) mixing and conversion of solar neutrinos
  in the core of the Sun 
\item 2002 - SNO confirms that solar 
$\nu_{e}$ are converted to a linear combination of $\nu_{\mu}$ and $\nu_{\tau}$
with approximate trimaximal mixing $|U_{e 2}|\approx |U_{\mu 2}|\approx |U_{\tau 2}| \approx 1/\sqrt{3}$
\item 2004 - Reactor antineutrinos $\overline{\nu}_{e}$ are observed by KamLAND
to oscillate with a probability consistent with the solar neutrino oscillations
\item 2006 - Accelerator neutrinos $\nu_{\mu}$ from Fermilab are observed over a long baseline (LBL) at MINOS 
with a disappearance probability consistent with the atmospheric oscillation results, providing a high precision confirmation of a similar observation from KEK to SK (K2K) in 2004
\item  2010 - LBL accelerator neutrinos $\nu_{\mu}$ from CERN appear at OPERA as $\nu_{\tau}$ 
\item 2011 - T2K and MINOS observe and excess of accelerator neutrinos $\nu_{\mu}$ appearing as  $\nu_{e}$, consistent with non-zero $U_{e3}$
\item 2012 - Daya Bay, RENO and Double Chooz  observe the disappearance of reactor antineutrinos $\overline{\nu}_{e}$ and measure  $|U_{e3}|\approx 0.15$
\end{itemize}

\subsection{The challenges ahead for experiment}
Despite the above observations, neutrinos remain the least understood particles.
Of the (at least) 7 new parameters which must be present due to neutrino mass and mixing, 
only 5 are currently measured,
namely the three mixing angles and two mass squared differences.
For example none of the CP violating phases are currently measured,
although there are plans to measure one of these phases in next generation neutrino oscillation experiments.
However, since the neutrino oscillations are only sensitive to mass squared differences,
the lightest neutrino mass cannot be measured by oscillation experiments.
Also the present experiments are not sensitive enough to uniquely determine  
the ordering of the neutrino square masses $m_1^2$, $m_2^2$, $m_3^2$,
although it is known from the solution to the solar neutrino problem
that  $m_2^2>m_1^2$.
The neutrino mass scale (i.e. the mass of the lightest neutrino) is not known, although,
as discussed later, there are cosmological reasons to believe that none of the neutrino
masses can exceed about 0.3 eV. Hence the lightest neutrino mass should be somewhere between zero
and 0.3 eV. However cosmology is not sensitive to whether neutrino mass is of
the Dirac or Majorana kind.\footnote{See Subsection~\ref{subsec:origin} for
the definition of Dirac and Majorana neutrino masses.}
In principle, if neutrinoless double beta decay were observed, it could simultaneously 
be used to measure both the lightest neutrino mass and show that it is Majorana (and future
measurements could shed light on the additional phases associated with Majorana masses).
However, despite our ignorance, we know that neutrino masses are much smaller than the other charged
fermion masses, and this already represents something of a puzzle. 

From the experimental perspective, the main known unknowns of neutrino mass and mixing may be summarised as:
\begin{itemize}
\item The neutrino mass squared ordering (normal or inverted)
\item The neutrino mass scale (i.e. the mass of the lightest neutrino, presumably between zero and 0.3 eV)
\item The nature of neutrino mass (Dirac or Majorana)
\item The CP violating phase measurable in neutrino oscillations (the so-called Dirac phase~$\delta$, although it is also present if neutrino mass is Majorana)
\item The two possible further CP violating phases associated with Majorana neutrino masses (not present if neutrino mass is Dirac)
\end{itemize}
Neutrino physics has now entered the precision era, at least as far as the  measured parameters are concerned.
T2K is presently running~\cite{Nakayama:2012kc} and will provide accurate 
measurements of the atmospheric neutrino mass squared difference and mixing angle,
while NO$\nu$A~\cite{Patterson:2012zs}, presently under construction, will provide complementary information
about the mass ordering.  
Future neutrino oscillation experiments, under discussion~\cite{Bertolucci:2012fb},
will give more accurate information about the mass squared splittings
$\Delta m_{ij}^2\equiv m_i^2-m_j^2$, mixing angles, the mass squared ordering
(commonly but incorrectly referred to as the ``mass hierarchy''),
and the neutrino mass scale (i.e. the mass of the lightest neutrino mass
eigenstate, which will indeed decide if neutrino masses involve
a significant mass hierarchy).
The ultimate goal of oscillation experiments, however, is to measure
the so far undetermined CP violating oscillation phase $\delta$
and there is considerable activity in this area~\cite{Coloma:2012wq}
to determine the best way to do this.

\subsection{The nature and scale of neutrino mass}

Oscillation experiments are not by themselves capable of telling
us anything about the nature or mass scale of neutrino mass.
They can, however, shed light on the neutrino mass ordering as mentioned above.
There are basically four ways to elucidate the mysteries surrounding neutrino masses.
\begin{enumerate}
\item Neutrinoless double beta decay experiments (for a recent review see
  e.g.~\cite{Rodejohann:2012xd}) effectively measure the 1-1 element
of the Majorana neutrino mass matrix corresponding to
\beq
m_{\beta\beta}\equiv \Big|\sum_iU_{ei}^2m_i\Big| ,\label{nulessma}
\eeq
and can validate the Majorana nature of neutrinos.
There was a claim of a signal in neutrinoless double
beta decay corresponding to $m_{\beta \beta}=0.11-0.56$ eV 
at 95\% C.L.~\cite{Klapdor-Kleingrothaus:2001ke}.
However this claim was criticised by two
groups~\cite{Feruglio:2002af}, and in turn this 
criticism has been refuted~\cite{Klapdor-Kleingrothaus:2002kf}.
Experiments such as GERDA should report soon and decide this question~\cite{Rodejohann:2012xd}.

\item Oscillation experiments can measure the sign of $\Delta m_{32}^2$
and resolve normal from inverted mass squared orderings.
\item Independently of whether neutrinos are Dirac or Majorana, 
the Tritium beta decay experiment KATRIN~\cite{Fischer:2011za} will tell us about the
absolute scale of neutrino mass down to about 0.35 eV.
Such experiments measure the ``electron neutrino mass'' defined by
\beq
m_{\nu_e}^2\equiv \sum_i|U_{ei}|^2|m_i|^2.
\eeq
\item More model dependently, cosmology can in principle probe the sum of neutrino
masses, and hence the lightest neutrino mass $m_{\mathrm{lightest}}$, down to
very small values~\cite{Lesgourgues:2006nd}. In future detection of energetic
neutrinos from gamma ray bursts, neutrino telescopes could also
provide important astrophysical information, and may provide another
means of probing neutrino mass, and even quantum
gravity~\cite{Choubey:2002bh}. 
\end{enumerate}

\begin{figure}[ht!]
\centering
\includegraphics{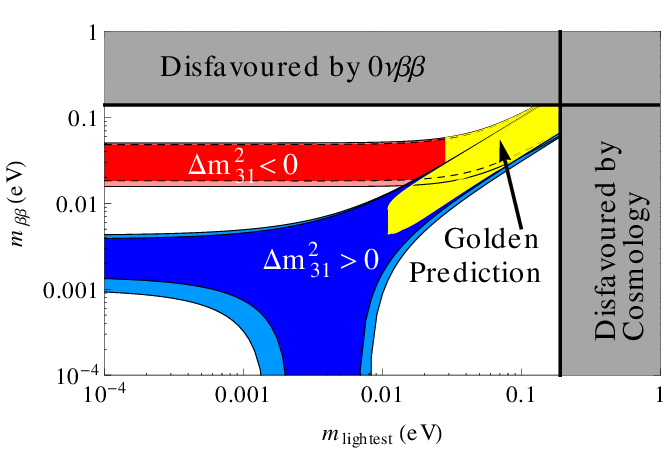}
\caption{\label{doublebeta2}\small{$m_{\beta\beta}$ vs. $m_{\rm lightest}$:  The red and light red regions represent the model independent values that the inverted neutrino mass ordering can take based on the central value and 1$\sigma$ deviation of
a recent global fit of neutrino parameters.  The blue and light blue regions are the analogue of this for the normal neutrino mass ordering. The gold regions correspond to the golden ratio prediction for $m_{\beta\beta}$ in both the normal and inverted orderings, resulting from the $A_5$ inverse mass sum rule.}}
\end{figure}
In Fig.~\ref{doublebeta2} we show the allowed range of the effective mass
parameter for neutrinoless double beta decay, $m_{\beta\beta}$, see
Eq.~(\ref{nulessma}), as a function 
of the lightest neutrino mass $m_{\rm lightest}$
for both the normal (blue) and inverted (red) neutrino mass squared orderings
consistent with the one sigma range of parameters taken from a recent global
fit as discussed in~\cite{Cooper:2012bd} from which this figure is
taken.\footnote{Fig.~\ref{doublebeta2} is generated using a code whose original version
was developed in~\cite{Lindner:2005kr}.}  
Also shown (gold) is the restricted region in a model
based on a discrete family symmetry $A_5$ which involves the golden ratio (GR)~\cite{Cooper:2012bd}.
Such restricted regions follow from relations between the neutrino masses which can generally arise in models based on discrete family symmetry. For example, the neutrino masses 
may be related by a sum rule of the form,
\beq
\alpha m_1+\beta  m_2 = m_3, \label{masssum}
\eeq
where $\alpha, \beta $ are model dependent constants. 
If the model involves a see-saw mechanism, then the right-handed neutrino masses may be similarly related
leading to inverse relationships between light physical neutrino masses of the form,
\beq
\frac{\gamma}{m_1}+\frac{\delta}{m_2} = \frac{1}{m_3},  \label{inversemasssum}
\eeq
where $\gamma, \delta $ are model dependent constants.\footnote{Note that $\delta$ here is nothing to do with the CP violating phase denoted by the same Greek letter.}
In certain models analogous relations may arise with the neutrino masses replaced by their square roots.
All these mass sum rules have been recently studied in~\cite{Dorame:2011eb}. 
The $A_5$ GR model~\cite{Cooper:2012bd} mentioned above involves an inverse sum rule as in 
Eq.~(\ref{inversemasssum})
with $\gamma = \delta = 1$. This may be compared to the $A_4$ model in~\cite{Altarelli:2005yx} which 
involves an inverse mass sum rule with $\gamma = 1$ and $\delta = -2$, or the 
$\Delta(96)$ model in~\cite{King:2012in} with $\gamma = 1$ and $ \delta = \pm 2i$, and so on.
The appearance of the complex constant reminds us that in all the (inverse) mass sum rules there are CP violating phases associated with Majorana neutrino masses which are implicit.

\subsection{\label{subsec:origin}The origin of neutrino mass}

It is worth recalling why the observation of non-zero neutrino mass and mixing
is evidence for new physics beyond the SM. The most intuitive way to
understand why neutrino mass is forbidden in the Standard Model, is to
understand that the Standard Model predicts that neutrinos
always have a ``left-handed'' spin - rather like rifle bullets which
spin counter clockwise to the direction of travel.
In fact this property was first experimentally measured in 1958,
two years after the neutrino was discovered, by
Maurice Goldhaber, Lee Grodzins and Andrew Sunyar.  More accurately,
the ``handedness'' of a particle describes the direction of its spin vector
along the direction of motion, and the neutrino being ``left-handed''
means that its spin vector always points in the opposite direction to its
momentum vector.  The fact that the neutrino is left-handed, written as 
$\nu_L$, implies that it
must be massless. If the neutrino has mass then, according to special
relativity, it can never travel at the speed of light. In principle,
a fast moving observer could therefore overtake the
spinning massive neutrino and would see it moving in the opposite
direction. To the observer, the massive neutrino would therefore
appear right-handed. Since the Standard Model predicts that neutrinos must
be strictly left-handed, it follows that neutrinos are massless
in the Standard Model. It also follows that the discovery of neutrino
mass implies new physics beyond the SM, with
profound implications for particle physics and cosmology.

Neutrinos are massless in the Standard Model for three independent reasons:
\begin{itemize}
\item There are no right-handed neutrinos $\nu_R$
\item There are only Higgs doublets (and no Higgs triplets) of $SU(2)_L$ 
\item There are only renormalisable term
\end{itemize}
In the SM, the three massless neutrinos $\nu_e$, $\nu_{\mu}$, $\nu_{\tau}$ are
distinguished by separate lepton numbers $L_e$, $L_{\mu}$,
$L_{\tau}$. Neutrinos and antineutrinos are distinguished by total conserved
lepton number 
$L=L_e+L_{\mu}+L_{\tau}$. To generate neutrino mass we must relax
one or more of the above three conditions. For example, by adding
right-handed neutrinos the Higgs mechanism of the Standard Model
can give neutrinos the same type of mass as the Dirac electron mass or
other charged lepton and quark masses, which would generally break the separate
lepton numbers $L_e$, $L_{\mu}$, $L_{\tau}$, but preserve the total lepton
number $L$. However it is also possible for neutrinos to have a new type of mass
of a type first proposed by Majorana, which would also break $L$. 
There exists a special case where total lepton number $L$ is broken, but the
combination $L_e-L_{\mu}-L_{\tau}$ is conserved; such a symmetry would give
rise to a neutrino mass matrix with an inverted mass spectrum.

From the theoretical perspective, the main unanswered question is the origin 
of neutrino mass, and in particular the smallness of neutrino mass. 
The simplest possibility is that neutrinos have Dirac mass
just like the electron mass in the SM, namely due to a term like
$y_D\overline{L}H\nu_R$, where $L$ is a lepton doublet containing $\nu_L$, $H$
is a Higgs doublet and $\nu_R$ is a right-handed neutrino. 
The observed smallness of neutrino masses implies that the Dirac Yukawa coupling $y_D$ must be 
of order $10^{-12}$ to achieve a Dirac neutrino mass of about 0.1~eV. 
Advocates of Dirac masses point out that the electron mass already requires a
Yukawa coupling $y_e$ of about $10^{-6}$, so we are used to such small Yukawa
couplings. In this case, 
all that is required is to add right-handed neutrinos $\nu_R$ to the SM and we are done.
Well, almost. It still needs to be explained why the $\nu_R$ have zero Majorana mass,
after all they are gauge singlets and so nothing prevents them acquiring
(large) Majorana mass terms $M_{RR} \nu_R \nu_R$ where $M_{RR}$ could be as
large as the Planck scale. 
Moreover, Majorana masses offer a unique (and testable) way to generate
neutrino masses (since neutrinos do not carry electric charge) even without
right-handed neutrinos. 
The simplest way to generate Majorana mass is via $y_M \Delta L L $
where $\Delta$ is a Higgs triplet and $y_M$ is a Yukawa coupling associated with Majorana mass. 
Alternatively,
at the effective level, Majorana neutrino mass can result from some additional
dimension 5 operators which couple two lepton doublets $L$ to two Higgs
doublets $H$  first proposed by Weinberg~\cite{Weinberg:1980bf},
\beq
-\frac{1}{2}HL^T\kappa HL ,
\label{dim5}
\eeq
where $\kappa$ has dimension $[\mathrm{mass}]^{-1}$. 
This is a non-renormalisable operator, so it violates one of the tenets of the SM. 
In order to account for a neutrino mass of order 0.1 eV requires $\kappa \sim 10^{-14}$ GeV$^{-1}$.
This suggests a new high energy mass scale $M$ in physics, a small dimensionless coupling associated
with $\kappa$, or both. There are basically five different proposals for the origin of neutrino mass:
\begin{itemize}
\item The see-saw mechanisms~\cite{seesaw,type2,Foot:1989type3} 
(Weinberg operator e.g. from large Majorana mass $M=M_{RR}$ for right-handed neutrinos $\nu_R$) 
\item $R$-parity violating supersymmetry~\cite{Drees:1997id}
(Weinberg operator from TeV scale Majorana mass for neutralinos $\chi$)  
\item TeV scale loop mechanisms~\cite{Zee:1980ai,Ma:2012ez}
(Majorana mass from extra Higgs doublets and singlets at the TeV scale)
\item Extra dimensions~\cite{Arkani-Hamed:1998vp} 
(Dirac mass with small $y_D$ due to right-handed neutrinos $\nu_R$ in the bulk)
\item String theory~\cite{Mohapatra:bd,Mohapatra:1986bd} (new mechanisms for generating large Majorana mass for right-handed neutrinos $\nu_R$
from Planck or string scale physics) 
\end{itemize}
These different mechanisms are reviewed in~\cite{Bandyopadhyay:2007kx}.
In this review we shall mainly focus on the see-saw mechanism which may be
incorporated into a theory of flavour. 

It has been one of the long standing goals of theories of particle
physics beyond the Standard Model to predict quark and lepton
masses and mixings. With the discovery of neutrino mass and
mixing, this quest has received a massive impetus. Indeed, perhaps
the greatest advance in particle physics over the past decade has
been the discovery of neutrino mass and mixing involving large
mixing. The largeness of the lepton mixing angles contrasts with the smallness of the
quark mixing angles, and this observation, together with the
smallness of neutrino masses, provides new and tantalising clues
in the search for the origin of quark and lepton flavour.
For example, it is amusing to note that the smallest lepton mixing may be related to the largest quark mixing,
$|U_{e3}|\approx \theta_C/\sqrt{2}$ where $\theta_C$ is the Cabibbo angle.
The quest to understand the origin of the three families of quarks and leptons
and their pattern of masses and mixing parameters is called the flavour puzzle,
and motivates the introduction of family symmetry. In particular, as we shall see,
lepton mixing provides a motivation for discrete family symmetry, which will
form the central part of this review. As we shall also see, such theories demand
a high precision knowledge of the lepton mixing angles, beyond that currently achieved.

\subsection{About this review}
It should be mentioned at the outset that at the time of writing this review
there were already good and fairly up to date reviews in the literature, for
example~\cite{Altarelli:2010gt,Grimus:2011fk,morisivalle}, although only the last one was
written after Daya Bay and RENO. 
Subsequently, two further reviews have appeared~\cite{King:2014nza,King:2015aea}.
It should be remarked that the signal of another independent mass
splitting from the LSND accelerator experiment~\cite{Athanassopoulos:1997pv}
would either require a further light neutrino state with no weak
interactions (a so-called ``light sterile neutrino'') or some other
non-standard physics.
This effect has not been confirmed by a similar
experiment KARMEN~\cite{Eitel:2000by}, and a subsequent 
experiment MiniBooNE~\cite{AguilarArevalo:2012va} has not decisively resolved the issue.
Since there is no solid evidence for light sterile neutrinos,
in this review we shall not discuss this subject any further, but
refer the interested reader to a recent dedicated discussion in~\cite{Abazajian:2012ys}.
Instead, in this review, we shall exclusively focus on the three active neutrino paradigm.

The starting point for the present review is the one that was written by one of us 
about ten years ago~\cite{King:2003jb}.
At that time it had just become apparent, after the first SNO results in 2002, that the solar mixing angle was large, 
which together with the large atmospheric mixing angle, meant that there were two large mixing angles in the
lepton sector. The solar mixing being large effectively killed many neutrino mass models that had previously
been proposed consistent with small solar mixing. 
This was actually the second great extinction of models, the first being 
after the discovery of a large atmospheric mixing angle
by Super-Kamiokande in 1998. A decade after the last review, in 2012 we are now in an analogous position,
namely that Daya Bay and RENO have just measured the reactor angle and shown it to be quite sizeable,
killing neutrino mass models consistent with a very small reactor angle.
Just as the review article in~\cite{King:2003jb}
was written shortly after the second great extinction in 2002,
so the present review article is being written just after the third great extinction in 2012,
so once again it is an opportune moment to
identify new and surviving model species which may come to dominate the theory landscape over the
coming years. In particular the fate of discrete family symmetry, which was to some extent
motivated by the possibility of the reactor mixing being zero, or very small, will be fully discussed.
We emphasise that we are now in the unique position in the history of
neutrino physics of knowing not only that neutrino mass is real,
and hence the Standard Model at least in its minimal formulation
is incomplete, but also we finally have information on all three mixing angles. 

As in the previous review, we focus on theoretical
approaches to understanding neutrino masses and mixings
in the framework of the {\em see-saw mechanism}, 
assuming three active neutrinos.
The goal of such models is to account for two very large mixing angles,
and one Cabibbo-sized mixing angle, and a pattern of neutrino masses consistent
with observation. We give a strong emphasis to classes of models where
large mixing angles can arise naturally and consistently
with a neutrino mass hierarchy.
We show that if one of the right-handed neutrinos 
contributes dominantly in the type I see-saw mechanism to the heaviest
neutrino mass, and a second right-handed neutrino contributes
dominantly to the second heaviest neutrino mass, then large
atmospheric and solar mixing angles may be interpreted
as simple ratios of Yukawa couplings. This is of course the 
sequential dominance (SD) mechanism~\cite{King:1998jw,King:2002nf,King:2002qh}. 
Sequential dominance is not a model, it is a mechanism
in search of a model. The conditions for sequential
dominance, such as ratios of Yukawa couplings being of order unity
for large mixing angles, and the required pattern of right-handed
neutrino masses are put in by hand and require further
theoretical input such as family symmetry. 
It is interesting to note that, without further constraints,  SD generically has long
predicted hierarchical neutrinos with a normal ordering and large reactor angle of order 
$|U_{e3}| \sim {\cal O} (m_2/m_3)\sim 0.2$.
However, in order to achieve more precise predictions, SD needs to be combined with family symmetry.

The use of {\em discrete} family symmetry was mainly motivated by the
hypothesis of exact tri-bimaximal (TB) mixing~\cite{Harrison:2002er} defined by:
\bea
\label{TB-pre}
 |U_{e3}| &=& 0,  \nonumber \\
|U_{\mu 3}|&=& |U_{\tau 3}|= 1/\sqrt{2}, \nonumber \\
|U_{e 2}|&=& |U_{\mu 2}|= |U_{\tau 2}| =1/\sqrt{3}.
\eea
The TB mixing and discrete family symmetry approach gained much impetus over the past decade.
Given that the measurement of the reactor mixing $|U_{e3}|\approx 0.15$ 
by Daya Bay and RENO kills exact TB mixing,
the obvious question is what is the impact on the discrete family symmetry approach, which was largely
inspired by TB mixing. This timely question will be addressed by the present review.
There are a huge number of proposals in the literature,
but not so many surviving the measurement of the reactor angle.
The simple answer to the question is the discrete family symmetry approach is alive and kicking after 
Daya Bay and RENO. However, theorists have been forced to work harder to go beyond the 
simple mixing pattern of TB mixing which is now excluded. The simple discrete family symmetries
proposed to account for TB mixing may still be viable as a leading order (LO) approximation,
but higher order (HO) corrections may play a more important role than anticipated in many models.
Alternatively, perhaps larger finite groups are relevant where the LO approximation already predicts 
a non-zero reactor angle. Yet another possibility is that the discrete family symmetry may be implemented
indirectly, as in the sequential dominance approach, in new ways.
All these interesting possibilities will be discussed.

The layout of the remainder of the review article is as follows.
In Section~\ref{sec:2} we introduce and review the current status of
neutrino masses and mixing angles. We also parameterise the
PMNS mixing matrix in two different ways, whose equivalence
is discussed in Appendix~\ref{equivalence}. In Section~\ref{sec:patterns} we discuss patterns
of lepton mixing that have been proposed, starting with 
simple mixing patterns such as bimaximal (BM), tri-bimaximal (TB),
bi-trimaximal (BT) and golden ratio (GR) mixing.
These all may apply to the neutrino mixing, which is then corrected by charged
lepton mixing corrections, to give acceptable PMNS mixing, leading to solar
mixing angle sum rules. 
The closeness of the TB mixing pattern to the data suggests a parametrisation
of the PMNS matrix in terms of deviations from TB mixing, which is also introduced.
Using these deviations, several TB variants are introduced and discussed,
including tri-bimaximal-reactor (TBR) mixing and trimaximal (TM) mixing in two forms,
namely where the first and second columns of the PMNS matrix take the TB values,
called TM1 and TM2, respectively.
Section~\ref{sec:seesaw} is devoted to the see-saw mechanisms, which are
central to this review, in both the simplest versions, called the type~I, and
including other types~II and~III, as well as alternative versions. We show how
the type~I see-saw mechanism may be applied to the hierarchical case in a very
natural way using sequential dominance. 
Section~\ref{sec:nut} contains a mini-review of finite group theory and may be skipped 
by those readers who are already familiar with this subject. In
Section~\ref{sec:FS} we give a pedagogical introduction to discrete family
symmetry, and its direct or indirect implementation in model building.
Section~\ref{sec:DmB} is devoted to the direct model building approach in
which different subgroups of the discrete family symmetry are preserved in the
neutrino and charged lepton sectors, and discusses the associated 
vacuum alignments arising from the breaking of the discrete family symmetry
using flavons. We also discuss the model building strategies following Daya
Bay and RENO. 
Section~\ref{sec:ImB} contains an analogous discussion for the indirect
approach in which the discrete family symmetry is completely broken by
flavons, but special vacuum alignments lead to particular mixing patterns,
including new viable patterns with a non-zero reactor angle. 
In Section~\ref{sec:gut} we briefly review grand unified theories (GUTs) such
as $SU(5)$ and how they may be combined 
with discrete family symmetry in order to account for all quark and lepton
masses and mixing.  
In Section~\ref{sec:examp} we discuss three model examples which combine an
$SU(5)$ GUT with the discrete family symmetries $A_4$, $S_4$ and
$\Delta(96)$. Section~\ref{sec:conclude} concludes the review.
We also present appendices dealing with more
technical issues which may provide useful model building tools.
Appendix~\ref{equivalence}
proves the equivalence between different parametrisations
of the neutrino mixing matrix and gives a useful dictionary.
Appendix~\ref{deviations} gives the full three family
neutrino oscillation formula in terms of deviations from TB mixing.
Appendix~\ref{app:CGs} catalogues the generators and Clebsch-Gordan
coefficients of $A_4$, $S_4$ and $T_7$.

\section{\label{sec:2}Neutrino masses and mixing angles}

The history of neutrino oscillations dates back to the work of
Pontecorvo who in 1957~\cite{Pontecorvo:1957qd}
proposed $\nu \rightarrow \bar{\nu}$ oscillations in 
analogy with $K \rightarrow \bar{K}$ oscillations,
described as the mixing of two Majorana neutrinos.
Pontecorvo was the first to
realise that what we call the ``electron neutrino'' for example is
really a linear combination of mass eigenstate neutrinos, and that
this feature could lead to neutrino oscillations
of the kind $\nu_e \rightarrow \nu_{\mu}$. Later on MSW proposed that such
neutrino oscillations could be resonantly enhanced in the Sun~\cite{MSW}. The
present section introduces the basic formalism 
of neutrino masses and mixing angles, and gives an up-to-date
summary of the current experimental status of this fast moving field.

\subsection{Three neutrino mixing ignoring phases}
The minimal neutrino sector required to account for the
atmospheric and solar neutrino oscillation data consists of
three light physical neutrinos with left-handed flavour eigenstates,
$\nu_e$, $\nu_\mu$, and $\nu_\tau$, defined to be those states
that share the same electroweak doublet as the corresponding left-handed
charged lepton mass eigenstates.
Within the framework of three-neutrino oscillations,
the neutrino flavour eigenstates $\nu_e$, $\nu_\mu$, and $\nu_\tau$ are
related to the neutrino mass eigenstates $\nu_1$, $\nu_2$, and $\nu_3$
with mass $m_1$, $m_2$, and $m_3$, respectively, by a $3\times3$
unitary matrix called the lepton mixing matrix $U_{\mathrm{PMNS}}$ introduced in Eq.~(\ref{MNS0}).

Assuming the light neutrinos are Majorana,
$U_{\mathrm{PMNS}}$ can be parameterised in terms of three mixing angles
$\theta_{ij}$ and three complex phases $\delta_{ij}$.
A unitary matrix has six phases but three of them are removed
by the phase symmetry of the charged lepton masses.
Since the neutrino masses are Majorana there is no additional
phase symmetry associated with them, unlike the case of quark
mixing where a further two phases may be removed.

To begin with, let us suppose that the phases are zero. Then
the lepton mixing matrix may be parametrised by
a product of three Euler rotations,
\begin{equation}
U_{\mathrm{PMNS}}=R_{23}R_{13}R_{12},
\label{euler}
\end{equation}
where
\begin{equation}
R_{23}=
\left(\begin{array}{ccc}
1 & 0 & 0 \\
0 & c_{23} & s_{23} \\
0 & -s_{23} & c_{23} \\
\end{array}\right), \qquad
R_{13}=
\left(\begin{array}{ccc}
c_{13} & 0 & s_{13} \\
0 & 1 & 0 \\
-s_{13} & 0 & c_{13} \\
\end{array}\right), \qquad
R_{12}=
\left(\begin{array}{ccc}
c_{12} & s_{12} & 0 \\
-s_{12} & c_{12} & 0\\
0 & 0 & 1 \\
\end{array}\right),
\end{equation}
with $c_{ij} = \cos\theta_{ij}$ and $s_{ij} = \sin\theta_{ij}$.
Note that the allowed range of the angles is
$0\leq \theta_{ij} \leq \frac{\pi}{2}$.

\begin{figure}[htb]
\begin{center}
\includegraphics[width=5in]{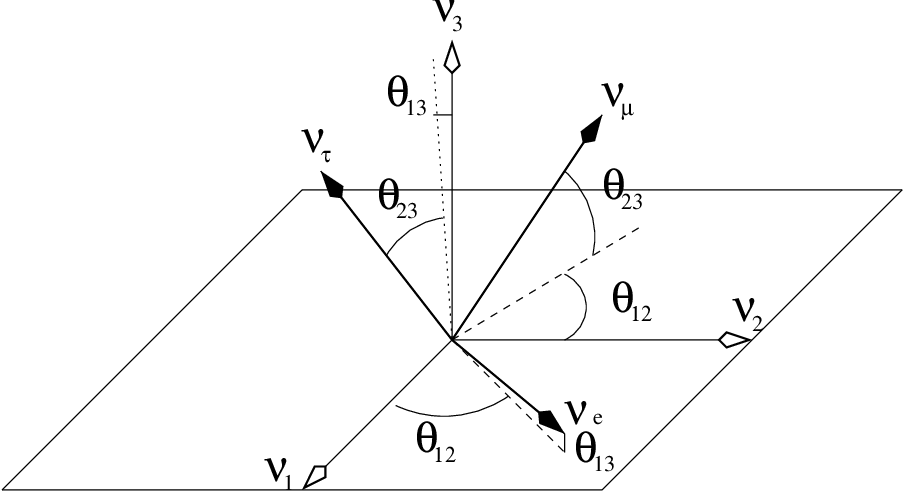}
\end{center}
\caption{\label{fig2}\small{The relation between
the neutrino flavour eigenstates $\nu_e$, $\nu_\mu$, and $\nu_\tau$ and
the neutrino mass eigenstates $\nu_1$, $\nu_2$, and $\nu_3$
in terms of the three mixing angles $\theta_{12}$,
$\theta_{13}$, $\theta_{23}$.}}
\end{figure}

Ignoring phases, the relation between the
neutrino flavour eigenstates $\nu_e$, $\nu_\mu$, and $\nu_\tau$ and the
neutrino mass eigenstates $\nu_1$, $\nu_2$, and $\nu_3$ is therefore given
as a product of three Euler rotations in Eq.~(\ref{euler}) as depicted in
Fig.~\ref{fig2}.

\subsection{Atmospheric neutrino mixing}

In 1998, the Super-Kamiokande experiment published a
paper~\cite{SK} 
which represents a watershed in the history of neutrino physics.
Super-Kamiokande measured the number of
electron and muon neutrinos that arrive at the Earth's surface as a
result of cosmic ray interactions in the upper atmosphere, which are
referred to as ``atmospheric neutrinos''. While the number and
angular distribution of electron neutrinos is as expected,
Super-Kamiokande showed that the number of muon neutrinos is
significantly smaller than expected and that the flux of muon
neutrinos exhibits a strong dependence on the zenith angle. These
observations gave compelling evidence that muon neutrinos undergo
flavour oscillations and this in turn implies that at least one
neutrino has a non-zero mass. The standard interpretation is
that muon neutrinos are oscillating into tau neutrinos.

As a first approximation, one can set the reactor angle $\theta_{13}$ to zero, 
and assume that $|\Delta m_{32}^2|\gg |\Delta m_{21}^2|$.
Current atmospheric neutrino oscillation data can then approximately be described
by simple two-state mixing,
\begin{equation}
\left(\begin{array}{c} \nu_\mu \\ \nu_\tau \end{array} \\ \right)=
\left(\begin{array}{cc}
 \cos \theta_{23} & \sin \theta_{23} \\
 -\sin \theta_{23} & \cos \theta_{23} \\
\end{array}\right)
\left(\begin{array}{c}
\nu_2 \\ \nu_3 \end{array} \\ \right)
\; ,
\end{equation}
and the two-state oscillation formula
describing the probability that a $\nu_{\mu }$ converts to a $\nu_{\tau}$,
\be
P(\nu_{\mu }\rightarrow \nu_{\tau})=\sin^22\theta_{23}
\sin^2(1.27\Delta m_{32}^2{L}/{E}),\label{eq:atm-osci}
\ee
where
\begin{equation}
\Delta m^2_{ij} \equiv m^2_i - m^2_j \; ,
\end{equation}
and $m_i$ are the physical neutrino mass eigenvalues associated
with the mass eigenstates $\nu_i$.
$\Delta m_{32}^2$ is in units of eV$^2$, the baseline $L$ is in km and
the beam energy $E$ is in GeV. 
Note that the sign of $\Delta m_{32}^2$, and thus the mass ordering,
cannot be determined from Eq.~\eqref{eq:atm-osci}.

The data can be well described by approximately maximal atmospheric mixing
$|U_{\mu 3}|\approx  |U_{\tau 3}|\approx1/\sqrt{2}$.
This corresponds to 
\beq
\sin \theta_{23}\approx 1/\sqrt{2},
\eeq
which means that the angle $\theta_{23}\approx \pi/4 \ \mathrm{or} \ 45^\circ $ and
we identify the heavy ``atmospheric neutrino'' of mass $m_3$ as being approximately
\beq
\nu_3\approx \frac{\nu_{\mu}+\nu_{\tau}}{\sqrt{2}}.
\eeq

\subsection{Solar neutrino mixing}

Super-Kamiokande was also sensitive to the electron neutrinos arriving
from the Sun, the ``solar neutrinos'', and independently confirmed
the reported deficit of such solar neutrinos long reported by other
experiments.  For example Davis's Homestake Chlorine experiment
which began data taking in 1970 consisted of 615 tons of
tetrachloroethylene, and uses radiochemical techniques to determine
the $^{37}$Ar production rate. The SAGE and Gallex
experiments contained large amounts of $^{71}$Ga which is converted to $^{71}$Ge
by low energy electron neutrinos arising from the dominant $pp$ reaction
in the Sun.  The combined data from these and other experiments
implied an energy dependent suppression of solar neutrinos which was
interpreted as due to flavour oscillations. Taken together with the
atmospheric data, this required that a second neutrino has a
non-zero mass. The standard interpretation is that the electron
neutrinos $\nu_e$ disappear due to an oscillation 
formula which involves a ``solar neutrino'' of mass $m_2$ given approximately by
\beq
\nu_2\approx \frac{\nu_{e}+\nu_{\mu}-\nu_{\tau}}{\sqrt{3}},
\eeq
consistent with trimaximal solar mixing 
$|U_{e 2}|\approx |U_{\mu 2}|\approx |U_{\tau 2}| \approx 1/\sqrt{3}$.
This corresponds to 
\beq
\sin \theta_{12}\approx 1/\sqrt{3},
\eeq
or $\theta_{12}\approx 35^\circ$.

SNO measurements of charged current (CC) reaction on deuterium were
sensitive exclusively to $\nu_e$, while the neutral current (NC) reaction as well
as the elastic scattering (ES) off electrons were also sensitive to
$\nu_{\mu}$ and $\nu_{\tau}$. 
The neutrino flux derived from the CC reactions was 
significantly smaller than the one obtained from NC and ES.
This immediately disfavoured oscillations of $\nu_e$ to
sterile neutrinos which would lead to a diminished flux of electron
neutrinos, but equal CC, NC and ES fluxes.  On the other hand the
observations were consistent with oscillations of $\nu_e$ to
active neutrinos $\nu_{\mu}$ and $\nu_{\tau}$ since this would
lead to a larger NC and ES rate. The SNO analysis was nicely consistent with
both the hypothesis that electron neutrinos from the Sun oscillate into other
active flavours, and with the Standard Solar Model prediction. The
results from SNO including the data taken with salt inserted
into the detector to boost the efficiency of detecting the NC
events~\cite{Ahmed:2003kj},  
strongly favoured the large mixing angle (LMA) MSW
solution.  In other words, after SNO, there was no longer
any solar neutrino problem: we had instead solar neutrino mass $m_2$!

KamLAND was a more powerful reactor experiment that measured
$\bar{\nu}_e$ produced by surrounding nuclear reactors. KamLAND observed
a signal of neutrino oscillations over the LMA MSW mass range,
confirming the LMA MSW region ``in the laboratory''~\cite{Eguchi:2002dm}.
KamLAND and SNO results when combined
with other solar neutrino data especially that of Super-Kamiokande
uniquely specify the LMA MSW~\cite{MSW} solar solution
with three active light neutrino states, and approximately trimaximal solar
mixing.
This solution, which requires a careful treatment of the
matter effects and the resulting asymmetry between the coherent forward
scattering of the different neutrino flavour states, furthermore determines the sign of
the mass squared splitting $\Delta m^2_{21}=m^2_2-m^2_1$  to be positive.
KamLAND has thus not only confirmed solar neutrino oscillations,
but has also uniquely specified the LMA solar
solution, heralding a new era in neutrino physics.

\subsection{Reactor neutrino mixing}
Until recently, the reactor angle $\theta_{13}$ was not measured, only limited by 
CHOOZ, a reactor experiment that failed to see any signal of
neutrino oscillations over the Super-Kamiokande mass range.  CHOOZ
data from $\bar{\nu}_{e}\rightarrow \bar{\nu}_{e}$ disappearance not
being observed provided a significant constraint on $\theta_{13}$ over
the Super-Kamiokande preferred range 
of $\Delta m_{32}^2$~\cite{Apollonio:1999ae}:
\begin{equation}
\sin^22\theta_{13}<0.2.
\end{equation}

Direct evidence for $\theta_{13}$ was first provided by T2K, MINOS and Double
Chooz~\cite{Abe:2011sj}. 
Recently the Daya Bay~\cite{DayaBay}, RENO~\cite{RENO}, and Double
Chooz~\cite{DCt13} collaborations have measured $\sin^2(2\theta_{13})$:
\begin{eqnarray}
\label{t13-exp}
\begin{array}{cl}
\mathrm{Daya \ Bay: } & \sin^2(2\theta_{13})=0.084\pm 0.005
\mathrm{(stat.~\&~syst.)}
\ ,\\
\mathrm{RENO: }    & \sin^2(2\theta_{13})= 0.082\pm 0.009 \mathrm{(stat.)}\pm 0.006
\mathrm{(syst.)\ ,}\\
\mathrm{Double \ Chooz: }~~& \sin^2(2\theta_{13})=0.090^{+0.032}_{-0.029}
\mathrm{(syst.~\&~stat.)}\ .\\
\end{array}
\end{eqnarray}
This corresponds to 
\beq
|U_{e3}|=\sin \theta_{13}\approx 0.15,
\eeq
or a reactor angle $\theta_{13}\approx 8.5^\circ$.

\subsection{Three neutrino mixing including phases}
If the reactor angle were zero then there would be no CP violation in neutrino oscillations.
The measurement of the reactor angle means that we cannot ignore the presence
of phases any more.
Including the phases, assuming the light neutrinos are Majorana,
$U_{\mathrm{PMNS}}$ can be parameterised in terms of three mixing angles
$\theta_{ij}$, a Dirac phase $\delta$, together with
two Majorana phases $\beta_1,\beta_2$, as follows~\cite{Nakamura:2010zzi},
\begin{equation}
U_{\mathrm{PMNS}}=R_{23}U_{13}R_{12}P_{12},
\end{equation}
where
\be
U_{13} ~ =~
\left(\begin{array}{ccc}
c_{13} & 0 & s_{13}e^{-i\delta} \\
0 & 1 & 0 \\
-s_{13}e^{i\delta} & 0 & c_{13} \\
\end{array}\right), \qquad
P_{12}  ~=~  
\left(\begin{array}{ccc}
e^{i\beta_1} & 0 & 0 \\
0 & e^{i\beta_2} & 0\\
0 & 0 & 1 \\
\end{array}\right),
\label{MNS}
\ee
and $R_{23}$ and $R_{12}$ were defined below Eq.~(\ref{euler}), giving,

\bea
 \label{eq:matrix}
U_{\mathrm{PMNS}} & = &
\left(\begin{array}{ccc}
    c_{12} c_{13}
    & s_{12} c_{13}
    & s_{13} e^{-i\delta}
    \\
    - s_{12} c_{23} - c_{12} s_{13} s_{23} e^{i\delta}
    & \hphantom{+} c_{12} c_{23} - s_{12} s_{13} s_{23}
    e^{i\delta}
    & c_{13} s_{23} \hspace*{5.5mm}
    \\
    \hphantom{+} s_{12} s_{23} - c_{12} s_{13} c_{23} e^{i\delta}
    & - c_{12} s_{23} - s_{12} s_{13} c_{23} e^{i\delta}
    & c_{13} c_{23} \hspace*{5.5mm}
    \end{array}\right)P_{12}.
\eea

Alternatively the lepton mixing matrix
may be expressed as a product of three complex Euler rotations~\cite{Schechter:1979bn},
\begin{equation}
\label{123}
U_{\mathrm{PMNS}}=U_{23}U_{13}U_{12},
\end{equation}
where
\begin{equation}
U_{23}=
\left(\begin{array}{ccc}
1 & 0 & 0 \\
0 & c_{23} & s_{23}e^{-i\delta_{23}} \\
0 & -s_{23}e^{i\delta_{23}} & c_{23} \\
\end{array}\right),
\end{equation}

\begin{equation}
U_{13}=
\left(\begin{array}{ccc}
c_{13} & 0 & s_{13}e^{-i\delta_{13}} \\
0 & 1 & 0 \\
-s_{13}e^{i\delta_{13}} & 0 & c_{13} \\
\end{array}\right),
\end{equation}

\begin{equation}
U_{12}=
\left(\begin{array}{ccc}
c_{12} & s_{12}e^{-i\delta_{12}} & 0 \\
-s_{12}e^{i\delta_{12}} & c_{12} & 0\\
0 & 0 & 1 \\
\end{array}\right).
\end{equation}

The equivalence of different parametrisations of the 
lepton mixing matrix, and
the relation between them is discussed in~\cite{King:2002nf}
with the results summarised in Appendix~\ref{equivalence}.
If the neutrinos are Dirac, then the phases $\beta_1=\beta_2=0$,
but the phase $\delta$ remains.

\subsection{\label{sec:glob}Global fits}

\begin{table}[t]
\centering
   \catcode`?=\active \def?{\hphantom{0}}
  \begin{tabular}{|c|c|c|c|}
    \hline
    parameter & Forero et al   &  Capozzi et al  &  Gonzalez-Garcia et al
    \\
    \hline &&& \\[-4mm]
$\Delta m^2_{21}  [10^{-5}\mathrm{eV^2}]$
& $7.60^{+0.19}_{-0.18}$ 
&  $7.37^{+0.17}_{-0.16}$ 
&  $7.50^{+0.19}_{-0.17}$  \\[3.5mm] 
$\Delta m^2_{31}   [10^{-3}\mathrm{eV^2}]$
    &
    \begin{tabular}{c}
      $2.48^{+0.05}_{-0.07}$\\[1mm]
      $-2.38^{+0.06}_{-0.05}$ \\    \end{tabular}
    &
    \begin{tabular}{c}
      $2.50^{+0.04}_{-0.04}$\\[1mm]
      $-2.46^{+0.04}_{-0.05}$ \\
    \end{tabular}
    &
    \begin{tabular}{c}
     $2.457^{+0.047}_{-0.047}$\\[1mm]
      $-2.449^{+0.048}_{-0.047}$
    \end{tabular}
    \\[6.8mm] 
$\sin^2\theta_{12}$
& $0.323^{+0.016}_{-0.016}$ 
& $0.297^{+0.017}_{-0.016}$ 
& $0.304^{+0.013}_{-0.012}$ \\[3.8mm]  
$\sin^2\theta_{23}$
    &
    \begin{tabular}{c}
      $0.567^{+0.032}_{-0.124}$\\[1mm]
      $0.573^{+0.025}_{-0.039}$ \\    
    \end{tabular}
    &
\begin{tabular}{c}
      $0.437^{+0.033}_{-0.020}$ \\[1mm]
      $0.569^{+0.028}_{-0.051}$\\    
    \end{tabular}
    &
    \begin{tabular}{c}
       $0.452^{+0.052}_{-0.028}$\\[1mm]
      $0.579^{+0.025}_{-0.037}$ \\        \end{tabular}
    \\[6.8mm] 
 $\sin^2\theta_{13}$
    &
    \begin{tabular}{c}
      $0.0226^{+0.0012}_{-0.0012}$\\[1mm]
      $0.0229^{+0.0012}_{-0.0012}$ \\      
               \end{tabular}
    &
    \begin{tabular}{c}
     $0.0214^{+0.0011}_{-0.0009}$\\[0.8mm]
      $0.0218^{+0.0009}_{-0.0012}$ \\         \end{tabular}
    &
    \begin{tabular}{c} 
    $0.0218^{+0.0010}_{-0.0010}$\\[1mm]
     $0.0219^{+0.0011}_{-0.0010}$\\ \end{tabular}
\\[6mm]
             $\delta$
   &
  \begin{tabular}{c}
      $(1.41^{+0.55}_{-0.40} )\pi$\\[0.8mm]
      $(1.48^{+0.31}_{-0.31})\pi$ \\     
   \end{tabular}
   &
   \begin{tabular}{c}
      $(1.35^{+0.29}_{-0.22})\pi$\\[1mm]
      $(1.32^{+0.35}_{-0.25})\pi$ \\     
   \end{tabular}
   &
  \begin{tabular}{c}
      $(1.70^{+0.22}_{-0.39})\pi$\\[1mm]
      $(1.41^{+0.35}_{-0.34})\pi$ \\     
   \end{tabular}
   \\[4mm]
           \hline
     \end{tabular}
\caption{\label{tab:summary}\small{Neutrino oscillation parameters
    summary. For $\Delta m^2_{31}$, $\sin^2\theta_{23}$, $\sin^2\theta_{13}$,
    and $\delta$ the upper (lower) row corresponds to normal (inverted)
    neutrino mass ordering. The best fit values and 1$\sigma$ errors are
    shown. The subtleties associated with these numbers are discussed in the
    respective references Forero et al~\cite{Tortola:2012te}, Capozzi et al~\cite{Fogli:2012ua} and Gonzalez-Garcia et al~\cite{GonzalezGarcia:2012sz}. In
    particular~\cite{GonzalezGarcia:2012sz} quotes two different global fits,
    depending on the assumptions made about reactor fluxes, where we only
    quote the first (``free flux"). Furthermore, the precise definition of the
    atmospheric neutrino mass splitting $\Delta m^2_{31}$ differs slightly
    between the three global fits.}}
\end{table}

\begin{figure}[t]
\centerline{
\includegraphics[height=37mm]{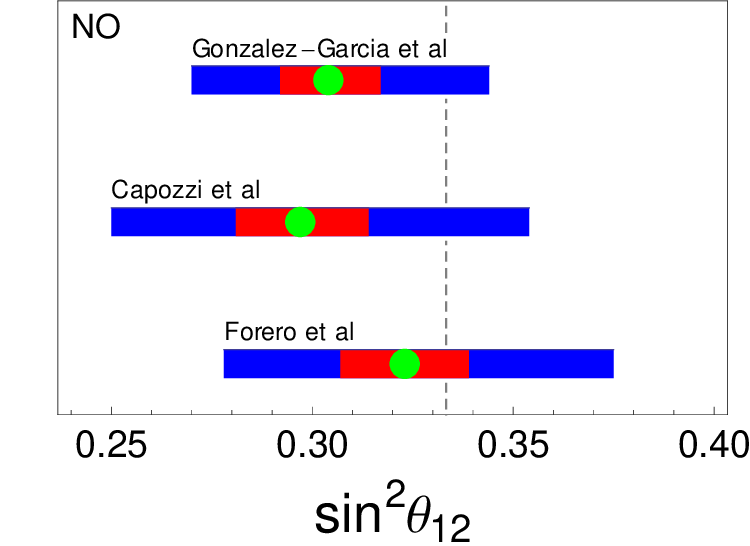}
\includegraphics[height=37mm]{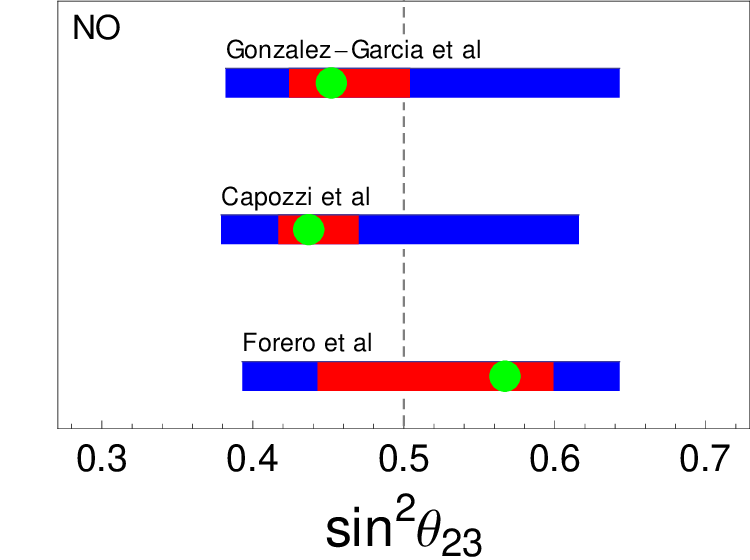}
\includegraphics[height=37mm]{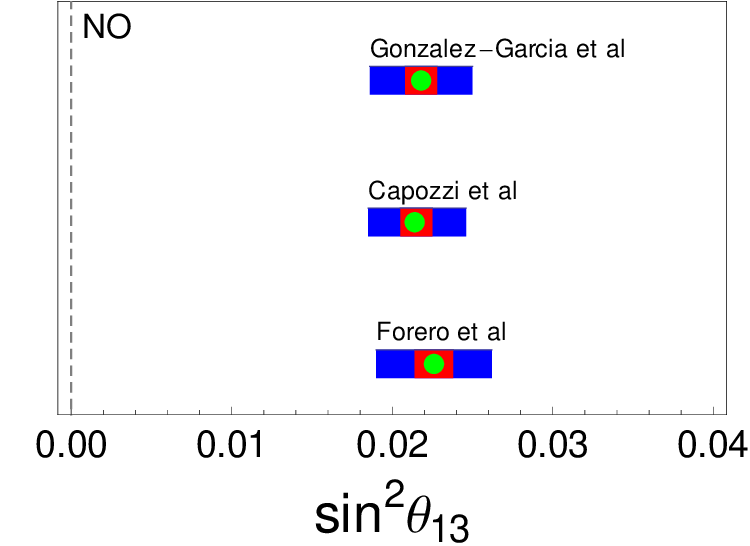}~~~}

\vspace{4mm}

\centerline{
\includegraphics[height=37mm]{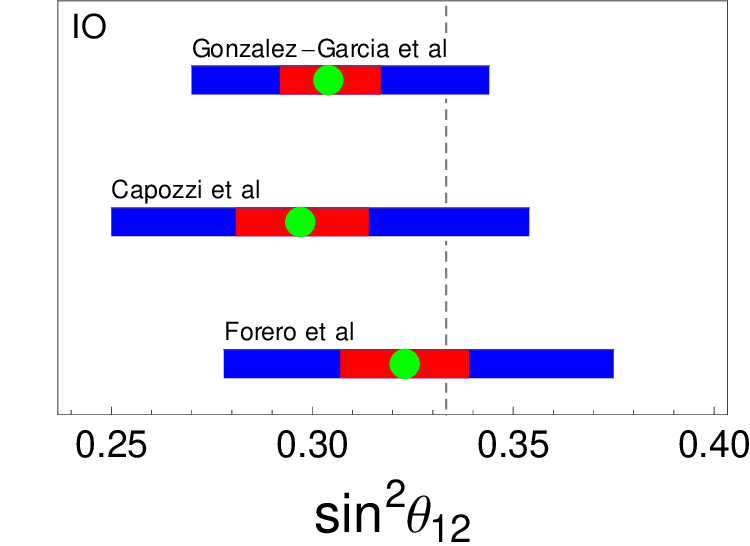}
\includegraphics[height=37mm]{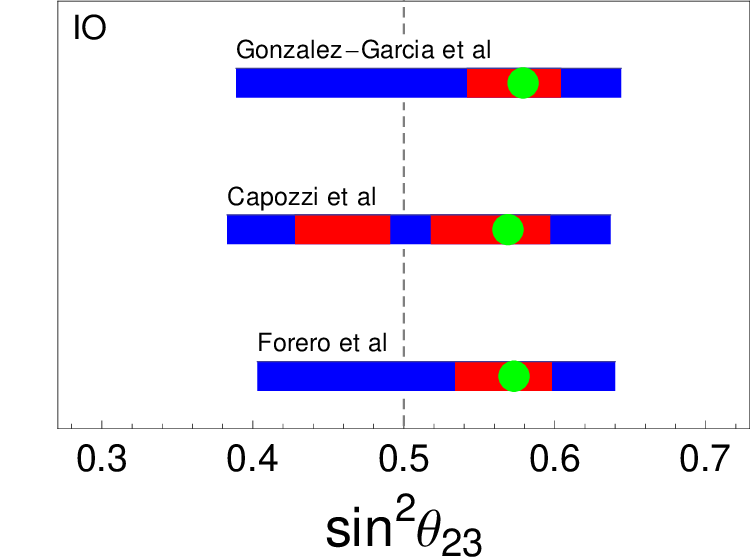}
\includegraphics[height=37mm]{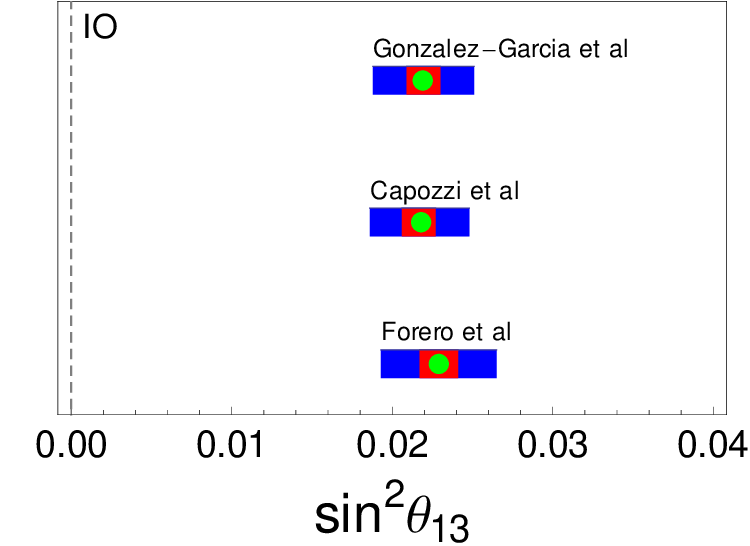}~~~}
\caption{\label{fig:mixingfits}\small{The mixing angles obtained from the three
  global fits~\cite{Tortola:2012te,Fogli:2012ua,GonzalezGarcia:2012sz}. The
  upper three panels correspond to the results for normal 
  neutrino mass ordering (NO), while the lower three panels are for an inverted
  mass ordering (IO). Shown are the best fit values (green) as well as the 1$\sigma$
  (red) and 3$\sigma$ (blue) intervals. Note that the solar angle is
  insensitive to the mass ordering.}}
\end{figure}

 In Table~\ref{tab:summary} we give the global fits of the neutrino mixing
 parameters. For $\Delta m^2_{31}$, $\sin^2\theta_{23}$, $\sin^2\theta_{13}$, 
and $\delta$ the upper (lower) row corresponds to normal (inverted)
neutrino mass ordering. The best fit values and 1$\sigma$ errors are shown from 
Forero et al~\cite{Tortola:2012te},
Capozzi et al~\cite{Fogli:2012ua}
and Gonzalez-Garcia et al~\cite{GonzalezGarcia:2012sz}.
The results for the mixing angles are graphically contrasted in
Fig.~\ref{fig:mixingfits}. 
We emphasise that this compilation is predominantly meant to illustrate
some possibilities arising from present global fits. The reader is referred to the
respective references for the subtleties associated with these numbers.

For the normal mass ordering, we shall take the average values and errors
which approximately encompass the one sigma ranges of all three global fits
(ignoring the best fit value of $\theta_{23}$ in the second octant found by
Forero et al~\cite{Tortola:2012te} in favour of the local minimum at
$\sin^2\theta_{23}=0.473$): 
 \bea
\sin^2\theta_{12}&=&0.31\pm 0.02, \\
\sin^2\theta_{23}&=&0.45\pm 0.05, \\
\sin^2\theta_{13}&=& 0.022\pm 0.002.
\eea
These values may be compared to the tri-bimaximal predictions 
$\sin^2\theta_{12}=0.33$, $\sin^2\theta_{23}=0.5$ and $\sin^2\theta_{13}=0$,
showing that TB mixing is excluded by the reactor angle.
Alternatively we may write, remembering that these are one sigma ranges in the squares of the sines and not the
sines themselves,
 \bea
\sin \theta_{12}&=&0.56 \pm 0.02, \label{s12}\\
\sin \theta_{23}&=&0.67 \pm 0.04,  \label{s23}\\
\sin \theta_{13}&=& 0.15 \pm 0.01  \label{s13}.
\eea
In terms of the angles themselves we have, approximately, in round figures,
 \bea
\theta_{12}&=&34^\circ\pm 1^\circ, \label{t12}\\
\theta_{23}&=&42^\circ\pm 3^\circ, \label{t23}\\
\theta_{13}&=& 8.5^\circ\pm 0.5^\circ \label{t13}.
\eea
A few comments are relevant about these angles. Firstly the errors are not
linear, since, for one thing, the global fits are made in terms of the squares
of the sines of the angles. 
Having said this, in the case of normal neutrino mass ordering, the
atmospheric and solar mixing angles are still consistent with their TB values,
but there is a preference for both to be slightly smaller (i.e. $\theta_{23}
\lesssim45^\circ$ and $\theta_{12} \lesssim35.26^\circ$).

\section{\label{sec:patterns}Patterns of lepton mixing and sum rules}

\subsection{\label{sec:patterns-1}Simple forms of neutrino mixing}
Below we give three  examples of simple patterns of mixing in the neutrino sector which all
have $s_{13}=0$ and $s_{23}=c_{23}=1/\sqrt{2}$.
Inserting these values in Eq.~(\ref{euler}) we obtain
a PMNS matrix of the form,
\begin{eqnarray}
U_0 =
\left( \begin{array}{ccc}
c_{12} & s_{12} & 0\\
 -\frac{s_{12}}{\sqrt{2}}  &  \frac{c_{12}}{\sqrt{2}} & \frac{1}{\sqrt{2}} \\
\frac{s_{12}}{\sqrt{2}}  &  -\frac{c_{12}}{\sqrt{2}} & \frac{1}{\sqrt{2}} 
\end{array}
\right),
\label{GR}
\end{eqnarray}
where the zero subscript reminds us that this form has $\theta_{13}=0$ (and $\theta_{23}=45^\circ$).

For golden ratio (GR) mixing~\cite{Datta:2003qg}, 
the solar angle is given by
$\tan \theta_{12}=1/\phi$, where $\phi = (1+\sqrt{5})/2$ is the golden ratio
which implies $\theta_{12}=31.7^\circ$. 
There is an alternative version where 
$\cos \theta_{12} =\phi/2$ and $\theta_{12}=36^\circ$~\cite{Rodejohann:2008ir},
which we refer to as GR$'$ mixing. 

For bimaximal (BM) mixing (see e.g.~\cite{Davidson:1998bi,Altarelli:2009gn,Meloni:2011fx} and references therein), 
we insert $s_{12}=c_{12}=1/\sqrt{2}$ ($\theta_{12}=45^\circ$) into Eq.~(\ref{GR}), 
\begin{eqnarray}
U_{\mathrm{BM}} =
\left( \begin{array}{ccc}
\frac{1}{\sqrt{2}} & \frac{1}{\sqrt{2}} & 0\\
-\frac{1}{2}  & \frac{1}{2} & \frac{1}{\sqrt{2}} \\
\frac{1}{2}  & -\frac{1}{2} & \frac{1}{\sqrt{2}} 
\end{array}
\right).
\label{BM}
\end{eqnarray}

For tri-bimaximal (TB) mixing~\cite{Harrison:2002er}, alternatively we use
$s_{12}=1/\sqrt{3}$, $c_{12}=\sqrt{2/3}$ ($\theta_{12}=35.26^\circ$) in Eq.~(\ref{GR}),
\begin{eqnarray}
U_{\mathrm{TB}} =
\left( \begin{array}{ccc}
\sqrt{\frac{2}{3}} & \frac{1}{\sqrt{3}} & 0 \\
-\frac{1}{\sqrt{6}}  & \frac{1}{\sqrt{3}} & \frac{1}{\sqrt{2}} \\
\frac{1}{\sqrt{6}}  & -\frac{1}{\sqrt{3}} & \frac{1}{\sqrt{2}}
\end{array}
\right).
\label{TB}
\end{eqnarray}

These simple examples of neutrino mixing are now all excluded by the data.
However they may still play a role in model building and we will revisit them 
when we consider charged lepton and other corrections below.

\subsection{\label{sec:patterns-2}Deviations from tri-bimaximal mixing}

From a theoretical or model building point of view, one significance of the
global fits is that they exclude the tri-bimaximal lepton mixing
pattern~\cite{Harrison:2002er} in which the solar mixing angle is trimaximal, the
atmospheric angle is maximal and the reactor angle
vanishes. 
When comparing global fits to TB mixing it is convenient to express the $s$olar,
$a$tmospheric and $r$eactor angles in terms of deviation parameters ($s$, $a$ and
$r$) from TB mixing~\cite{King:2007pr}: 
\be
\label{rsadef}
\sin \theta_{12}=\frac{1}{\sqrt{3}}(1+s),\ \ \ \ 
\sin\theta_{23}=\frac{1}{\sqrt{2}}(1+a), \ \ \ \ 
\sin \theta_{13}=\frac{r}{\sqrt{2}}.
\ee
A related expansion is given in~\cite{Pakvasa:2007zj}.
From these definitions we can write,
\be
\label{rsadef2}
s= \sqrt{3} \sin \theta_{12} -1,\ \ \ \ 
a= \sqrt{2} \sin \theta_{23} -1,\ \ \ \ 
r= \sqrt{2} \sin \theta_{13}.
\ee
Using this last form, and Eqs.~(\ref{s12},\ref{s23},\ref{s13}), we find the
following values and ranges for the TB deviation parameters:
\beq
s=-0.03\pm 0.03, \ \ \ \ a=-0.05 \pm 0.05, \ \ \ \ r=0.21\pm 0.01,
\label{rsafit}
\eeq
assuming a normal neutrino mass ordering.
As well as showing that TB is excluded by the reactor angle being non-zero, 
Eq.~(\ref{rsafit}) shows a slight preference (at the one sigma level) for the
atmospheric angle to be below its maximal value and also a slight preference
(at the one sigma level) for the solar angle to be below its trimaximal value. 
In general, this parametrisation shows that both the solar and the atmospheric
angles must be quite close to their TB values, while the reactor angle is
necessarily very far from zero. In any expansion in terms of these parameters,
it should be a good approximation to work to first order in $s$ and $a$,
although it is worth working to second order to obtain the most accurate
results. In Appendix~\ref{deviations} the PMNS matrix is expanded to second
order in $r,s,a$, and the neutrino oscillation formulae including matter
effects are given to a similar level of approximation (results taken
from~\cite{King:2007pr}). Note that the global fit values in
Eq.~(\ref{rsafit}) are consistent with,
\beq
s=0\,, \ \ \ \ a=0\,, \ \ \ \ r=R\lambda\,,
\label{rsabetter}
\eeq
at the one sigma level, where $\lambda$ is the Wolfenstein parameter and $R$ a real
number of order unity.

\subsection{\label{sec:patterns-3}Tri-bimaximal-Cabibbo mixing}
The recent data  is consistent with the remarkable relationship~\cite{Giunti:2002ye},
\beq
s_{13}= \frac{ \sin \theta_C }{\sqrt{2}} = \frac{\lambda}{\sqrt{2}}, 
\label{rem1}
\eeq
where $\lambda  = 0.2253\pm 0.0007$~\cite{Nakamura:2010zzi} 
is the Wolfenstein parameter. 
The above ansatz implies a reactor angle of 
\beq
\theta_{13}\approx \frac{\theta_C}{\sqrt{2}}\approx 9.2^\circ,
\label{rem2}
\eeq
where $\theta_C\approx 13^\circ$ is the Cabibbo angle.
One may combine the relation
in Eq.~(\ref{rem1})
with TB mixing to yield tri-bimaximal-Cabibbo (TBC) mixing~\cite{King:2012vj}:
\be
s_{13} = \frac{\lambda}{\sqrt{2}}, \ \ s_{12} = \frac{1}{\sqrt{3}},
\ \ s_{23} = \frac{1}{\sqrt{2}}.
\label{TBC0}
\ee
This corresponds to $s=a=0$ and $r=\lambda$ and 
leads to the following approximate form of the mixing matrix~\cite{King:2012vj}, 
\begin{eqnarray}
U_{\mathrm{TBC}} \approx
\left( \begin{array}{ccc}
\sqrt{\frac{2}{3}}(1-\frac{1}{4}\lambda^2)  & \frac{1}{\sqrt{3}}(1-\frac{1}{4}\lambda^2) 
& \frac{1}{\sqrt{2}}\lambda e^{-i\delta } \\
-\frac{1}{\sqrt{6}}(1+\lambda e^{i\delta })  & \frac{1}{\sqrt{3}}(1- \frac{1}{2}\lambda e^{i\delta })
& \frac{1}{\sqrt{2}}(1-\frac{1}{4}\lambda^2) \\
\frac{1}{\sqrt{6}}(1- \lambda e^{i\delta })  & -\frac{1}{\sqrt{3}}(1+ \frac{1}{2}\lambda e^{i\delta })
 & \frac{1}{\sqrt{2}}(1-\frac{1}{4}\lambda^2)
\end{array}
\right)P_{12} + \mathcal{O}(\lambda^3),
\label{TBC}
\end{eqnarray}
corresponding to the mixing angles,
\beq
\theta_{13}\approx 9.2^\circ, \ \ 
\theta_{12}= 35.26^\circ, \ \
\theta_{23}= 45^\circ. \ \
\label{TBangles}
\eeq
This is consistent with the data at the three sigma level.

\subsection{\label{sec:patterns-4}Charged lepton contributions to neutrino masses and mixing angles}

The mixing matrix in the lepton sector, the PMNS matrix
$U_{\mathrm{PMNS}}$, is defined as the matrix which appears in the
electroweak coupling to the $W$ bosons expressed in terms of lepton
mass eigenstates. With the mass matrices of charged leptons
$M_{e}$ and neutrinos $m^{\nu}_{LL}$ written as\footnote{Although we
have chosen to write a Majorana mass matrix, all relations in the
following are independent of the Dirac or Majorana nature of neutrino
masses.}
\begin{eqnarray}
{\cal L}=-  \ol{e_L} M_{e} e_R  
- \frac{1}{2}\ol{\nu_L} m^{\nu}_{LL} \nu_{L}^c 
+ {H.c.}\; ,
\end{eqnarray}
and performing the transformation from flavour to mass basis by
 \begin{eqnarray}\label{eq:DiagMe}
V_{e_L} \, M_{e} \,
V^\dagger_{e_R} =
\mbox{diag}(m_e,m_\mu,m_\tau)
 , \quad~
V_{\nu_L} \,m^{\nu}_{LL}\,V^T_{\nu_L} =
\mbox{diag}(m_1,m_2,m_3),
\label{mLLnu}
\end{eqnarray}
the PMNS matrix is given by
\begin{eqnarray}\label{Eq:PMNS_Definition}
U_{\mathrm{PMNS}}
= V_{e_{L}} V^\dagger_{\nu_{L}} \,.
\end{eqnarray}
Here it is assumed implicitly that unphysical phases are removed by
field redefinitions, and $U_\mathrm{PMNS}$ contains one Dirac phase
and two Majorana phases. The latter are physical only in the case of
Majorana neutrinos, for Dirac neutrinos the two Majorana phases can be
absorbed as well.

As shown in~\cite{King:2002nf,King:nf,King:nf-s,sumrule-1,sumrule-2} the lepton mixing matrix 
can be expanded
in terms of neutrino and charged lepton mixing angles and phases
to leading order in the charged lepton mixing angles which
are assumed to be small,
\bea
s_{23}e^{-i\delta_{23}}
& \approx &
s_{23}^{\nu}e^{-i\delta_{23}^{\nu}}
-\theta_{23}^{e}
c_{23}^{\nu}e^{-i\delta_{23}^{e}} \ ,
\label{chlep23}
\\
\theta_{13}e^{-i\delta_{13}}
& \approx &
\theta_{13}^{\nu}e^{-i\delta_{13}^{\nu}}
-\theta_{13}^{e}c_{23}^{\nu}e^{-i\delta_{13}^{e}} 
- \theta_{12}^{e}s_{23}^{\nu}e^{i(-\delta_{23}^{\nu}-\delta_{12}^{e})} \ ,
\label{chlep13}
\\
s_{12}e^{-i\delta_{12}}
& \approx &
s_{12}^{\nu}e^{-i\delta_{12}^{\nu}}
%+\theta_{23}^{e}s_{12}^{\nu}e^{-i\delta_{12}^{\nu}}
+\theta_{13}^{e}
c_{12}^{\nu}s_{23}^{\nu}e^{i(\delta_{23}^{\nu}-\delta_{13}^{e})}
-\theta_{12}^{e}
c_{23}^{\nu}c_{12}^{\nu}e^{-i\delta_{12}^{e}}\ ,
\label{chlep12}
\eea
where we have dropped the subscripts $L$ for simplicity.
Clearly $\theta_{13}$
receives important contributions not just from $\theta_{13}^{\nu}$,
but also from the charged lepton angles
$\theta_{12}^{e}$, and $\theta_{13}^{e}$.
In models where $\theta_{13}^{\nu}$ is
extremely small, $\theta_{13}$ may originate almost entirely from
the charged lepton sector.
Charged lepton contributions could also be important in models
where $\theta_{12}^{\nu}=\pi /4$, since charged lepton mixing angles
may allow consistency with the LMA MSW solution.

Note that it is useful and possible to be able to diagonalise the mass matrices
analytically, at least to first order in the small 13 mixing angle,
but allowing the 23 and 12 angles to be large, while retaining
all the phases. The procedure for doing this is discussed for a hierarchical
and an inverted hierarchical neutrino mass matrix in~\cite{King:2002nf}.
The analytic results enable the
separate mixing angles and phases associated with each of the
unitary transformations $V_{e_L}$ and $V_{\nu_L}^{\dagger}$
to be obtained in many useful cases of interest.

\subsection{\label{sec:patterns-5}Solar mixing sum rules}
In many models the neutrino mixing matrix has a simple form $U_0$ in Eq.~(\ref{GR}),
where $ s_{23}^{\nu}= c_{23}^{\nu}= 1/\sqrt{2}$ and $ s_{13}^{\nu}=0$, while 
the charged lepton mixing
matrix has a CKM-like structure, in the sense that $V_{e_L}$
is dominated by a 12 mixing, i.e.\ its elements
$(V_{e_L})_{13}$, $(V_{e_L})_{23}$,
$(V_{e_L})_{31}$ and $(V_{e_L})_{32}$ are
very small compared to $(V_{e_L})_{12}$ and $(V_{e_L})_{21}$,
where in practice we take them to be zero.
In this case we are led to a solar sum rule~\cite{sumrule-1,sumrule-2,King:nf-s} derived from
$U_{\mathrm{PMNS}} = V_{e_L} U_0$, which takes the form,
\begin{eqnarray}
U_{\mathrm{PMNS}} &=& \left(\begin{array}{ccc}
\!c^e_{12}& -s^e_{12}e^{-i\delta^e_{12}}&0\!\\
\!s^e_{12}e^{i\delta^e_{12}}&c^e_{12} &0\!\\
\!0&0&1\!
\end{array}
\right)\left( \begin{array}{ccc}
c^{\nu}_{12} & s^{\nu}_{12} & 0\\
 -\frac{s^{\nu}_{12}}{\sqrt{2}}  &  \frac{c^{\nu}_{12}}{\sqrt{2}} & \frac{1}{\sqrt{2}}  \\
\frac{s^{\nu}_{12}}{\sqrt{2}}  &  -\frac{c^{\nu}_{12}}{\sqrt{2}} & \frac{1}{\sqrt{2}} 
\end{array}
\right) 
= \left(\begin{array}{ccc}
\! \cdots & \ \ 
\! \cdots&
\! -\frac{s^e_{12}}{\sqrt{2}}e^{-i\delta^e_{12}} \\
\! \cdots
& \ \
\! \cdots
&
\! \frac{c^e_{12}}{\sqrt{2}}
\!\\
\!\frac{s^{\nu}_{12}}{\sqrt{2}} 
& \ \
\!-\frac{c^{\nu}_{12}}{\sqrt{2}}
&
\! \frac{1}{\sqrt{2}} 
\end{array}
\right) \! . ~~~~~~~
\label{Ucorr}
\end{eqnarray}
The important point to notice is that the 3-1, 3-2 and 3-3 elements of
$U_{\mathrm{PMNS}}$ in Eq.~(\ref{Ucorr}) are uncorrected by charged lepton
corrections and are the same as those of $U_0$, and also the 1-3 element of
$U_{\mathrm{PMNS}}$ has a simple form. By comparing Eq.~(\ref{Ucorr}) to the
PDG parametrisation of $U_{\mathrm{PMNS}}$ in Eq.~(\ref{eq:matrix}) we find
the relations,  
\bea
\label{Eq:Sumrule4} s_{13} &=&  \frac{s^e_{12}}{\sqrt{2}}  \; , \\
\label{Eq:Sumrule3a}  s_{23}c_{13} &=&  \frac{c^e_{12}}{\sqrt{2}} \;, \\
\label{Eq:Sumrule3}  c_{23}c_{13} &=&  \frac{1}{\sqrt{2}} \;, \\
\label{Eq:Sumrule1} 
  |s_{23}s_{12}-s_{13}c_{23}c_{12}e^{i\delta} |    &=&      \frac{s^{\nu}_{12}}{\sqrt{2}}\;,\\
\label{Eq:Sumrule2} 
  | s_{23}c_{12}+s_{13}c_{23}s_{12}e^{i\delta} |    &=&   \frac{c^{\nu}_{12}}{\sqrt{2}} \;.
\eea
Using Eq.~(\ref{Eq:Sumrule3}) we see that, to leading order in $\theta_{13}$,
the atmospheric angle is unchanged from its maximal value by the assumed form
of the charged lepton corrections. 
To this approximation, it is then straightforward to expand these results to obtain the more useful approximate form of the sum rule~\cite{sumrule-1,sumrule-2,King:nf-s},
\be
\theta_{12}\approx \theta_{12}^{\nu}+ \theta_{13}\cos \delta .
\label{sumrule}
\ee
Given the accurate determination of the reactor angle in Eq.~(\ref{t13}) ($\theta_{13}\approx 8.5^\circ\pm 0.5^\circ$)
and the solar angle Eq.~(\ref{t12}) ($\theta_{12}\approx 34^\circ\pm 1^\circ$)
the sum rule in Eq.~(\ref{sumrule}) yields
a favoured range of $\cos \delta $ for 
each of the cases  $\theta_{12}^{\nu}=35.26^\circ, 45^\circ, 31.7^\circ, 36^\circ$ for the cases of TB, BM, GR, GR$'$,
namely $\cos \delta \approx -0.2, -1,0.2, -0.2,$
or $\cos \delta \approx -\lambda, -1, \lambda, -\lambda$,  respectively.
For example, for TB neutrino mixing, the sum rule in Eq.~(\ref{sumrule}) may be written compactly as,
\beq
s\approx r\,\cos \delta.
\label{sumrule2}
\eeq

This approach relies on a Cabibbo-sized charged lepton mixing angle as is clear from Eq.~(\ref{Eq:Sumrule4})
which, together with Eq.~(\ref{rem1}), shows that we need $s^e_{12}\approx \lambda$ in order to 
account for the observed reactor angle, starting from one of the simple classic patterns of neutrino mixing.
This is not straightforward to achieve in realistic models~\cite{King:2012vj,Antusch:2012fb},
which would typically prefer smaller charged lepton mixing angles such as $s^e_{12}\approx \lambda /3$.
This suggests that the neutrino mixing angle $\theta_{13}^{\nu}$ is not zero, but has some non-zero value
closer to the observed reactor angle. In the next subsection we consider this possibility.

\subsection{\label{sec:patterns-6}Atmospheric mixing sum rules}
In this subsection we consider simple alternative patterns related to TB mixing which allow a non-zero reactor angle.  
When looking for variants of TB mixing, it is useful to start from the general expansion around TB mixing in
Eq.~(\ref{rsadef})~\cite{King:2007pr}, which to leading order, gives a PMNS mixing matrix, as in Eq.~(\ref{MNS1}),
\begin{eqnarray}
U_{\mathrm{PMNS}} \approx
\left( \begin{array}{ccc}
\sqrt{\frac{2}{3}}(1-\frac{1}{2}s)  & \frac{1}{\sqrt{3}}(1+s) & \frac{1}{\sqrt{2}}re^{-i\delta } \\
-\frac{1}{\sqrt{6}}(1+s-a + re^{i\delta })  & \frac{1}{\sqrt{3}}(1-\frac{1}{2}s-a- \frac{1}{2}re^{i\delta })
& \frac{1}{\sqrt{2}}(1+a) \\
\frac{1}{\sqrt{6}}(1+s+a- re^{i\delta })  & -\frac{1}{\sqrt{3}}(1-\frac{1}{2}s+a+ \frac{1}{2}re^{i\delta })
 & \frac{1}{\sqrt{2}}(1-a)
\end{array}
\right)P_{12}.~~~~
\label{MNS1-main}
\end{eqnarray}

Clearly TB mixing in Eq.~(\ref{TB}) corresponds to $s=a=r=0$.
If we set $s=a=0$ but retain a non-zero value of $r$ then this defines
tri-bimaximal-reactor mixing (TBR)~\cite{King:2009qt},
\begin{eqnarray}
U_{\mathrm{TBR}} \approx
\left( \begin{array}{ccc}
\sqrt{\frac{2}{3}}  & \frac{1}{\sqrt{3}} & \frac{1}{\sqrt{2}}re^{-i\delta } \\
-\frac{1}{\sqrt{6}}(1+ re^{i\delta })  & \frac{1}{\sqrt{3}}(1- \frac{1}{2}re^{i\delta })
& \frac{1}{\sqrt{2}} \\
\frac{1}{\sqrt{6}}(1- re^{i\delta })  & -\frac{1}{\sqrt{3}}(1+ \frac{1}{2}re^{i\delta })
 & \frac{1}{\sqrt{2}}
\end{array}
\right)P_{12}.
\label{TBR}
\end{eqnarray}
This is very constrained, in particular it requires maximal atmospheric mixing $a=0$.
We can maintain trimaximal (TM) mixing defined by $s=0$ but relax maximal atmospheric mixing,
allowing both a non-zero $a$ and $r$,
\begin{eqnarray}
U_{\mathrm{TM}} \approx
\left( \begin{array}{ccc}
\sqrt{\frac{2}{3}}  & \frac{1}{\sqrt{3}} & \frac{1}{\sqrt{2}}re^{-i\delta } \\
-\frac{1}{\sqrt{6}}(1-a + re^{i\delta })  & \frac{1}{\sqrt{3}}(1-a- \frac{1}{2}re^{i\delta })
& \frac{1}{\sqrt{2}}(1+a) \\
\frac{1}{\sqrt{6}}(1+a- re^{i\delta })  & -\frac{1}{\sqrt{3}}(1+a+ \frac{1}{2}re^{i\delta })
 & \frac{1}{\sqrt{2}}(1-a)
\end{array}
\right)P_{12}.
\label{TM}
\end{eqnarray}
There are two interesting special cases of TM mixing in which the first or second column
of the mixing matrix reduce to those of the  first or second column
of the TB mixing matrix, referred to as TM1 and TM 2 mixing, namely,
\begin{eqnarray}
U_{\mathrm{TM1}} \approx
\left( \begin{array}{ccc}
\sqrt{\frac{2}{3}}  & \frac{1}{\sqrt{3}} & \frac{1}{\sqrt{2}}re^{-i\delta } \\
-\frac{1}{\sqrt{6}}  & \frac{1}{\sqrt{3}}(1- \frac{3}{2}re^{i\delta })
& \frac{1}{\sqrt{2}}(1+a) \\
\frac{1}{\sqrt{6}}  & -\frac{1}{\sqrt{3}}(1+ \frac{3}{2}re^{i\delta })
 & \frac{1}{\sqrt{2}}(1-a)
\end{array}
\right)P_{12},
\label{TM1}
\end{eqnarray}
with
\beq
a=r\cos \delta ,
\label{TM1r}
\eeq
and 
\begin{eqnarray}
U_{\mathrm{TM2}} \approx
\left( \begin{array}{ccc}
\sqrt{\frac{2}{3}}  & \frac{1}{\sqrt{3}} & \frac{1}{\sqrt{2}}re^{-i\delta } \\
-\frac{1}{\sqrt{6}}(1+  \frac{3}{2}re^{i\delta })  & \frac{1}{\sqrt{3}}
& \frac{1}{\sqrt{2}}(1+a) \\
\frac{1}{\sqrt{6}}(1- \frac{3}{2} re^{i\delta })  & -\frac{1}{\sqrt{3}}
 & \frac{1}{\sqrt{2}}(1-a)
\end{array}
\right)P_{12},
\label{TM2}
\end{eqnarray}
with
\beq
a=-(r/2)\cos \delta .
\label{TM2r}
\eeq
These TM1 and TM2 relations, both with $s=0$,
are examples of atmospheric sum rules to first order in $\lambda$.

The above atmospheric sum rules are valid to leading order in $\lambda$. 
The exact  TM1 relations (for both $a$ and $s$) are obtained by
equating PMNS elements to the first column of the TB mixing matrix:
\bea
c_{12}c_{13}=\sqrt{\frac{{2}}{{3}}} ,\label{a1} \\
|c_{23}s_{12}+s_{13}s_{23}c_{12}e^{i\delta}|=\frac{1}{\sqrt{6}} , \label{a2}\\
|s_{23}s_{12}-s_{13}c_{23}c_{12}e^{i\delta}|=\frac{1}{\sqrt{6}} , \label{a3}
\eea
where these lead to Eq.~(\ref{TM1r}) when expanded to leading order.

The exact sum rule relations for TM2 are obtained by equating
PMNS elements to the second column of the TB mixing matrix:
\bea
s_{12}c_{13}=\frac{1}{\sqrt{3}} ,\label{a4}\\
|c_{23}c_{12}-s_{13}s_{23}s_{12}e^{i\delta}|=\frac{1}{\sqrt{3}},  \label{a5}\\
|s_{23}c_{12}+s_{13}c_{23}s_{12}e^{i\delta}|=\frac{1}{\sqrt{3}},  \label{a6}
\eea
where these lead to Eq.~(\ref{TM2r}) when expanded to leading order.

From Eqs.~(\ref{a1},\ref{a4}) we see that to leading order in $s_{13}$ the solar angle
is unchanged from its TB value for both TM1 and TM2, corresponding to $s=0$ as
discussed earlier, but to second order in $s_{13}$, the solar angle deviates
and this deviation is different for TM1 and TM2.

\section{\label{sec:seesaw}The see-saw mechanisms}
The starting point of the see-saw mechanisms is the dimension 5 operator in 
Eq.~(\ref{dim5}) which we repeat below,
\beq
-\frac{1}{2}HL^T\kappa HL.
\label{dim52}
\eeq
One might wonder if it is possible to simply write down an operator
by hand similar to Eq.~(\ref{dim5}), without worrying about its origin.
In fact, historically, such an operator was introduced suppressed by
the Planck scale (rather than the right-handed neutrino mass scales)
by Weinberg in order to account for small neutrino
masses~\cite{Weinberg:sa}. The problem is that such a Planck scale
suppressed operator would lead to neutrino masses of order
$10^{-5}{\rm eV}$ which are too small to account for the two heavier neutrino
masses (though it could account for the lightest neutrino mass). To account for
the heaviest neutrino mass requires a dimension 5 operator suppressed by a
mass scale of order $3 \times 10^{14}$ GeV if the dimensionless coupling
of the operator is of order unity, and the Higgs vacuum expectation value
(VEV) is equal to that of the Standard Model.

There are several different kinds of see-saw mechanism in the
literature. In this review we shall focus on the
simplest type I see-saw mechanism, which we shall introduce below.
However for completeness we shall also discuss the type II and III see-saw
mechanisms and the double see-saw mechanism, as well as the inverse and linear 
see-saw mechanisms. 

\subsection{\label{sec:seesaw-1}Type I see-saw}
Before discussing the see-saw mechanism it is worth first reviewing
the different types of neutrino mass that are possible. So far we
have been assuming that neutrino masses are Majorana masses of the form
\beq
m^{\nu}_{LL}\ol{\nu_L}\nu_L^c \ ,
\label{mLL}
\eeq
where $\nu_L$ is a left-handed neutrino field and $\nu_L^c$ is
the CP conjugate of a left-handed neutrino field, in other words
a right-handed antineutrino field. Such Majorana masses are possible
since both the neutrino and the antineutrino
are electrically neutral and so
Majorana masses are not forbidden by electric charge conservation.
For this reason a Majorana mass for the electron would
be strictly forbidden. However such Majorana neutrino masses
violate lepton number conservation. Assuming only Higgs {\it doublets}, they are
forbidden in the Standard Model at the renormalisable level by gauge invariance.
The idea of the simplest version of the see-saw mechanism is to assume
that such terms are zero to begin with, but are generated effectively,
after right-handed neutrinos are introduced~\cite{seesaw}.

If we introduce right-handed neutrino fields then there are two sorts
of additional neutrino mass terms that are possible. There are
additional Majorana masses of the form
\beq
M_{RR}\ol{\nu_R^c}\nu_R^{} \ ,
\label{MRR}
\eeq
where $\nu_R$ is a right-handed neutrino field and $\nu_R^c$ is
the CP conjugate of a right-handed neutrino field, in other words
a left-handed antineutrino field. In addition there are
Dirac masses of the form
\beq
m_{LR}\ol{ \nu_L}\nu_R\ .
\label{mLR}
\eeq
Such Dirac mass terms conserve lepton number, and are not forbidden
by electric charge conservation even for the charged leptons and
quarks.

Once this is done then the types of neutrino mass discussed
in Eqs.~(\ref{MRR},\ref{mLR}) (but not Eq.~(\ref{mLL}) since we
assume no Higgs triplets)
are permitted, and we have the mass matrix
\begin{equation}
\left(\begin{array}{cc} \ol{\nu_L} & \ol{\nu^c_R}
\end{array} \\ \right)
\left(\begin{array}{cc}
0 & m_{LR}\\
m_{LR}^T & M_{RR} \\
\end{array}\right)
\left(\begin{array}{c} \nu_L^c \\ \nu_R \end{array} \\ \right).
\label{matrix}
\end{equation}
Since the right-handed neutrinos are electroweak singlets
the Majorana masses of the right-handed neutrinos $M_{RR}$
may be orders of magnitude larger than the electroweak
scale. In the approximation that $M_{RR}\gg m_{LR}$
the matrix in Eq.~(\ref{matrix}) may be diagonalised to
yield effective Majorana masses of the type in Eq.~(\ref{mLL}),
\beq
m^{\nu}_{LL}=-m_{LR}M_{RR}^{-1}m_{LR}^T\ .
\label{seesaw}
\eeq
The effective left-handed Majorana masses $m_{LL}^\nu$ are naturally
suppressed by the heavy scale $M_{RR}$.
In a one family example if we take $m_{LR}\sim M_W$ (where $M_W$ is the mass
of the $W$ boson) and $M_{RR}\sim M_{\mathrm{GUT}}$
then we find $m_{LL}^\nu\sim 10^{-3}$ eV which looks good for solar
neutrinos.
Atmospheric neutrino masses would require
a right-handed neutrino with a mass below the GUT scale.

\begin{figure}[tb]
\centering
\includegraphics[width=0.36\textwidth]{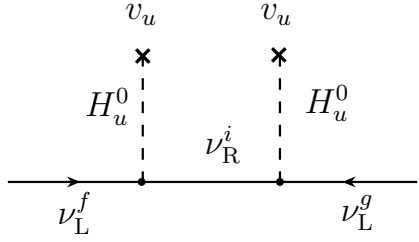}
 \caption{\label{fig:TypeIDiagrams}\small{
 Diagram illustrating the type I see-saw mechanism.}} 
\end{figure}

The type I see-saw mechanism is illustrated diagrammatically in 
Fig.~\ref{fig:TypeIDiagrams}.
It can be formally derived from the following Lagrangian
\be
{\cal L}=-\ol{\nu_L}m_{LR}\nu_R-\frac{1}{2}\nu_R^TM_{RR}\nu_R+H.c. \ ,
\ee
where $\nu_L$ represents left-handed neutrino fields
(arising from electroweak doublets), $\nu_R$ represents right-handed
neutrino fields (arising from electroweak singlets), in a matrix
notation where the $m_{LR}$ matrix elements
are typically of order the charged lepton masses,
while the $M_{RR}$ matrix elements may be much larger than the
electroweak scale, and maybe up to the Planck scale.
The number of right-handed neutrinos is not fixed, but the number
of left-handed neutrinos is equal to three.
Below the mass scale of the right-handed neutrinos we can
integrate them out using the equations of motion
\be
\frac{d{\cal L}}{d\nu_R}=0 \ ,
\ee
which gives
\be
\nu_R^T=-\ol{\nu_L}m_{LR}M_{RR}^{-1}\ ,\ \quad
\nu_R=-M_{RR}^{-1}m_{LR}^T\ol{\nu_L}^T \ .
\ee
Substituting back into the original Lagrangian we find
\be
{\cal L}=-\frac{1}{2}\ol{\nu_L}m^{\nu}_{LL}\nu_L^c+H.c. \ ,
\ee
with $m^{\nu}_{LL}$ as in Eq.~(\ref{seesaw}).

\subsection{\label{sec:seesaw-2}Minimal see-saw extension of the Standard Model}

We now briefly discuss what the Standard Model looks like,
assuming a minimal see-saw extension.
In the Standard Model Dirac mass terms for charged leptons and quarks
are generated from Yukawa type couplings to Higgs doublets whose vacuum
expectation value gives the Dirac mass term.
Neutrino masses are zero in the Standard Model because
right-handed neutrinos are not present, and also
because the Majorana mass terms in Eq.~(\ref{mLL})
require Higgs triplets in order to be generated at the renormalisable
level. The simplest way to generate neutrino masses
from a renormalisable theory is to introduce right-handed neutrinos,
as in the type I see-saw mechanism, which we assume here.
The Lagrangian for the lepton sector of the Standard Model
containing three right-handed neutrinos with heavy Majorana masses
is\footnote{We introduce two Higgs doublets to pave the way for 
the supersymmetric Standard Model.
For the same reason we express the Standard Model Lagrangian
in terms of left-handed fields, replacing right-handed fields $\psi_R$
by their CP conjugates $\psi^c$, where the subscript $R$ has been dropped.
In the case of the minimal standard see-saw model with only one Higgs doublet,
the other Higgs doublet in Eq.~(\ref{SM}) is obtained by charge
conjugation, i.e. $H_d\equiv H_u^c$.}  
\be
{\cal
  L}_{\mathrm{mass}}=-\left[\epsilon_{\alpha\beta}\tilde{Y}_e^{ij}H_d^\alpha L_i^\beta
  e^c_j -\epsilon_{\alpha\beta}\tilde{Y}_{\nu}^{ij}H_u^\alpha L_i^\beta \nu^c_j +
\frac{1}{2} \nu^c_i\tilde{M}_{RR}^{ij}\nu^c_j
\right] +H.c. \ ,
\label{SM}
\ee
where $\epsilon_{\alpha \beta}=-\epsilon_{\beta\alpha}$, $\epsilon_{12}=1$,
and the remaining notation is standard except that
the $3$ right-handed neutrinos $\nu_R^i$ have been replaced by their
CP conjugates $\nu^c_i$, and $\tilde{M}_{RR}^{ij}$ is a complex symmetric
Majorana matrix.
When the two Higgs doublets get their
VEVs $<H_u^2>=v_u$, $<H_d^1>=v_d$, where the ratio of VEVs
is defined to be $\tan \beta \equiv v_u/v_d$,
we find the terms
\be
{\cal L}_{\mathrm{mass}}= -v_d\tilde{Y}_e^{ij}e_ie^c_j
-v_u\tilde{Y}_{\nu}^{ij}\nu_i\nu^c_j -
\frac{1}{2}\tilde{M}_{RR}^{ij}\nu^c_i\nu^c_j +H.c. \ .
\ee
Replacing CP conjugate fields we can write in a matrix notation
\be
{\cal L}_{\mathrm{mass}}=-\ol{e_L}v_d {\tilde{Y}_e}^\ast e_R
-\ol{\nu_L}v_u{\tilde{Y}_\nu}^\ast \nu_R -
\frac{1}{2}\nu^T_R\tilde{M}_{RR}^\ast \nu_R +H.c. \ . \label{eq:Mast-MRR}
\ee
It is convenient to work in the diagonal charged lepton basis
\be
{\rm diag}(m_e,m_{\mu},m_{\tau})
= V_{e_L}v_d\tilde{Y}_e^\ast V_{e_R}^{\dag} \ ,
\ee
and the diagonal right-handed neutrino basis
\be
{\rm diag}(M_{1},M_{2},M_{3})=
V_{\nu_R}\tilde{M}^\ast_{RR}V_{\nu_R}^{T} \ ,
\ee
where $V_{e_L},V_{e_R},V_{\nu_R}$ are unitary transformations.
In this basis the neutrino Yukawa couplings are given by
\be
Y_{\nu}=V_{e_L} \tilde{Y}_\nu^\ast V_{\nu_R}^{T} \ ,
\ee
and the Lagrangian in this basis is
\bea
{\cal L}_{\mathrm{mass}}&=-&(\ol{e_L} \;\ol{\mu_L} \;\ol{\tau_L})
{\rm diag}{(m_e,m_{\mu},m_{\tau})} ({e}_R \,{\mu}_R\, {\tau}_R)^T
\nonumber \\
&-&(\ol{\nu_{e L}}\; \ol{\nu_{\mu L}} \; \ol{\nu_{\tau L}})v_uY_{\nu}
(\nu_{R1}\, \nu_{R2} \, \nu_{R3} )^T
\nonumber \\
&-&\frac{1}{2} (\nu_{R1} \,\nu_{R2}\, \nu_{R3} ){\rm diag}{(M_{1},M_{2},M_{3})}
(\nu_{R1} \,\nu_{R2} \,\nu_{R3} )^T
+H.c. \ .
\eea
After integrating out the right-handed neutrinos (the see-saw mechanism)
we find
\bea
{\cal L}_{\mathrm{mass}}&=-&(\ol{e_L} \;\ol{\mu_L} \;\ol{\tau_L})
{\rm diag}{(m_e,m_{\mu},m_{\tau})} ({e}_R \,{\mu}_R \,{\tau}_R)^T
\nonumber \\
&-&\frac{1}{2}(\ol{\nu_{eL}}\; \ol{\nu_{\mu L}} \;\ol{\nu_{\tau L}} )
m^{\nu}_{LL} ({\nu_e}_L^c \,{\nu_\mu}_L^c \,{\nu_\tau}_L^c)^T
+H.c. \ ,
\label{Lmass2}
\eea
where the light effective left-handed Majorana
neutrino mass matrix in the above basis is given by the
following see-saw formula which is equivalent to Eq.~(\ref{seesaw}),
\be
m^{\nu}_{LL}=-v_u^2\,Y_{\nu}\,
{\rm diag}{(M_{1}^{-1},M_{2}^{-1},M_{3}^{-1})}  \,Y_{\nu}^T \ . 
\label{seesaw1}
\ee
In this basis the type I see-saw mechanism reproduces the dimension 5 operator in Eq.~(\ref{dim52}) with 
\beq
\kappa = Y_{\nu}\,{\rm diag}{(M_{1}^{-1},M_{2}^{-1},M_{3}^{-1})} \,Y_{\nu}^T \ .
\label{kappa}
\eeq

\subsection{\label{sec:seesaw-3}Sequential right-handed neutrino dominance}
In this subsection we show how the type I see-saw mechanism may lead to a neutrino mass hierarchy
with large solar and atmospheric mixing angles, and a reactor angle of order $m_2/m_3$
via a simple and natural mechanism known as sequential dominance
(SD).\footnote{SD should not be confused with an alternative mechanism 
proposed by Smirnov which is based on the premise that there are no large mixing angles
in the Yukawa sector, and does not discuss any mechanism for achieving this involving right-handed neutrinos~\cite{Smirnov:af}. }
First consider the 
case of single right-handed neutrino dominance where only one right-handed neutrino $\nu_3^c$
of heavy Majorana mass $M_3$
is present in the see-saw mechanism, namely the one responsible for 
the atmospheric neutrino mass $m_3$~\cite{King:1998jw,King:2002nf}.
We work in the basis of the previous subsection where the right-handed neutrinos and charged lepton mass matrices 
are diagonal. If the single right-handed neutrino
couples to the three lepton doublets $L_i$ in the diagonal charged lepton mass basis as,
\be
H_u (dL_e + eL_{\mu}+ f L_{\tau}) \nu_3^c,\label{eq:SDdef}
\ee
where $d,e,f$ are Yukawa couplings (assumed real for simplicity\footnote{The
  full results including phases are discussed
  in~\cite{King:1998jw,King:2002nf} 
and summarised for the case of $d=0$ in~\cite{King:2002qh}.}), and $H_u$ is the Higgs doublet,
where it is assumed that $d\ll e,f$. so that the see-saw mechanism yields
the atmospheric neutrino mass, 
\be
m_3\approx (e^2+f^2)\frac{v_u^2}{M_3},
\label{m3}
\ee
where 
$v_u=\langle H_u\rangle$.
Then the reactor and atmospheric angles are approximately given by 
simple ratios of Yukawa couplings~\cite{King:1998jw,King:2002nf},
\be
\theta_{13}\approx \frac{d}{\sqrt{e^2+f^2}}, \qquad  \tan \theta_{23}\approx \frac{e}{f}.
\label{angles1}
\ee

According to SD~\cite{King:1998jw,King:2002nf}
the solar neutrino mass and mixing are accounted for by introducing 
a second right-handed neutrino $\nu_2^c$ with mass $M_2$ which
couples to the three lepton doublets $L_i$ in the diagonal charged lepton mass basis as,
\be
H_u (aL_e + bL_{\mu}+ c L_{\tau}) \nu_2^c,\label{eq:SDabc}
\ee
where $a,b,c$ are Yukawa couplings (assumed real for simplicity). Then the second right-handed neutrino
is mainly responsible for the solar neutrino mass, providing 
\be
(a,b,c)^2/M_2\ll (e,f)^2/M_3,
\label{SD}
\ee
which is the basic SD condition. Assuming this, then the see-saw mechanism leads to the solar
neutrino mass,
\be
m_2\approx \left(a^2+ (c_{23}b-s_{23}c)^2\right)\frac{v_u^2}{M_2},
\label{m2}
\ee
and the solar neutrino mixing is approximately given by 
a simple ratios of Yukawa couplings~\cite{King:1998jw,King:2002nf},
\be
\tan \theta_{12}\approx \frac{a}{(c_{23}b-s_{23}c)}.
\label{angles2}
\ee
There is an additional contribution to the reactor angle of the form~\cite{King:1998jw,King:2002nf},
\be
\Delta \theta_{13}  \approx 
\frac{a(eb+fc)}{(e^2+f^2)^{3/2}}
\frac{M_3}{M_2} \sim {\cal O}(m_2/m_3).
\label{Delta}
\ee
There may also be a third right-handed neutrino but with completely
subdominant contributions to the see-saw mechanism, and hence it may ignored to
leading order.\footnote{The contributions of the third sub-subdominant
  right-handed neutrino to the mixing angles has been considered
  in~\cite{Antusch:2010tf}.} 

Let us summarise what SD achieves. With the assumption in Eq.~(\ref{SD}),
SD predicts a neutrino mass hierarchy, together with solar and atmospheric mixing angles which are
independent of neutrino mass. Since they only involve ratios of Yukawa couplings they may easily be large.
On the other hand the reactor angle has two contributions, one proportional to a ratio of Yukawa couplings
which may be small if $d\ll e$, while the other one gives a contribution $ {\cal O}(m_2/m_3)$ which is by itself of the correct magnitude to account for the reactor angle (even if $d=0$).
The origin of these conditions and assumptions may be due to family symmetry as we will discuss.

\subsection{\label{sec:seesaw-4}Other see-saw mechanisms}

One might also wonder if the see-saw mechanism with right-handed
neutrinos is the only possibility? In fact it is possible to
generate the dimension 5 operator in Eq.~(\ref{dim5}) by the exchange
of heavy Higgs triplets of $SU(2)_L$, referred to as the type II see-saw
mechanism~\cite{type2} or the exchange of heavy $SU(2)_L$ triplet fermions,
referred to as the type III see-saw mechanism~\cite{Foot:1989type3}.

In the type II see-saw the general neutrino mass matrix is given by
\begin{equation}
\left(\begin{array}{cc} \ol{\nu_L} & \ol{\nu^c_R}
\end{array} \\ \right)
\left(\begin{array}{cc}
m^{II}_{LL} & m_{LR}\\
m_{LR}^T & M_{RR} \\
\end{array}\right)
\left(\begin{array}{c} \nu_L^c \\ \nu_R \end{array} \\ \right) .
\label{matrix-II}
\end{equation}

\begin{figure}[tb]
\centering
\includegraphics[width=0.26\textwidth]{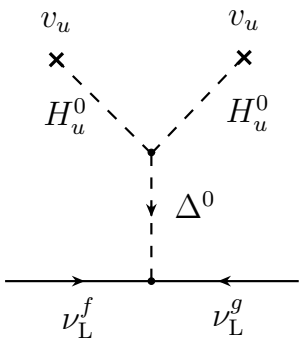}
 \caption{\label{fig:TypeIIDiagrams}\small{Diagram leading to a type II
     contribution $m^{{II}}_{{LL}}$ to the neutrino mass matrix
     via an induced VEV of the neutral component of a triplet Higgs $\Delta$.}}
\end{figure}

Under the assumption that the mass eigenvalues $M_{i}$ of
$M_{{RR}}$ are very large compared to the components of  $m^{{II}}_{LL}$
and $m_{{LR}}$, the mass matrix can approximately be 
diagonalised yielding effective Majorana masses  
\begin{eqnarray}\label{eq:TypIIMassMatrix}
m^\nu_{{LL}} \approx 
m^{{II}}_{{LL}} + m^{{I}}_{{LL}} \ ,
\end{eqnarray} 
with 
\begin{eqnarray}
m^{{I}}_{{LL}} = - m_{{LR}}
\,M^{-1}_{{RR}}\,m^{ T}_{{LR}}\; ,
\end{eqnarray}
for the light neutrinos. 
The direct mass term  
$m^{{II}}_{{LL}}$ can also provide a naturally small contribution to the 
light neutrino masses if it stems e.g.~from a see-saw suppressed induced VEV,
see Fig.~\ref{fig:TypeIIDiagrams}. 
The general case,  where both possibilities are allowed, is also referred to
as the type II see-saw mechanism. 

Alternatively the see-saw can be implemented in a two-stage process
by introducing additional neutrino singlets beyond the three right-handed
neutrinos that we have considered so far. It is useful to distinguish
between ``right-handed neutrinos'' $\nu_R$ which carry $B-L$ and perhaps
form $SU(2)_R$ doublets with right-handed charged leptons,
and ``neutrino singlets'' $S$ which have no Yukawa couplings
to the left-handed neutrinos, but which may couple to $\nu_R$.
If the singlets have Majorana masses $M_{SS}$, but the right-handed neutrinos
have a zero Majorana mass $M_{RR}=0$, the see-saw mechanism may
proceed via mass couplings of singlets to right-handed neutrinos
$M_{RS}$. In the basis $(\nu_L^c,\nu_R,S)$ the mass matrix is
\beq
\label{double}
\left( \begin{array}{ccc}
0& m_{LR} & 0    \\
m_{LR}^T & 0 & M_{RS} \\
0 & M_{RS}^T & M_{SS}
\end{array}
\right).
\eeq
There are two different cases often considered:

(i) Assuming $M_{SS}\gg M_{RS}$ 
the light physical left-handed
Majorana neutrino masses are then,
\beq
m^{\nu}_{LL} = m_{LR}M_{RR}^{-1}m_{LR}^T \ ,
\eeq
where 
\beq
M_{RR} = M_{RS}M_{SS}^{-1}M_{RS}^T.
\eeq
This is called the double see-saw mechanism~\cite{Mohapatra:bd}.
It is often used in GUT or string inspired neutrino mass
models to explain why $M_{RR}$ is below the GUT or string scale.

(ii) Assuming $M_{SS}\ll M_{RS}$, the matrix has one light and two heavy
quasi-degenerate states for each generation. The mass matrix of the light
physical left-handed Majorana neutrino masses is,
\beq
m^{\nu}_{LL} = m_{LR}{M^{T\,-1}_{RS}} M_{SS}M_{RS}^{-1}m_{LR}^T,
\eeq
which has a double suppression. 

In the limit that $M_{SS}\rightarrow 0$ all neutrinos become massless and
lepton number symmetry is restored. Close to this limit one may have
acceptable light neutrino masses for $M_{RS}\sim~1$~TeV, allowing a testable
low energy see-saw mechanism referred to as the inverse see-saw mechanism. 
If one allows the 1-3 elements of Eq.~(\ref{double}) to be filled
in~\cite{Akhmedov:1995vm} then one obtains another version of the low 
energy see-saw mechanism called the linear see-saw mechanism.

\section{\label{sec:nut}Finite group theory in a nutshell}
In this section, we give a first introduction to finite group theory, using the permutation
group of three objects $S_3$ as an example, and later generalising the discussion to include all finite groups 
with triplet representations. Readers who are familiar with finite group theory may wish to skip this section.

\subsection{Group multiplication table}
Non-Abelian discrete symmetries appear to play an important role in
understanding the physics of flavour. In order for this pedagogical review to
be self-contained, we give a brief introduction into the main
mathematical concepts of finite group theory. Many more details can be found,
for instance, in the recent textbook by Ramond~\cite{Ramond-book} which provides
an excellent survey of the topic particularly aimed at physicists.

Finite groups $G$ consist of a finite number of elements $g$ together with a binary
operation that maps two elements onto one element of $G$. In the following we
use the term multiplication for such an operation. By definition, a group must
include an identity element $e$ as well as the inverse $g^{-1}$ of a given element
$g$. Furthermore, the multiplication must be associative, meaning that the product
of three elements satisfies $(g_1 g_2)g_3=g_1(g_2g_3)$. Groups are called
Abelian if $g_1g_2=g_2g_1$ for all group elements, while the elements of
non-Abelian groups do not satisfy this trivial commutation relation in
general. We shall only be interested in non-Abelian groups from now on.

The most basic way of defining a group is given in terms of the multiplication
table, where the result of each product of two elements is listed. In the case
of the smallest non-Abelian finite group, the permutation group $S_3$, we have:
$$\begin{array}{c|cccccc}
S_3& e & a_1 & a_2 & b_1 & b_2 & b_3 \\\hline
e & e & a_1 & a_2 & b_1 & b_2 & b_3        \\ 
a_1 & a_1 & a_2 & e & b_2 & b_3 & b_1       \\ 
a_2 & a_2 & e & a_1 & b_3 & b_1 & b_2    \\
b_1 & b_1 & b_3 & b_2 & e & a_2 & a_1   \\ 
b_2 & b_2 & b_1 & b_3 & a_1 & e & a_2 \\ 
b_3 & b_3 & b_2 & b_1 & a_2 & a_1 & e
\end{array}$$
The six elements are classified into the identity element $e$, elements $b_i$ whose
square is $e$ and finally elements $a_i$ for which the square
does not yield $e$ but, as can be seen easily, the cube does. It is generally
true for any finite group that there exists some exponent $n$ for each element
$g$ such that $g^n=e$. The smallest exponent for which this holds is called
the order of the element $g$. 
This is not to be confused with the order of a group $G$ which simply means
the number of elements contained in $G$.

\subsection{Group presentation}
Clearly, the definition of a finite group in terms of its multiplication table
becomes cumbersome very quickly with increasing order of $G$. It is therefore
necessary to find a more compact way of defining $G$. Noticing that all six
elements of $S_3$ can be obtained by multiplying only a subset of all
elements, we arrive at the notion of generators of a group. Denoting $a_1=a$
and $b_1=b$, we obtain $a_2=a^2$ as well as $b_2=ab$ and $b_3=ba$. In other
words, $a$ and $b$ generate the group $S_3$. Being the group of permutations
on three objects which is isomorphic to the group of symmetry transformations of
an equilateral triangle, $a$ corresponds to a $120^\circ$ rotation and $b$ to
a reflection. This observation leads to the definition of $S_3$ using the
so-called presentation 
\be
< a\, , \, b ~|~ a^3 \,=\,b^2 \,=\,e \, , \, bab^{-1} \,=\, a^{-1}
> \ ,\label{eq:presentation}
\ee
where the generators have to respect the rules listed on the right. Depending
the these presentation rules, a group can be defined uniquely in a compact
way. Unfortunately, such an abstract definition of a group is not very useful
for physical applications as it does not show the possible irreducible
representations of the group. We therefore quickly continue our journey through the
fields of finite group theory towards the important notion of character tables.

\subsection{Character table}
In order to understand the meaning of a character table, is it mandatory to
introduce the idea of conjugacy classes and irreducible representations. 
Conjugacy classes are subsets of elements of $G$ which are obtained from
collecting all those elements related to a given element $g_i$ by
conjugation $gg_ig^{-1}$, for all $g\in G$. The union of all possible
conjugacy classes is nothing but the set of all elements of $G$. 
In the case of $S_3$ we find three different classes,
\begin{eqnarray}
1C^{1}(1) &=& \{ g\,1\,g^{-1} \,|\,  g\in  S_3 \}~=~\{ 1 \} \ ,\nonumber\\
2C^{3}(a) &=& \{ g\,a\,g^{-1} \,|\,  g\in  S_3 \}~=~\{ a,\,a^2 \}\ , \nonumber\\
3C^{2}(b) &=& \{ g\,b\,g^{-1} \,|\,  g\in  S_3 \}~=~\{ b,\, ab,\,ba\} \ .
\end{eqnarray}
Here we have used the notation $N_i C^{n_i} (g_i)$, where $g_i$ is an element
of the class, $N_i$ gives the number of different elements contained 
in that class, and $n_i$ denotes the order of these elements, which is
identical for all $gg_ig^{-1}$ with $g\in G$.

The other ingredient for constructing a character table is the set of possible
irreducible representations of the group $G$. In general non-Abelian groups can be
realised in terms of $r\times r$ matrices, where the positive integers $r$
depend on the group. Then, the abstract generators of a group are promoted to
matrices which satisfy the presentation rules. Such matrix representations are
called reducible if there exists a basis in which the $r\times r$ matrices of
all generators of $G$ can be brought into the same block diagonal form. If this
is not possible, the representation is called irreducible. Clearly, the
trivial singlet representation ${\bf 1}$, where all generators of
$G$ are identically 1, satisfies any presentation rule and is thus an
irreducible representation of all groups. This trivial example shows that irreducible
representations do not necessarily have to be faithful, i.e. multiplying the
matrices corresponding to the group generators can give a smaller number of
different matrices than the order of $G$. In the case of $S_3$, the
irreducible representations compatible with the presentation rules
of~Eq.~(\ref{eq:presentation}) take the form
\begin{eqnarray}
&{\bf 1}: & a=1\ , \qquad b=1 \ ,\nonumber \\
&{\bf 1'}: & a=1\ , \qquad b=-1 \ ,\nonumber\\[1mm]
&{\bf 2}: & a=\begin{pmatrix}e^{\frac{2\pi i}{3}} & 0 \\ 
0 & e^{-\frac{2\pi i}{3}} \end{pmatrix}\ ,  \qquad
b=\begin{pmatrix}0&1\\ 1&0\end{pmatrix} \ .\label{eq:s3irreps}
\end{eqnarray}
The fact that $S_3$ has three irreducible representations and also three
conjugacy classes is not a coincidence. It is generally true that the number
of irreducible representations of a finite group is equal to the number of its
conjugacy classes. Moreover, summing up the squares of the dimensions of all
irreducible representations always yields the order of the group $G$. For
example, in $S_3$ we get $1^2+1^2+2^2=6$. These two facts can be used to work
out all irreducible representations of a given group $G$.

In the case of irreducible representations ${\bf r}$ with $r>1$, the explicit
matrix form of the generators depends on the basis. In order to obtain a basis
independent quantity, one defines the character $\chi^{[{\bf r}]}_{g_i}$ of
the matrix representation of a group element $g_i$ to be its trace. Since the
elements of a conjugacy class are all related by $gg_ig^{-1}$ with $g\in G$,
it is meaningful to speak of the character $\chi^{[{\bf r}]}_{i}$ of the
elements of a conjugacy class $i$. Therefore one can define the (quadratic) 
character table where the rows list the irreducible representations and the
columns show the conjugacy classes. Using~Eq.~(\ref{eq:s3irreps}), we easily find
the following character table of $S_3$.
$$ 
\begin{array}{c|ccc}
{S_3} & ~1C^{1}(1)~ & ~2C^{3}(a)~ & ~3C^{2}(b)~ \\ \hline
\phantom{\Big|}\chi^{[{\bf 1}]}_i \phantom{\Big|}& 1&1&1 \\
\phantom{\Big|}\chi^{[{\bf 1'}]}_i \phantom{\Big|}& 1&1&-1 \\
\phantom{\Big|}\chi^{[{\bf 2}]}_i \phantom{\Big|}& 2&-1&0 
\end{array}
$$
Defining a group in terms of  its character table is much more suitable for
physical applications than the previous two definitions. First, it immediately
lists all possible irreducible representations which might be used in
constructing particle physics models. Secondly, it is also straightforward to
extract the Kronecker products of a finite group $G$ from its character table.  

\subsection{Kronecker products and Clebsch-Gordan coefficients}
Multiplying arbitrary irreducible representations ${\bf r}$ and ${\bf s}$
\bea
{\bf r\otimes s} ~=~\sum_{\bf t} d({\bf r,s,t})  \; {\bf t} \ ,
\eea
one can calculate the multiplicity $d({\bf r,s,t})$ with which the irreducible
representation ${\bf t}$ occurs in the product by
\bea
d({\bf r,s,t})~=~ \frac{1}{N} \sum_{i} N_i \cdot
 \chi^{[{\bf r}]}_i\chi^{[{\bf s}]}_i {\chi^{[{\bf t}]}_i}^\ast  \ ,
\eea
where the sum is over all classes. $N$ denotes the order of the group $G$ and
the asterisk indicates complex conjugation. With this, we obtain the following
non-trivial Kronecker products from the $S_3$ character table,
\begin{eqnarray}
{\bf 1' \otimes 1'} &=&{\bf 1} \ , \nonumber \\
{\bf 1' \otimes 2~} &=&{\bf 2}  \ , \nonumber\\
{\bf 2 \:\otimes 2~} &=&{\bf 1~+~1'~+~ 2} \ .\nonumber
\end{eqnarray}
The Kronecker products are necessarily independent of the bases of the
irreducible representations ${\bf r}$ with $r>1$. 
When formulating and spelling out the details of a model, particular bases
have to be chosen by hand. With the bases fixed, it is possible to work out
the basis dependent Clebsch-Gordan coefficients of a group. Denoting the
components of the two multiplet of a product by $\alpha_i$ and $\beta_j$, the
resulting representation with components $\gamma_k$ are obtained from,
\bea
\gamma_k = \sum_{i,j}c^k_{ij} ~\alpha_i\,\beta_j \ ,
\eea
where $c_k^{ij}$ are the Clebsch-Gordan coefficients. These are determined by
the required transformation properties of the components $\gamma_k$ under the
group generators. In the case of $S_3$, using the basis of
Eq.~(\ref{eq:s3irreps}), one gets,
\be\begin{array}{lll}
{\bf 1' \otimes 1'}~ \rightarrow ~ {\bf 1} &~~~~~& \alpha \beta \ ,\\[2mm]
{\bf 1' \otimes 2}~ \rightarrow ~ {\bf 2}  && \alpha \begin{pmatrix}\beta_1\\
 - \beta_2\end{pmatrix} \ , \\[5mm]
{\bf 2 \otimes 2}~ \rightarrow ~ {\bf 1}  && \alpha_1\beta_2+
 \alpha_2 \beta_1 \ ,\\[2mm]
{\bf 2 \otimes 2}~ \rightarrow ~ {\bf 1'}  && \alpha_1\beta_2-
 \alpha_2 \beta_1 \ ,\\[2mm]
{\bf 2 \otimes 2}~ \rightarrow ~ {\bf 2}  && \begin{pmatrix}\alpha_2\beta_2\\
 \alpha_1 \beta_1\end{pmatrix} \ , 
\end{array}\nonumber\ee
where $\alpha_i$ refers to the first factor of the Kronecker product and
$\beta_j$ to the second.
We conclude our discussion of the most important concepts in finite group
theory by pointing out that -- due to the choice of convenient
bases -- a representation which is real (that is for which there exists a basis
where all generators are explicitly real) may have complex generators. This is
for instance the case for the doublet of $S_3$ in the basis of~Eq.~(\ref{eq:s3irreps}).

\subsection{Finite groups with triplet representations}
For applications in flavour physics, we are interested in finite
groups with triplet representations. They can be found among the subgroups of
$SU(3)$ and fall into four classes~\cite{Miller-book,Grimus:2011fk}:\footnote{Subgroups of $U(3)$ can be derived from $SU(3)$~\cite{Grimus:2011fk},
however, a complete classification is still lacking~\cite{Ludl:2010bj}.} 
\begin{itemize}
\item Groups of the type $(Z_n\times Z_m)\rtimes S_3$
\item Groups of the type $(Z_n\times Z_m)\rtimes Z_3$
\item The simple groups $A_5$ and $PSL_2(7)$~\cite{Luhn:2007yr} plus a few more
  ``exceptional'' groups~\cite{Fairbairn:1964}
\item The double covers of the tetrahedral ($A_4$), octahedral ($S_4$) and
  icosahedral ($A_5$) groups
\end{itemize}
The latter are subgroups of $SU(2)$, whose triplet representations are
identical to the triplets of the respective rotation groups (which in turn are
already included in the other classes). Many of the physically useful symmetries
are special cases within these general classes. For instance, $S_4$, the natural
symmetry of tri-bimaximal mixing in direct models, see
Subsection~\ref{sec:dirappr}, is isomorphic to 
$\Delta(6n^2) = (Z_n\times Z_n)\rtimes S_3$ with $n=2$. The presentation rules
of $\Delta(6n^2)$ can be given in terms of four\footnote{In principle, the presentation 
can be easily formulated with only three generators
by expressing either $c$ or $d$ in terms of the other three generators.} 
generators, $a,b,c,d$~\cite{Escobar:2008vc},
\be
a^3=b^2=(ab)^2=c^n=d^n=1 \ , \qquad cd=dc \ ,\nonumber
\ee
\be
aca^{-1} = c^{-1}d^{-1} \ , ~\quad ada^{-1} = c \ ,\qquad
bcb^{-1} = d^{-1} \ , ~\quad bdb^{-1} = c^{-1} \ .\label{eq:delta6n2}
\ee
The dimensions of all irreducible representations can only take values 1, 2, 3
or 6.\footnote{As shown in~\cite{Zwicky:2009vt}, this is also true for the
more general series of groups $(Z_n   \times Z_m) \rtimes S_3$.} 
A faithful triplet representation is found, e.g., in the following set
of matrices~\cite{Escobar:2008vc}, 
\be
a=\begin{pmatrix}
0&1&0 \\
0&0&1 \\
1&0&0 
\end{pmatrix}, \quad
b=-\begin{pmatrix}
0&0&1 \\
0&1&0 \\
1&0&0 
\end{pmatrix}, \quad
c=\begin{pmatrix}
\eta&0&0 \\
0&\eta^{-1}&0 \\
0&0&1 
\end{pmatrix}, \quad
d=\begin{pmatrix}
1&0&0 \\
0&\eta&0 \\
0&0&\eta^{-1} 
\end{pmatrix}, \label{eq:6n2irrep}
\ee
where $\eta=e^{\frac{2\pi i}{n}}$. 
With $n=2$ this triplet representation is explicitly real, and therefore does
not correspond to the basis in which the $S_4$ order three generator $T$ is diagonal
and complex, cf. Section~\ref{sec:FS}. To make connection to the $S_4$ triplet
generators $S$, $U$ and $T$ as 
listen in Appendix~\ref{app:CGs}, we have to perform the basis
transformation~\cite{King:2009mk}, 
\be
S=w \,d\, w^{-1} \ , \qquad
U=w \,(aba^{-1})\, w^{-1} \ , \qquad
T=w \,a\,w^{-1} \ ,\label{eq:s4irrep}
\ee
where
\be
w=\frac{1}{\sqrt{3}} \begin{pmatrix}
1&1&1\\
1&\omega&\omega^2\\
1&\omega^2&\omega 
\end{pmatrix} , \quad ~\mathrm{with}~~ \quad \omega=e^{\frac{2\pi i}{3}}\ .
\label{eq:diagonalize-a}
\ee
This shows how the tri-bimaximal Klein symmetry $Z_2\times Z_2$ of the
neutrino mass matrix in the diagonal charged lepton basis, generated by $S$ and $U$
of Eq.~(\ref{eq:TBsu}), is inherited from $\Delta(24)=(Z_2\times Z_2)\rtimes S_3$: 
one $Z_2$ factor (namely $S$) originates from the first factor, $Z_2\times Z_2$,
and the other (namely $U$) is derived from the second, $S_3$. We remark in passing that the
smallest group within the series $\Delta(6n^2)$ containing sextets is
$\Delta(96)$. 

Another series of groups can be obtained from the presentation of
Eq.~(\ref{eq:delta6n2}) by simply dropping the generator $b$, and consequently
all conditions involving $b$~\cite{Luhn:2007uq}. This results in the groups
$\Delta(3n^2)=(Z_n\times Z_n)\rtimes Z_3$ which only allow for irreducible
representations of dimension 1 and 3. The case with $n=2$ generates the
tetrahedral group $A_4$, and the faithful triplet representation is the same
as in the case of $S_4$ only without the $b$ or $U$ generator,
cf. Eqs.~(\ref{eq:6n2irrep},\ref{eq:s4irrep}). With $n=3$ we obtain the group
$\Delta(27)$ which has also been applied successfully as a family symmetry in
indirect models~\cite{deMedeirosVarzielas:2006fc,Ma:2006ip,Ma:2007wu,survey-SM-Delta27}.

\begin{figure}[t]
\begin{center}
\includegraphics[width=0.76\textwidth]{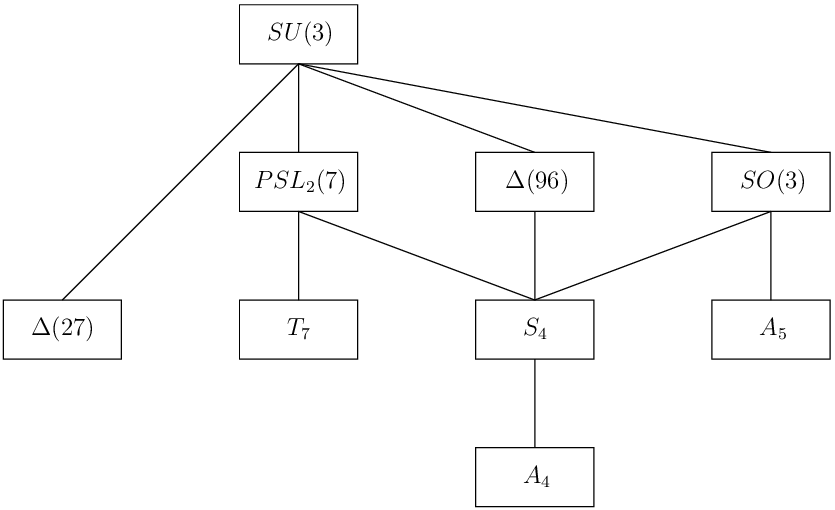}
\end{center}
\caption{\label{fig:family}\small{Examples of subgroups of $SU(3)$ with triplet
    representations discussed in this review. A line connecting two groups indicates that the smaller
    is a subgroup of the bigger one.}}
\end{figure}

A third series is obtained from the second class of groups,
$(Z_n\times Z_m)\rtimes Z_3$, by setting $m=1$. The presentation of this series
of groups $T_n = Z_n \rtimes Z_3$ reads~\cite{Bovier:1980ga}
\be
a^3=c^n=1 \ , \qquad aca^{-1} = c^k \ ,
\label{eq:pres-tn}
\ee
where the integer $k$ must satisfy $1+k+k^2 = 0~\mathrm{mod}~n$. One can
easily check that, with $\eta=e^{\frac{2\pi i}{n}}$, 
a faithful triplet representation is given by
\be
a=\begin{pmatrix}
0&1&0 \\
0&0&1 \\
1&0&0 
\end{pmatrix}, \quad
c=\begin{pmatrix}
\eta&0&0 \\
0&\eta^{k}&0 \\
0&0&\eta^{k^2}
\end{pmatrix}.  \label{eq:tnirrep}
\ee
Popular examples of such groups include
$T_7$~\cite{Luhn:2007sy} and 
$T_{13}$~\cite{Kajiyama:2010sb},
both of which do not include a $Z_2$ subgroup so that the Klein symmetry of the neutrino mass
matrix cannot be obtained from these groups in a direct or semi-direct
way, see Subsection~\ref{sec:dirappr}. Yet, from the model building point of
view it can still be useful to change to a basis in which the order three
generator becomes diagonal~\cite{Luhn:2012bc}, analogously to 
the case of $S_4$. In Appendix~\ref{app:CGs} we list the generators and
Clebsch-Gordan coefficients of the groups $S_4$, $A_4$ and $T_7$ in the $T$
diagonal basis. Their relation to $SU(3)$ and some of its subgroups is
schematically illustrated in~Fig.~\ref{fig:family}.

\section{\label{sec:FS}Discrete family symmetries and model building approaches}

\subsection{\label{sec:FNs}Family symmetries and flavons}
The masses and mixings of the three families of quarks and leptons result from
the form of the respective Yukawa matrices formulated in the flavour basis. Is
there an organising principle which dictates the family structure of these
Yukawa couplings? While this review takes the view that the observed mass and
mixing patterns can be traced back to a family symmetry, we remark that some
authors answer this question negatively, referring to a landscape of parameter
choices out of which Nature has picked one that is compatible with the experimental
measurements. In particular, the observation of a large reactor angle has been
interpreted as a sign for an anarchical neutrino mass
matrix~\cite{Hall:1999sn}. 
Following the symmetry approach, it is clear that the family symmetry must be
broken in order to generate the observed non-trivial structures. This is
achieved by means of Higgs-type fields. These so-called flavon fields $\phi$
are neutral under the SM gauge group and break the family symmetry
spontaneously by acquiring a VEV. This VEV in turn introduces an expansion parameter
\be
\epsilon= \frac{\vev{\phi}}{\Lambda}\ ,
\ee
($\Lambda$ denotes a high energy mass scale) which
can be used to derive hierarchical Yukawa matrices, possibly with texture
zeroes. 

Family symmetries are sometimes also called horizontal symmetries, as
opposed to GUT symmetries which unify different members within a family.
It is possible to impose Abelian or non-Abelian family symmetries. The former
choice goes back to the old idea of Froggatt and
Nielsen~\cite{Froggatt:1978nt} to explain the hierarchies of the quark masses
and mixings in terms of an underlying $U(1)_{\mathrm{FN}}$ symmetry. 
In such a framework, the three generations of (left- and right-handed) quark
fields $q_{L,R}$ (where we do not distinguish up- from down-type quarks)
carry different charges under $U(1)_{\mathrm{FN}}$ such that the usual Yukawa
terms have positive integer charges. Depending on the involved generations,
this can be compensated by multiplying $n_{ij}$ factors of the flavon field
$\phi$ which conventionally has a $U(1)_{\mathrm{FN}}$ charge assignment of
$-1$. The resulting terms which give rise to the usual Yukawa interactions
then take the form 
\be
c_{ij} \left(\frac{\phi}{\Lambda}\right)^{n_{ij}} {\overline{q_L}^{}}_i\, {q_R}_j ~H \ ,
\ee
where $H$ is the Higgs doublet, $i$ and $j$ are family indices and $c_{ij}$ denote undetermined order
one coefficients. Inserting the flavon VEV then generates the Yukawa couplings
$Y_{ij} = c_{ij} \epsilon^{n_{ij}}$ which become hierarchical if the
$U(1)_{\mathrm{FN}}$ charges are chosen appropriately. 
We emphasise that this approach is mainly useful for explaining
hierarchical structures as the order one coefficients $c_{ij}$ remain
unspecified. Nonetheless, there have been recent proposals to adopt extensions of
the Froggatt-Nielsen idea, involving additional generation dependent $Z_n$
symmetries, in order to make qualitative predictions for the lepton sector as
well, in particular aiming to explain the so-called bi-large neutrino mixing
pattern~\cite{Boucenna:2012xb-1,Boucenna:2012xb-2,seidl}.

In order to accurately describe non-hierarchical family structures such as the
observed peculiar lepton mixing pattern, it is necessary to impose a
non-Abelian family symmetry. The three generations of quarks and leptons can
then be unified into suitable multiplets (i.e. irreducible representations) of
the family symmetry $G$. An intimate connection of all three families is
provided if $G$ contains triplet representations, $\psi =
(\psi_1,\psi_2,\psi_3)^T \sim {\bf 3}$. Requiring irreducible
triplet representations, the possible choices for $G$ are limited to $U(3)$ and
subgroups thereof. To illustrate the idea, we sketch the essential steps using
the example of $SU(3)$~\cite{King:2001uz}, applied to the Weinberg
operator $HL^T\kappa L H$, cf. Eq.~(\ref{dim5}). The three generations of lepton
doublets are unified into a triplet of $SU(3)$ while the Higgs doublet $H$ is
assumed to be a singlet of $G$. In order to construct an $SU(3)$ invariant
operator, a flavon field $\phi$ transforming in the ${\bf \ol{3}}$ of $SU(3)$
can be introduced, leading to the term 
\be
H L^T \phi \phi^T L H \ .\label{eq:su3-flavon}
\ee
The VEV of the flavon field $\phi$  will now correspond to a vector with a
particular alignment, i.e. $\vev{\phi} \propto (a,b,c)^T$, where $a,b,c$ take
numerical values dictated by the scalar potential. Inserting this vacuum
configuration into the factor $\phi \phi^T$ of Eq.~(\ref{eq:su3-flavon}) will
generate a contribution to the neutrino mass matrix which is proportional to 
\be
\begin{pmatrix}
a^2&ab&ac \\
ba & b^2 & bc \\
ca&cb&c^2
\end{pmatrix} \ .
\ee
Assuming simple flavon alignments, it is possible to relate all entries of the
neutrino mass matrix in a particular way so that special mixing patterns can
be explained. In practice, at least two  flavons with different alignments
have to be imposed in order to avoid degeneracies in the neutrino masses. 
Although it is possible to obtain simple and predictive flavon alignments in models
based on continuous family symmetries such as $SU(3)$ and $SO(3)$, the problem
of vacuum alignment can be solved in a significantly simpler and more natural way
by imposing a {\it discrete} family symmetry instead. In the following we therefore
focus our attention on non-Abelian discrete family symmetries with triplets.

\subsection{\label{sec:kleinse}The Klein symmetry of the neutrino mass matrix}
The PMNS mixing is dictated by the structure of charged and neutral lepton
mass matrices in a weak eigenstate basis. More precisely it is obtained as the
mismatch of the transformations on the two left-handed lepton states needed to
bring the charged lepton and the neutrino mass matrices into diagonal form. 
In order to easily reach a physical interpretation, it is convenient to work
in a basis in which the charged leptons are diagonal, or approximately
diagonal. The latter is useful in GUT model building where the non-diagonal
hierarchical down-type quark mass matrix, required for the observed CKM 
mixing, is directly related to the charged lepton mass matrix which, as a
consequence, is also not completely diagonal. The total PMNS mixing will then be
predominantly determined by the neutrino mass matrix, and small charged
lepton corrections, see Subsection~\ref{sec:patterns-4}, will have to be taken
into account separately, leading to characteristic mixing sum rules as
explained in Subsection~\ref{sec:patterns-5}.

With this in mind, one can hope to obtain clues on the nature of the underlying
family symmetry by studying the symmetry properties of the neutrino mass
matrix in the basis of (approximately) diagonal charged leptons. Assuming
neutrinos to be Majorana rather than Dirac particles, their mass matrix is
always symmetric under a Klein symmetry $ Z_2 \times  Z_2$. This follows from the
obvious observation that the diagonalised neutrino mass matrix
$m_{LL}^{\nu,\mathrm{diag}}$ is left invariant under the transformation
\be
\wt K_{p,q}^T 
\,m_{LL}^{\nu,\mathrm{diag}}\,
\wt K_{p,q}^{} ~=~ m_{LL}^{\nu,\mathrm{diag}} \ , \qquad
 \mathrm{with}\quad
\wt K_{p,q}~=~\begin{pmatrix}
(-1)^p & 0&0\\
0&(-1)^q&0\\
0&0&(-1)^{p+q}
\end{pmatrix}\ ,
\ee
where $p$ and $q$ take the integer values 0 and 1. The explicit form of the
Klein symmetry of the non-diagonalised neutrino mass matrix $m_{LL}^\nu$,
expressed in terms of $3\times 3$ matrices, can then be determined as 
\be
K_{p,q}~=~U^\ast_{\mathrm{PMNS}}\wt K_{p,q} U^T_{\mathrm{PMNS}}
\ ,
\label{eq:klein}
\ee
where $U_{\mathrm{PMNS}}$ is (approximately) the unitary PMNS mixing
matrix. The matrices $K_{p,q}$ form a group of four elements whose squares yield the identity
element $K_{0,0}$. The fact that the neutrino mass matrix is symmetric under a
transformation by $K_{p,q}$ can be easily verified using Eq.~(\ref{mLLnu}), 
\bea
K_{p,q}^T\,m_{LL}^\nu\,K_{p,q}^{} &=& 
U^{}_{\mathrm{PMNS}}\wt K_{p,q}^T 
(
U^\dagger_{\mathrm{PMNS}}
\,m_{LL}^\nu\,
U^\ast_{\mathrm{PMNS}}
)
\wt K_{p,q}^{} U^T_{\mathrm{PMNS}} \nonumber\\
&=& 
U^{}_{\mathrm{PMNS}}
(
\wt K_{p,q}^T 
\,m_{LL}^{\nu,\mathrm{diag}}\,
\wt K_{p,q}^{} 
)
U^T_{\mathrm{PMNS}} \nonumber\\
&=& 
U^{}_{\mathrm{PMNS}}
\,m_{LL}^{\nu,\mathrm{diag}}\,
 U^T_{\mathrm{PMNS}} \nonumber\\
&=& 
\,m_{LL}^{\nu} \ .
\eea
We point out that the $Z_2\times Z_2$ symmetry of Eq.~(\ref{eq:klein}) exists for
any choice of PMNS mixing. In the remainder of this review we will denote the
two generators of this Klein symmetry by $\sg$ and $\ug$. Particularly simple
forms of these generators are obtained when $U_{\mathrm{PMNS}}$ features a
simple mixing pattern. For instance, in the case of tri-bimaximal mixing,
$U_{\mathrm{PMNS}}=U_{\mathrm{TB}}$, we find 
\be
\sg = \frac{1}{3} \begin{pmatrix}-1&2&2\\ 2&-1&2\\ 2&2&-1\end{pmatrix} \ , \qquad
\ug = - \begin{pmatrix}1&0&0\\ 0&0&1\\ 0&1&0\end{pmatrix}  .
\label{eq:TBsu}
\ee

Such a symmetry of the neutrino mass matrix is only meaningful if the charged
leptons are (approximately) diagonal. Therefore, it is useful to consider also
the (approximate) symmetry of the charged lepton mass matrix $M_e$. As
charged leptons are Dirac particles, one has to consider the square
$M_e M_e^\dagger$ which -- if diagonal -- is symmetric under a general
phase transformation $\tg$,
\be
\tg^\dagger \, \Big(M_e M_e^\dagger\Big)\, \tg 
~=~M_eM_e^\dagger
\ ,
\qquad \mathrm{with}\quad
\tg~=~\begin{pmatrix}
1 & 0&0\\
0&e^{-\frac{2\pi i}{m}}&0\\
0&0&e^{\frac{2\pi i}{m}}
\end{pmatrix} 
\ ,
\label{eq:cl}
\ee 
and $m$ being an integer. The smallest value of $m$ which also enforces a diagonal 
$M_eM_e^\dagger$ is $m=3$. Choosing $m=2$ would leave room for
non-vanishing off-diagonal entries in the 2-3 block of the squared
charged lepton mass matrix, which however could be removed by imposing a
second $\tg$-type generator with permuted diagonal entries.

We conclude the discussion of the symmetries of the mass matrices by 
emphasising that the $\tg$ symmetry of the charged leptons can hold exactly in
models which are only concerned about the lepton sector. In GUT models, which
additionally include the quarks, such a $\tg$ symmetry is usually only valid
approximately.

\subsection{\label{sec:dirappr}The direct model building approach}
Family symmetry models can be classified according to the origin
of the symmetry of the neutrino mass matrix. The neutrino Klein symmetry can
arise as a residual symmetry of the underlying family symmetry~$G$, in other
words, the four elements $K_{p,q}$ of~Eq.~(\ref{eq:klein}) are also elements of
the imposed family symmetry. Models of this type are called  direct
models~\cite{King:2009ap}. 

\begin{figure}[t]
\begin{center}
\includegraphics[clip=true,trim=80mm 40mm 80mm 92mm,
height=60mm]{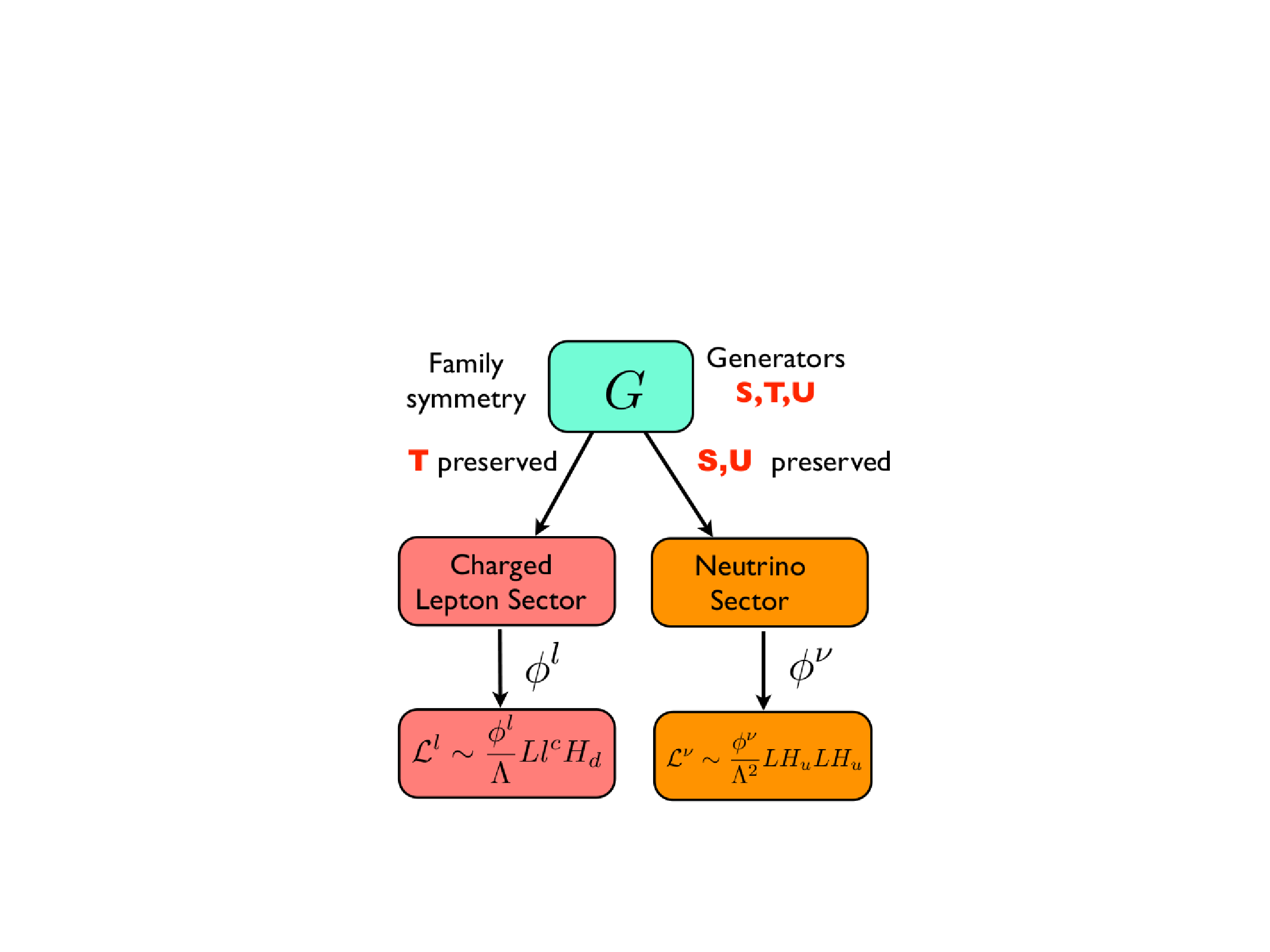}
\end{center}
\caption{\label{fig:directapp}\small{A sketch of the direct model building
approach. The charged lepton sector is (approximately) diagonal either due to a
remnant (approximate) $T$ symmetry or simply by construction.}}
\end{figure}

In such models, the neutrino mass term involves flavon fields $\phi^\nu$
whose vacuum alignments break the family symmetry $G$ down to the remnant
Klein symmetry of~Eq.~(\ref{eq:klein}).  Schematically this can be expressed as
\be
\ma L^\nu ~\sim~ \frac{\phi^\nu}{\Lambda^2} L H_u LH_u \ , \qquad 
\mathrm{with}~\quad
\sg \vev{\phi^\nu} = \ug \vev{\phi^\nu} =  \vev{\phi^\nu}  \ ,
\label{eq:dir-class}
\ee
where the flavon enters only linearly, 
and the lepton doublet $L$ with hypercharge $-1/2$ transforms as a triplet
${\bf 3}$ under the family symmetry, while the up-type Higgs doublet $H_u$
with hypercharge $+1/2$ is a singlet ${\bf 1}$ of $G$.\footnote{We use the
hypercharge convention such that $Q=T_3+Y$.} Depending on the family symmetry,
there are several neutrino-type flavons $\phi^\nu$ which contribute to the
neutrino mass matrix and furnish different representations of $G$, 
typically also including a triplet representation. Therefore,
$\sg$ and $\ug$ stand for the respective Klein generators in the 
representation of $\phi^\nu$. In particular, if the group theory allows
$LH_uLH_u$ to be contracted to a singlet ${\bf 1}$ of $G$, then it is
possible to introduce a flavon in the ${\bf 1}$ of $G$ with
$\sg=\ug=1$. Such a trivial flavon would clearly not break the Klein symmetry,
in fact, it does not even break $G$. 
The same Klein symmetry can be realised straightforwardly in direct models based on
the type~I see-saw mechanism. There, the right-handed neutrinos $\nu^c$
transform as a ${\bf \ol 3}$ of $G$ so that the Dirac neutrino term does not
involve a flavon field, and therefore does not break $G$ at all, whereas the
Majorana neutrino mass term involves the $S$ and $U$ preserving flavons
linearly, 
\be
\ma L^\nu \sim L\nu^cH_u + \phi^\nu \nu^c\nu^c \ .
\ee
Application of the type~I see-saw formula yields an effective light neutrino
mass matrix which is again symmetric under $\sg$ and $\ug$.
Analogously, the charged lepton sector often involves a flavon $\phi^\ell$ which
breaks $G$ (approximately) to the symmetry generated by~$\tg$,
\be
\ma L^\ell ~\sim~ \frac{\phi^\ell}{\Lambda} L \ell^c H_d \ , \qquad 
\mathrm{with}~\quad
\tg \vev{\phi^\ell} \approx   \vev{\phi^\ell}  \ .
\ee
The direct approach is schematically illustrated in
Fig.~\ref{fig:directapp}. However, we remark that according to our
classification, the charged lepton sector may be diagonal simply by
construction.

Assuming the $\tg$ symmetry to be exact, the class of direct models clearly requires
$G$ to contain both the Klein symmetry of the neutrino sector as well as the
symmetry of the charged leptons. The minimal symmetry group for which this is
satisfied can be determined by simply calculating all possible products of the
matrices $\sg$, $\ug$ and $\tg$. In the tri-bimaximal case, i.e. with the
generators of Eq.~(\ref{eq:TBsu}) for the neutrino Klein symmetry, and a charged
lepton symmetry $\tg$ of Eq.~(\ref{eq:cl}) with $m=3$ one obtains a total of 24
different matrices which form a finite group isomorphic to $S_4$, the
group of permutation on four objects. It is interesting to note that a
different choice of $m$ does not necessarily yield a finite group. In fact,
with $\sg$ and $\ug$ of~Eq.~(\ref{eq:TBsu}) one can easily show, using the
computer algebra programme GAP~\cite{GAP4}, that no finite group is obtained
for reasonably small $m\neq 3$. We have checked the validity of this statement
for $m\leq 30$, which is even true for the case with $m=2$. In that sense,
$S_4$ is the natural symmetry group of tri-bimaximal mixing in direct
models. Of course, any group that contains $S_4$ as a subgroup could be
applied equally well.

In addition to the pure class of direct models, there are semi-direct models in
which one $Z_2$ factor of the Klein symmetry can be identified as a subgroup
of~$G$, while the other $Z_2$ factor arises accidentally. The flavons of
semi-direct models appear linearly in the neutrino mass term, similar to
Eq.~(\ref{eq:dir-class}), and break $G$ down to one of its $Z_2$ subgroups. An
example of such a model is provided by the famous Altarelli-Feruglio $A_4$
model of tri-bimaximal mixing~\cite{Altarelli:2005yp,Altarelli:2005yx}. $A_4$
is the group of even permutations on four 
object, and as such a subgroup of $S_4$. It can be obtained from $S_4$ by
simply dropping the $\ug$ generator. Not being a part of the underlying family
symmetry, it is therefore evident that the $\ug$ symmetry of Eq.~(\ref{eq:TBsu})
must arise accidentally.

\subsection{\label{sec:indirappr}The indirect model building approach}
In the class of indirect models, no $Z_2$ factor of the Klein symmetry
of~Eq.~(\ref{eq:klein}) forms a subgroup of~$G$. Models of this class are
typically based on the type~I see-saw mechanism together with the assumption of
sequential dominance, see Subsection~\ref{sec:seesaw-3}. Here, the main role
of the family symmetry consists in relating the Yukawa couplings $d,e,f$ of
Eq.~(\ref{eq:SDdef}) as well as $a,b,c$ of Eq.~(\ref{eq:SDabc}) by introducing triplet
flavon fields which acquire special vacuum configurations. The directions of
the flavon alignments are determined by the $G$ symmetric operators of the
flavon potential~\cite{King:2009ap}.

\begin{figure}[t]
\begin{center}
\includegraphics[clip=true,trim=80mm 41mm 96mm 89mm,
height=60mm]{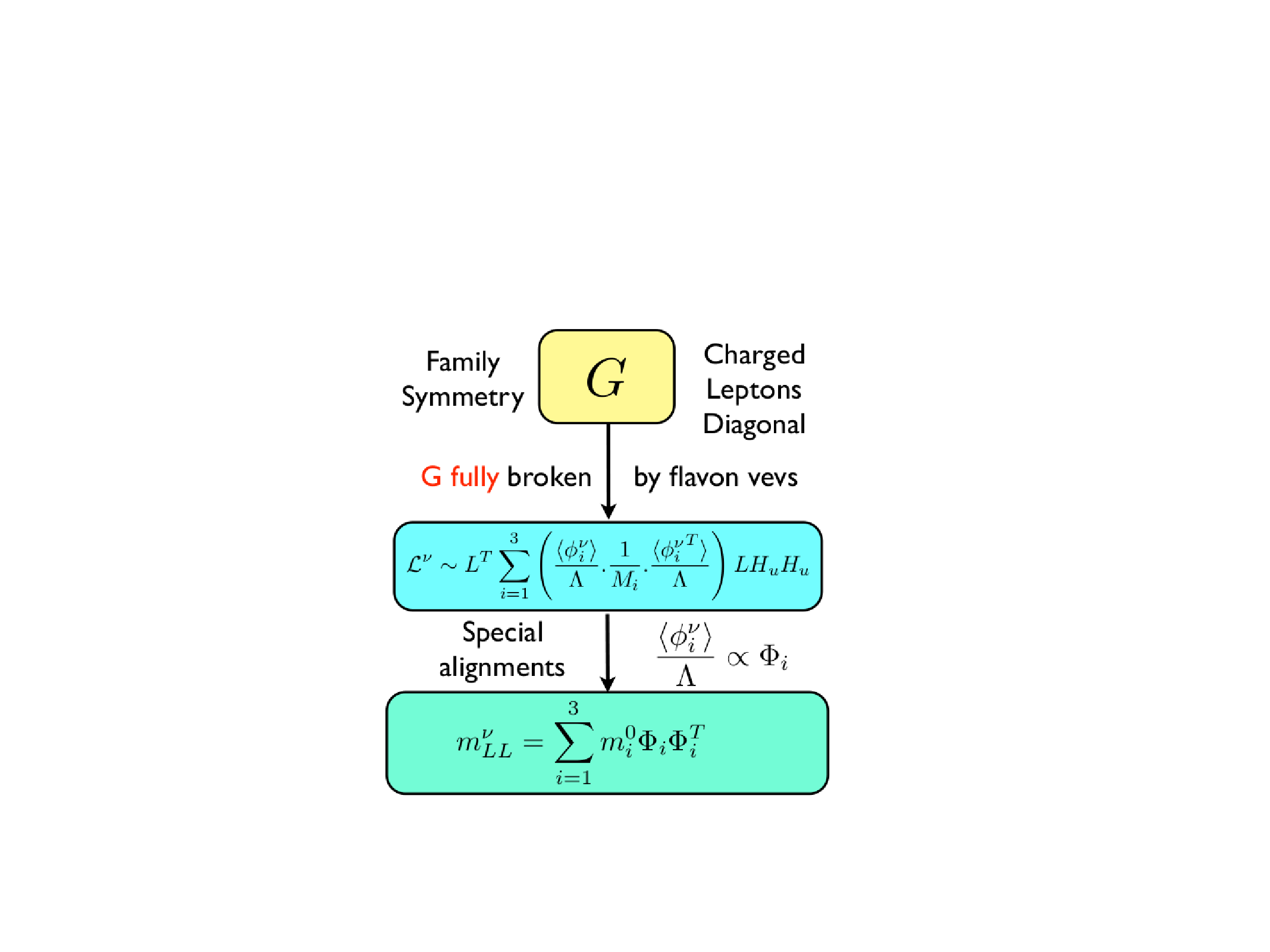}
\end{center}
\caption{\label{fig:indirectapp}\small{A sketch of the indirect model building
approach. The charged lepton sector is (approximately) diagonal by construction.}}
\end{figure}

Working in a basis where both the charged leptons as well as the right-handed
neutrinos are diagonal, the leptonic flavour structure is encoded in the Dirac
neutrino Yukawa operator. The triplet flavons $\phi^\nu_i$ of indirect models enter
linearly in this term,
\be
\ma L^\nu ~\sim~ \sum_i 
\frac{\phi^\nu_i}{\Lambda} L \nu^c_i H_u + M_i \nu^c_i\nu^c_i \ ,
\label{eq:ind-seesaw}
\ee
where $\Lambda$ is a cut-off scale and the sum is over the number of right-handed
neutrinos. The lepton doublet $L$ with hypercharge $-1/2$ transforms as a triplet of
$G$, while the right-handed neutrinos $\nu^c_i$ and the up-type Higgs
doublet with hypercharge $+1/2$ are all singlets of $G$. Adopting the notation
of Subsection~\ref{sec:seesaw-3}, extended to include a third right-handed
neutrino~$\nu^c_1$, we obtain the Dirac neutrino Yukawa matrix by inserting
the flavon VEVs into Eq.~(\ref{eq:ind-seesaw}). Suppressing the dimensionless
couplings of the Dirac neutrino terms for notational clarity, we get 
\be
Y^\nu ~=~ 
\begin{pmatrix} 
a'&a&d\\
b'&b&e\\
c'&c&f
\end{pmatrix}
\sim
\frac{1}{\Lambda}
\begin{pmatrix}
%y^\nu_1 
\vev{\phi^\nu_{1}}_1^{}& 
%y^\nu_2 
\vev{\phi^\nu_{2}}_1^{}& 
%y^\nu_3
\vev{\phi^\nu_{3}}_1^{}\\
%y^\nu_1 
\vev{\phi^\nu_{1}}_2^{}& 
%y^\nu_2 
\vev{\phi^\nu_{2}}_2^{}& 
%y^\nu_3
\vev{\phi^\nu_{3}}_2^{}\\
%y^\nu_1 
\vev{\phi^\nu_{1}}_3^{}& 
%y^\nu_2 
\vev{\phi^\nu_{2}}_3^{}& 
%y^\nu_3
\vev{\phi^\nu_{3}}_3^{}
\end{pmatrix}.\label{eq:indirectYUK}
\ee
The columns of the Dirac neutrino Yukawa matrix are therefore proportional to
the vacuum alignments of the flavons fields $\phi^\nu_i$. The effective
Majorana operators of the light neutrinos can be derived from 
this using the see-saw formula of Eq.~(\ref{seesaw}), yielding
\be
\ma L^\nu_{\mathrm{eff}} ~\sim~  L^T 
 \sum_{i=1}^3 \left(
\frac{
%y^\nu_i
\vev{\phi^\nu_i}}{\Lambda} \cdot 
\frac{1}{M_i}  \cdot
\frac{{
%y^\nu_i
\vev{\phi^\nu_i}}^T}{\Lambda} 
\right)
L  H_uH_u  \ .\label{eq:indirectSQ}
\ee
Note that the flavons enter the effective neutrino mass terms quadratically.
In models with sequential dominance, the three contributions to the effective light
neutrino mass matrix are hierarchical, and it is often possible to ignore one
term (e.g. $i=1$)
such that the sum contains only one dominant and one subdominant
contribution. In the class of indirect models, the PMNS mixing pattern thus
becomes a question of the alignment vectors $\vev{\phi^\nu_i}$. For instance,
a neutrino mass matrix that gives rise to tri-bimaximal mixing can be
obtained using the flavon alignments 
\begin{equation}
\label{Phi0} 
\frac{\vev{{\phi}^\nu_{1}}}{\Lambda}\propto\frac{1}{\sqrt{6}} \begin{pmatrix}-2 \\ 1 \\ 1\end{pmatrix}
, \qquad
\frac{\vev{{\phi}^\nu_{2}}}{\Lambda}\propto\frac{1}{\sqrt{3}} \begin{pmatrix}1 \\ 1 \\ 1\end{pmatrix}
, \qquad
\frac{\vev{{\phi}^\nu_{3}}}{\Lambda}\propto\frac{1}{\sqrt{2}} \begin{pmatrix}0 \\ 1 \\ -1\end{pmatrix}.
\end{equation}
Note that the resulting columns of the Dirac neutrino Yukawa matrix are
proportional to the columns of the unitary (in the present case tri-bimaximal)
mixing matrix. Such a property of the Dirac neutrino Yukawa matrix is
generally called form dominance~\cite{Chen:2009um}.\footnote{Exact form
dominance entails vanishing leptogenesis~\cite{Choubey:2010vs}.} 
Furthermore these alignments are left invariant under the action of the $\sg$
and  $\ug$ generators of Eq.~(\ref{eq:TBsu}), up to an irrelevant sign which
drops out due to the quadratic appearance of each flavon in
Eq.~(\ref{eq:indirectSQ}). Since the family symmetry $G$ does not contain the
neutrino Klein symmetry, its primary role is then to explain the origin of
these or similarly simple flavon alignments.  We schematically illustrated the
indirect approach in Fig.~\ref{fig:indirectapp}.  

For completeness we also mention that there are a few special models not based
on the type~I see-saw mechanism, in which the Klein symmetry of the neutrino sector
is not a subgroup of the family symmetry, see e.g.~\cite{Luhn:2012bc,Ma:2007wu}. 
In such models, the flavon enters linearly in the neutrino mass term, and the
Klein symmetry arises accidentally from a combination of the Clebsch-Gordan
coefficients and suitable flavon alignments of~$\phi^\nu$.

\subsection{Comments on the classification into direct and indirect models}
Before continuing to discuss the details of the family symmetry breaking, we
compare the direct approach with the indirect one, see 
Figs.~\ref{fig:directapp} and~\ref{fig:indirectapp}. This classification is
purely based on the origin of the $Z_2\times Z_2$ Klein symmetry of the
neutrino sector, formulated in a basis of (approximately) diagonal charged
leptons. In direct models, this symmetry, generated by the order two elements
$S$ and $U$, arises as a subgroup of $G$, whereas this is not the case for
indirect models.  In both approaches, the family symmetry $G$ has to be broken
spontaneously by flavon fields acquiring a VEV. The flavon vacuum
configuration of direct models is dictated by the requirement that $S$ and $U$
be preserved. In indirect models based on the type~I see-saw, the vacuum
alignment of the flavons  enters in the columns of the Dirac neutrino Yukawa
matrix, thereby generating contributions to the effective light neutrino mass
matrix of the form proportional to $\vev{\phi^\nu} \vev{\phi^\nu}^T$. 

We emphasise that the charged leptons have to be (approximately) diagonal by
construction for the purpose of this classification. In the framework of direct models,
this can be enforced by demanding a subgroup of $G$, generated by $T$, to be
(approximately) preserved in the charged lepton sector. However, in grand unified
models such a $T$ symmetry cannot be exact as it would then apply to the
quarks as well. This in turn would entail a phenomenologically unacceptable
quark sector without CKM mixing.

\section{\label{sec:DmB}Direct model building}

In models based on a family symmetry $G$, the Dirac Yukawa and the Majorana
couplings are typically generated dynamically from $G$ invariant operators
involving one or more flavon fields. In general, these flavons can transform
in any of the irreducible representations of $G$. For non-Abelian
discrete symmetries, the choice is limited to a finite set of representations.
With flavons transforming as multiplets of the family symmetry $G$, the
breaking of $G$ and with it the family structure of the Dirac Yukawa and the
Majorana couplings crucially depends on the {\it alignment} of the flavon
VEVs. In this section we discuss general strategies for identifying useful
flavons alignments in direct models where the family symmetry $G$ is broken to
a particular subgroup in the neutrino sector. Furthermore, we give explicit
examples which illustrate ways of deriving vacuum alignments from flavon potentials. 
We remark that throughout this section we assume a diagonal or approximately
diagonal charged lepton mass matrix which may or may not arise as a result of
an (approximately) unbroken subgroup of $G$.

\subsection{Flavon alignments in direct models}
In direct models, flavons enter linearly in the terms of the neutrino Lagrangian as
shown in Eq.~(\ref{eq:dir-class}). These operators are invariant under the full
family symmetry $G$. When $G$ is broken spontaneously by the flavon fields
acquiring a VEV, the smaller $Z_2\times Z_2$ Klein symmetry, generated
by the order two elements $S,U\in G$, still remains intact. The criteria on the
vacuum alignment of the involved flavon VEVs can therefore be formulated by
the condition
\be
S \vev{\phi^\nu} = U \vev{\phi^\nu} = \vev{\phi^\nu} \ .\label{eq:direct-flavons}
\ee
With flavons generally furnishing different representations ${\bf{r}}$ of the family
group $G$, the explicit matrix form of $S$ and $U$ clearly differs for
different ${\bf{r}}$. For a given representation, it is then straightforward to calculate
the form of the alignments which satisfy Eq.~(\ref{eq:direct-flavons}). This
procedure can be repeated for all other representations ${\bf{r}}$, but not
all of them will necessarily yield a solution to Eq.~(\ref{eq:direct-flavons}),
meaning that flavons transforming in such representations cannot be adopted
in the considered case.

We discuss this strategy of identifying the structure of the flavon alignments
in an explicit example for the purpose of illustration. Consider the case
of an $S_4$ family symmetry. The generators $S$ and $U$ of the tri-bimaximal Klein
symmetry are listed in Appendix~\ref{app:CGs} for all five irreducible
representations, cf. also Eq.~(\ref{eq:TBsu}). The $S$ generator of the
${\bf{1}}$, ${\bf{1'}}$ and ${\bf{2}}$ are all trivial, i.e. the identity
element. Therefore any vacuum configuration of flavons transforming in these
representations will leave invariant the $Z_2$ symmetry associated
with~$S$. The second $Z_2$ symmetry of the Klein symmetry, generated by $U$ is 
always broken by the VEV of a flavon transforming in the ${\bf{1'}}$ since
$U=-1$ in this case, while it is left intact by a flavon in the ${\bf{1}}$ of $S_4$.
For the two-dimensional representation, one quickly finds that the flavon
alignment has to be proportional to $(1,1)^T$ in order not to break the $U$
symmetry. Turning to the ${\bf{3}}$ of $S_4$, invariance under $U$ entails a
flavon alignment of the form $(0,1,-1)^T$. Applying the $S$ transformation on
such an alignment yields $(0,-1,1)^T$, hence, this alignment does not satisfy
Eq.~(\ref{eq:direct-flavons}) as it is not an eigenvector of $S$ with
eigenvalue~$+1$. Finally, we discuss flavons transforming in the ${\bf{3'}}$
representation of $S_4$. The most general alignment which is left invariant
under the $U$ transformation reads $(a,b,b)^T$. Demanding $S (a,b,b)^T =
(a,b,b)^T$ fixes $a=b$, showing that an alignment proportional to $(1,1,1)^T$
leaves invariant both $S$ and $U$. Collecting the results of this discussion,
we have shown that flavons transforming as a ${\bf{1'}}$ and ${\bf{3}}$ of
$S_4$ cannot be used to break the family symmetry $S_4$ down to the
tri-bimaximal Klein symmetry. On the other hand, flavon fields in the
${\bf{1}}$, ${\bf{2}}$ and ${\bf{3'}}$ representations can be adopted, where
the latter two have to be aligned as 
\be
\vev{\phi^\nu_{\bf{2}}} = \varphi^\nu_{\bf{2}}  \begin{pmatrix}1\\1\end{pmatrix}  \ , \qquad
\vev{\phi^\nu_{\bf{3'}}} = \varphi^\nu_{\bf{3'}}  \begin{pmatrix}1\\1\\1\end{pmatrix}  \ ,
\label{s4-tb-ali}
\ee 
in order to leave invariant the $Z_2$ symmetries associated with $S$ and
$U$. Here $\varphi$ denotes the overall VEV of a flavon $\phi$.
Inserting all three flavons into Eq.~(\ref{eq:dir-class}), assuming the
lepton doublets $L$ to transform in the ${\bf{3}}$ representation, we end up
with a neutrino mass matrix which comprises three terms,
\be
m_{LL}^\nu ~\approx~\left[ 
 \varphi^\nu_{\bf{3'}}\begin{pmatrix}
2&-1&-1\\
-1&2&-1\\
-1&-1&2
\end{pmatrix}
+
 \varphi^\nu_{\bf{1}} \begin{pmatrix}
1&0&0\\
0&0&1\\
0&1&0
\end{pmatrix}
+
 \varphi^\nu_{\bf{2}} \begin{pmatrix}
0&1&1\\
1&1&0\\
1&0&1
\end{pmatrix}
\right] \frac{v_u^2}{\Lambda^2} \ .\label{eq:tb-mass-ma}
\ee
Using the matrices $S$ and $U$ of Eq.~(\ref{eq:TBsu}), one can easily check
explicitly that $S^Tm_{LL}^\nu S = U^Tm_{LL}^\nu U = m_{LL}^\nu$ as required.
Clearly, the alignments of Eq.~(\ref{s4-tb-ali}) depend on the chosen basis. In
particular the basis of the doublet representation could have been chosen
differently without affecting the basis of the triplet representation (which
we fixed by demanding a diagonal $T$ generator). This, however, would also
change the Clebsch-Gordan coefficients such that the form of the neutrino mass
matrix in Eq.~(\ref{eq:tb-mass-ma}) remains unchanged. We emphasise that the same 
procedure of identifying the flavon alignments of direct models can be applied
to arbitrary choices of the Klein symmetry.

\subsection{\label{sec:vacuumdir}Vacuum alignment mechanism in direct models}
Having determined the alignments required in a given direct models, the next step is
to derive them from minimising a flavon potential. In the context of direct
models, the most popular and perhaps natural approach to tackle the problem of the
flavon alignment is provided by the so-called $F$-term alignment 
mechanism~\cite{Altarelli:2005yp,Altarelli:2005yx}. The idea is to couple the
flavons to so-called driving fields in a supersymmetric setup. Like flavons,
driving fields are neutral under the SM gauge group and transform in general
in a non-trivial way under the family symmetry $G$. Introducing a $U(1)_R$
symmetry under which the chiral supermultiplets containing the SM fermions
carry charge $+1$, allows to distinguish flavons from driving fields by
assigning a charge of $+2$ to the latter while keeping the former
neutral. With this $U(1)_R$ charge assignment, the driving fields can only
appear linearly in the superpotential and cannot couple to the SM
fermions. The set of superpotential operators involving the driving fields
$X_i$ is usually referred to as the driving or simply flavon potential
$W_{\mathrm{flavon}}$. 
Assuming that supersymmetry remains unbroken at the scale where the flavons
develop their VEVs, we can obtain the flavon alignments from the terms of
$W_{\mathrm{flavon}}$ by setting the $F$-terms of the driving fields to zero, i.e.
\be
- F_{X_i}^\ast = \frac{W_{\mathrm{flavon}}}{X_i} = 0 \ , \label{eq:ftermcond}
\ee
by which the scalar potential is minimised. 

To illustrate the $F$-term alignment mechanism we give two simple examples
based on the family symmetry group $S_4$. First, consider a driving field
$X_{\bf{1}}$ and a flavon field $\phi_{\bf{2}}$ transforming in the ${\bf{1}}$
and ${\bf{2}}$ representations of $S_4$, respectively. Expanding the
resulting term of the driving superpotential in terms of the component
fields $\phi_{{\bf{2}},i}$ we obtain
\be
X_{\bf{1}} \phi_{\bf{2}} \phi_{\bf{2}} ~=~ X_{\bf{1}} 
( \phi_{{\bf{2}},1} \phi_{{\bf{2}},2}  +  \phi_{{\bf{2}},2} \phi_{{\bf{2}},1})
= 2 X_{\bf{1}}  \phi_{{\bf{2}},1} \phi_{{\bf{2}},2}\ .\label{eq:01doub}
\ee
The $F$-term condition of Eq.~(\ref{eq:ftermcond}) then gives rise to the
following two solutions,
\be
\vev{\phi_{\bf{2}}} \propto \begin{pmatrix}1\\0 \end{pmatrix} \ ,\qquad
\vev{\phi_{\bf{2}}} \propto \begin{pmatrix}0\\1 \end{pmatrix} \ .\label{eq:01doublet}
\ee
Notice that these two alignments are related by the $S_4$ symmetry
transformation~$U$, while a transformation induced by $T$ does not change
the alignment but only the phase of the VEV. It is a general feature of any
$G$ symmetric theory that one particular solution for the flavon alignments
will automatically imply a whole set of solutions which are related by
symmetry transformations. However, the reverse need not be true, i.e. there
may be cases in which two or more solutions exist which are not related
through symmetry transformations. 

As a second example let us consider the alignments of Eq.~(\ref{s4-tb-ali}). One
possible way to derive these using the $F$-term alignment mechanism consists
in introducing two driving fields, one transforming in the ${\bf{3}}$ of
$S_4$, the other in the ${\bf{3'}}$~\cite{King:2011zj}.
The corresponding terms of the flavon superpotential then read
\be
g_0 X_{\bf{3}}  \phi^\nu_{\bf{3'}} \phi^\nu_{\bf{2}} ~+~
X_{\bf{3'}} \left( g_1 \phi^\nu_{\bf{3'}} \phi^\nu_{\bf{3'}} +
g_2 \phi^\nu_{\bf{3'}} \phi^\nu_{\bf{2}} +
g_3 \phi^\nu_{\bf{3'}} \phi^\nu_{\bf{1}} 
\right) \ , 
\ee
where $g_i$ are dimensionless coupling constants. Denoting the VEVs of
$\phi^\nu_{\bf{3'}}$, $\phi^\nu_{\bf{2}}$ and $\phi^\nu_{\bf{1}}$ by $c_i$,
$b_i$ and $a$, respectively, the $F$-term conditions take the form
\be
g_0\left[b_1 \begin{pmatrix} c_2\\c_3 \\c_1  \end{pmatrix} - 
b_2 \begin{pmatrix} c_3\\c_1 \\c_2  \end{pmatrix}\right] ~=~
\begin{pmatrix} 0\\0 \\0  \end{pmatrix} \ ,\label{eq:cons41}
\ee
\be
2g_1 \begin{pmatrix} c_1^2-c_2 c_3\\ c_3^2-c_1 c_2\\ c_2^2-c_3
  c_1 \end{pmatrix}+ 
g_2 
\left[b_1 \begin{pmatrix} c_2\\c_3 \\c_1  \end{pmatrix} +
b_2 \begin{pmatrix} c_3\\c_1 \\c_2  \end{pmatrix} \right]
+g_3 a
\begin{pmatrix}c_1 \\c_2 \\c_3  \end{pmatrix} ~=~
\begin{pmatrix} 0\\0 \\0  \end{pmatrix} \ .\label{eq:cons42}
\ee
Restricting to solutions in which all of the three flavons develop a VEV,
Eq.~(\ref{eq:cons41}) requires non-zero values for all $b_i$ and all $c_i$.
Using this, it is straightforward to find the most general solution to the set
of $F$-term equations. Up to symmetry transformations, we obtain
\be
\vev{\phi^\nu_{\bf{3'}}} = \varphi^\nu_{\bf{3'}} \begin{pmatrix} 1\\1
  \\1 \end{pmatrix} \ , \qquad
\vev{\phi^\nu_{\bf{2}}} = \varphi^\nu_{\bf{2}} \begin{pmatrix} 1\\1
 \end{pmatrix} \ , \qquad 
\varphi^\nu_{\bf{2}} = -\frac{g_3}{2g_2} \varphi^\nu_{\bf{1}} \ .
\ee

We remark that the trivial vacuum, that is the vacuum configuration where none of
the flavons develops a VEV, typically provides a solution to the $F$-term
equations as well. This can be eliminated by including soft supersymmetry
breaking effects. Then the scalar potential relevant for the flavon alignments
takes the general form 
\be
V_{\mathrm{flavon}}~=~\sum_i 
 \bigg| \frac{W_{\mathrm{flavon}}}{X_i} \bigg|^2 +
\bigg| \frac{W_{\mathrm{flavon}}}{\phi_i} \bigg|^2 +
m_{X_i}^2 |X_i|^2 + m_{\phi_i}^2|\phi_i|^2
 +~ \cdots \ , \label{eq:scalar-pot}
\ee
where $m_{X_i}^2$ and $m_{\phi_i}^2$ denote the soft breaking masses of the driving
fields $X_i$ and the flavons~$\phi_i$. The dots stand for additional soft
breaking terms. Assuming positive $m_{X_i}^2$, the driving fields do not
develop a VEV. As a consequence, the operators which involve a driving field,
i.e. those represented by the second term of Eq.~(\ref{eq:scalar-pot}),
vanish. The first term, on the other hand, only depends on the flavon
fields. This together with negative $m_{\phi_i}^2$ removes the trivial vacuum
configuration and the flavons acquire a VEV~\cite{Altarelli:2005yp,Altarelli:2005yx}. 

Alternatively, it is in principle also possible to add an explicit mass scale in the flavon
potential which will then drive the flavon VEVs to non-zero values. For
instance, the cube of the $S_4$ doublet flavon $\phi_{\bf{2}}$ of
Eq.~(\ref{eq:01doublet}) can be contracted to an $S_4$ singlet with a non-vanishing
VEV. Introducing a driving field $X'_{\bf{1}}$ one could therefore write down
the driving terms
\be
X'_{\bf{1}} \left[ \frac{(\phi_{\bf{2}})^3}{M} - M^2 \right] \ ,\label{eq:driv-term}
\ee
where $M$ denotes an explicit mass scale. Since $M$ is a pure (dimensionful)
number, the driving field $X'_{\bf{1}} $ and with it $(\phi_{\bf{2}})^3$  must be
completely neutral under any imposed symmetry. In particular, they must not
carry charges under extra so-called shaping symmetries which are typically
introduced in concrete models to separate the flavons of different sectors. In
the given example, a $Z_3$ shaping symmetry under which the flavon
$\phi_{\bf{2}}$ (as well as the driving field $X_{\bf{1}}$) carries charge
$+1$ allows for the coexistence of the alignment term of Eq.~(\ref{eq:01doub})
together with the term of Eq.~(\ref{eq:driv-term}) which explictly drives the
flavon VEV to non-zero values. Assuming a CP conserving high energy theory
where all parameters of the model can be chosen to be real, driving terms of
the form of Eq.~(\ref{eq:driv-term}) could generate spontaneous CP violation
where the values of the CP violating phases are constrained to a finite number
of choices~\cite{Antusch:2011sx}.

\subsection{Direct models after Daya Bay and RENO}
The method of identifying the flavon alignments of direct models using
Eq.~(\ref{eq:direct-flavons}) can be applied to any mixing pattern. Yet, until
recently, the main focus was limited to only a few simple cases, namely
tri-bimaximal, bimaximal and golden ratio mixing, see
Subsection~\ref{sec:patterns-1}, all of 
which predict $\theta_{13}=0^\circ$. The observation of a sizable reactor
neutrino mixing angle of about $8.5^\circ$ by the Daya Bay and RENO
collaborations in early~2012,  preceded by first hints for a non-zero
$\theta_{13}$ from the T2K, MINOS and Double Chooz experiments in~2011, has
now ruled out these simple mixing patterns. This fact seems to call into question the
direct model building approach. However, it is worth recalling that a
vanishing reactor angle has long been compatible with experimental data, and
hence there was no need to consider more complicated mixing patterns. The
situation has now changed, and new strategies for constructing direct models
have to be conceived. 

\begin{figure}[t]
\begin{center}
\includegraphics[clip=true,trim=50mm 0mm 50mm 60mm,
height=80mm]{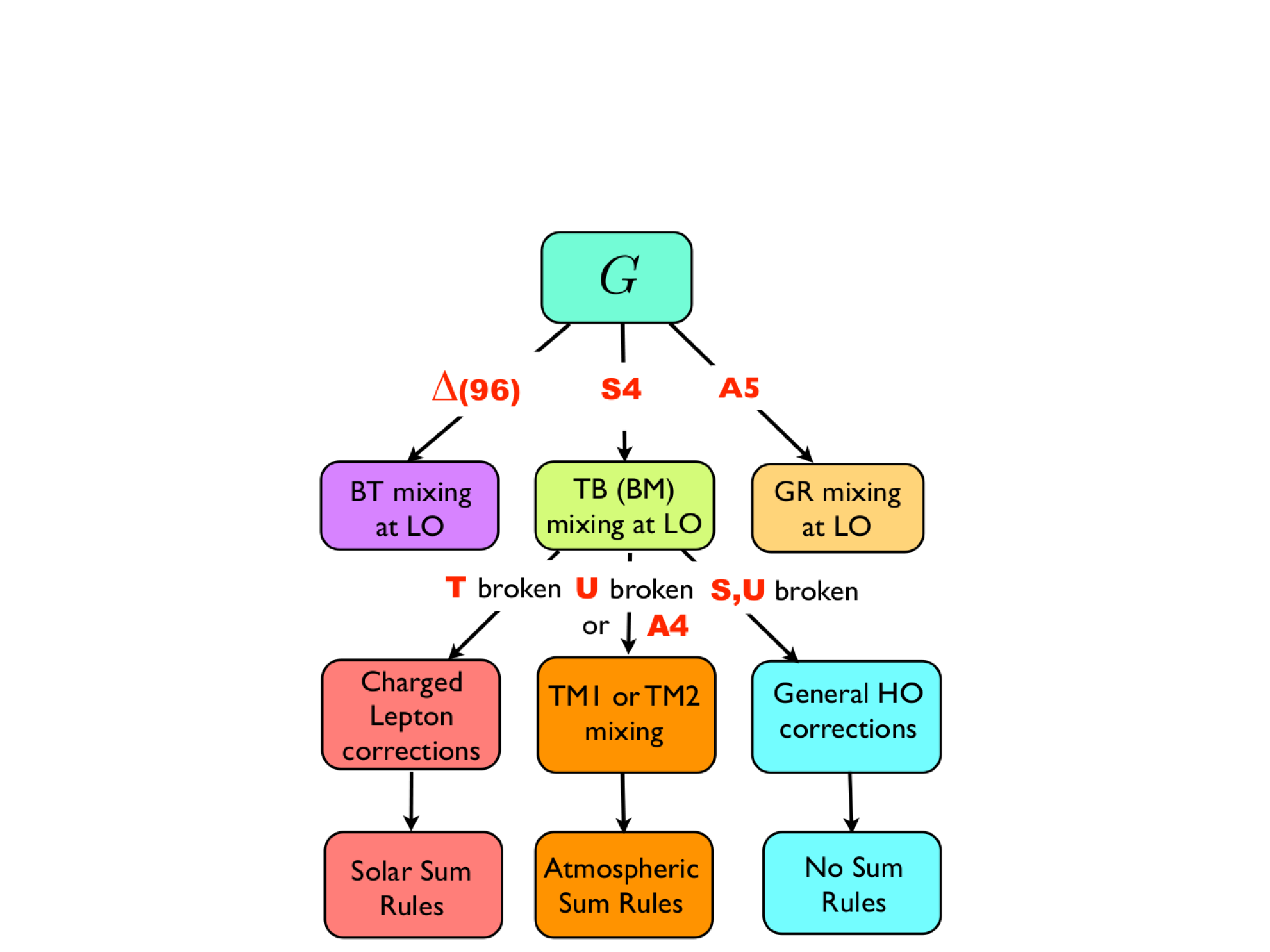}
\end{center}
\caption{\label{fig:direct_map}\small{Possible strategies for constructing
    direct models after Daya Bay and RENO. Adopting small family symmetries
    $G$  which predict simple leading order (LO) mixing patters with $\theta_{13}=0$
    (e.g. $S_4$, $A_5$), requires higher order (HO) corrections. Larger family
    symmetries can give rise to richer LO mixing patterns with non-zero $\theta_{13}$
    (e.g. $\Delta(96)$). The $A_4$ family symmetry refers to the semi-direct case as
    discussed in the text.
In this diagram, we have used the acronyms 
BT=bi-trimaximal,
TB=tri-bimaximal,
BM=bimaximal,
GR=golden ratio,
TM=trimaximal.
}}
\end{figure}

There are two main paths one can pursue. The first is based on {\it small
groups} like e.g. $S_4$ and $A_5$, and leads to the simple TB, BM or GR
mixing patterns with vanishing reactor angle at leading order. 
To render such models compatible with a sizable $\theta_{13}$ it is
critical to discuss higher order corrections which break the simple structure
of the mixing matrix. This situation is depicted in the central branch of
Fig.~\ref{fig:direct_map} for the family symmetry~$S_4$. Depending on which
group elements are selected for the symmetries of the charged lepton sector ($T$)
and the neutrino sector ($S,U$), it is possible to obtain either TB or BM
mixing from $S_4$ at leading order. Analogously, the leading order GR mixing
derived from $A_5$ can be perturbed by higher order effects (not shown
explicitly in Fig.~\ref{fig:direct_map}). In general, higher order corrections are
guaranteed to perturb the leading order structure by only small
contributions. The breaking of the leading order structure can happen either
in the charged lepton or the neutrino sector. The former entails charged
lepton corrections of the simple leading order mixing patterns, which give
rise to solar mixing sum rules as discussed in
Subsection~\ref{sec:patterns-5}. If the breaking occurs in the neutrino
sector, it is possible to break either one or both $Z_2$ factors of the
leading order Klein symmetry. As the $U$ symmetry typically enforces
$\theta_{13}=0$ in these models, it is necessary to break $U$ in either
case. Demanding $S$ to remain a good symmetry at higher order, gives rise to
atmospheric mixing sum rules, see Subsection~\ref{sec:patterns-6}, while
breaking also $S$ leads to arbitrary and unpredictive higher order corrections. In
Subsection~\ref{sec:s4su5} we will present a concrete $S_4\times SU(5)$ model
of tri-bimaximal mixing at leading order, augmented by higher order
corrections which break $U$ but not $S$. This model yields the trimaximal
neutrino mixing pattern TM2, see Eq.~(\ref{TM2}), which can accommodate a sizable
reactor angle. 

The second strategy of constructing direct models compatible with 
a sizable reactor angle makes use of {\it larger groups} such as $\Delta(96)$,
see left branch of Fig.~\ref{fig:direct_map}. Such groups are capable of
predicting richer leading order mixing patterns (e.g. bi-trimaximal
mixing~\cite{King:2012in}) as they contain non-standard 
Klein symmetries, generated by more complicated forms of the elements $S$
and/or $U$~\cite{Toorop:2011jn,deAdelhartToorop:2011re}. As 
before, higher order effects can correct these leading order mixing
patterns. Charged lepton corrections induced by a breaking of the $T$
generator give rise to new solar sum rules. Indeed, in the
$\Delta(96)\times SU(5)$ model discussed in Subsection~\ref{sec:delta96}, the
charged lepton corrections are essential in driving the resulting reactor
angle to a physically viable value. 
In the neutrino sector, it is generally possible to break either $S$ or $U$,
however, in typical models, the symmetry associated with $S$ stabilises the
solar angle at a phenomenologically viable value. In practice, it should
therefore be the $U$ generator which gets broken at higher order, leading to
new atmospheric sum rules.

Before turning to the breaking of the family symmetry in indirect models, we
comment on the case of semi-direct models. As mentioned at the end of
Subsection~\ref{sec:dirappr}, the Altarelli-Feruglio $A_4$
model~\cite{Altarelli:2005yp,Altarelli:2005yx} provides an example of a
semi-direct model. While the tri-bimaximal $S$ symmetry of the neutrino mass
matrix forms part of the family symmetry, the tri-bimaximal $U$ symmetry
arises accidentally due to the absence of flavons in the ${\bf 1'}$ and ${\bf 1''}$
representations of $A_4$. Introducing such neutrino type flavons in the non-trivial
one-dimensional representations, a situation which is tantamount to breaking
the $U$ generator in $S_4$ (see Fig.~\ref{fig:direct_map}), generates
contributions to the neutrino mass matrix which are not of tri-bimaximal
form~\cite{Brahmachari:2008fn,King:2011zj}. However,
as can be seen from Appendix~\ref{app:CGs}, they respect the original $S$
symmetry, thus enforcing the trimaximal mixing pattern TM2, see Eq.~(\ref{TM2}). 
The accidental tri-bimaximal $U$ symmetry, on the other hand, gets broken by
the VEVs of the flavons in the ${\bf   1'}$ and ${\bf 1''}$ representations,
which allows to accommodate arbitrary values of $\theta_{13}$. The relative
smallness of the reactor angle, compared to the solar and atmospheric
angles, remains unaccounted for and must therefore be understood as a result
of a mild tuning of parameters.

A similar semi-direct approach was taken by Hernandez and
Smirnov~\cite{Hernandez:2012ra} in an effort to accommodate a sizable reactor
angle. Focusing on the relevant von Dyck groups $A_4$, $S_4$ and $A_5$, they
demand the $T$ symmetry of the charged leptons and (only) one $Z_2$ factor of the
neutrino Klein symmetry\footnote{Assuming a diagonal charged lepton sector,
correlations of the neutrino mixing parameters arising from
requiring only one $Z_2$ factor were previously derived in~\cite{Ge:2011ih}.}
to arise as unbroken subgroups of the underlying family
symmetry. This strategy allows to identify viable mixing patterns in which a
given column of the PMNS matrix is completely determined by the properties of
the imposed symmetry group. For the successful cases, these columns are
identical (in some cases up to permutations of the rows) to either the first
or the second column of the bimaximal, the tri-bimaximal and the golden ratio mixing
patterns, cf.~Subsection~\ref{sec:patterns-1},~\cite{BalletPeter}. With the
other two  columns of the mixing matrix unspecified, the reactor angle can be
regarded as a free  parameter which, together with the CP phase~$\delta$,
gives rise to predictions for the other two mixing angles, expressed in the
form of (exact) sum rules.

\section{\label{sec:ImB}Indirect model building}

\subsection{\label{sec:ImB-1}Flavon alignments in indirect models}

The vast majority of indirect models is formulated in the framework of the
type I see-saw mechanism where the right-handed neutrino and the charged lepton
mass matrices are both diagonal, see Subsection~\ref{sec:indirappr}.
The lepton mixing arises from the structure of the Dirac neutrino Yukawa matrix, which
in turn originates from the alignment of the flavon fields~$\phi^\nu_i$.
With the lepton doublet~$L$ furnishing a triplet representation ${\bf 3}$ of
the family symmetry $G$, the neutrino flavons typically transform as 
a ${\bf \ol  3}$ of $G$.\footnote{If the triplet ${\bf 3}$ is real, $L$ and
$\phi^\nu_i$ transform in the same representation. In indirect models, the
basis of the triplet representation of $G$ must then be chosen explicitly
real. Note that for this reason the Clebsch-Gordan coefficients of $S_4$ and
$A_4$ given in Appendix~\ref{app:CGs} are not applicable in indirect
models. For a basis suitable for indirect models, see Eq.~(\ref{eq:6n2irrep})
and the discussion thereafter.} The family indices are then contracted to the $G$
singlet in the familiar $SU(3)$ way, showing that the columns of the Dirac
neutrino Yukawa matrix are proportional to the alignments of the flavon
fields~$\phi^\nu_i$, as presented in Eq.~(\ref{eq:indirectYUK}). 
Application of the see-saw formula gives rise to an effective light neutrino
mass matrix of the form
\be
m^\nu_{LL} ~=~ \sum_{i=1}^3 m_i^0 \Phi_i^{} \Phi_i^T  \ ,\label{eq:lightnuind}
\ee
where $\Phi_i \propto \frac{\vev{\phi^\nu_i}}{\Lambda}$ denotes a dimensionless
vector normalised to one. From the model building perspective, the direction of
these vectors in flavour space depends on the alignment of the flavon VEVs. In
general one can distinguish two cases. 
In this subsection we focus on the case where the flavon alignments are
orthogonal to each other. The situation where this is not the case will be
treated in Subsection~\ref{sec:indirect-flavon}.

Under the assumption that $\Phi_i$ and $\Phi_j$ are orthogonal for $i\neq j$, 
the light neutrino mass matrix $m^\nu_{LL}$ of Eq.~(\ref{eq:lightnuind}) is
diagonalised by a unitary PMNS mixing matrix with columns $\Phi_i^\ast$.
The resulting eigenvalues are simply $m^0_i$. This scenario, in which the
columns of the Dirac neutrino Yukawa matrix are proportional to the columns of
the PMNS mixing matrix, is called form dominance (FD)~\cite{Chen:2009um}. 
An example of this is provided by the alignments of Eq.~(\ref{Phi0}),
\be
\label{Phi-1} 
\Phi_1 =\frac{1}{\sqrt{6}} \begin{pmatrix}-2 \\ 1 \\ 1\end{pmatrix}
, \qquad
\Phi_2 = \frac{1}{\sqrt{3}} \begin{pmatrix}1 \\ 1 \\ 1\end{pmatrix}
, \qquad
\Phi_3 = \frac{1}{\sqrt{2}} \begin{pmatrix}0 \\ 1 \\ -1\end{pmatrix} ,
\ee
generating the famous tri-bimaximal mixing pattern. It is important
to notice that FD is a general concept which applies to arbitrary
orthogonal vectors $\Phi_i$. In principle, one could therefore choose the
$\Phi_i$ directions so as to yield the experimentally observed mixing
matrix. However, when it comes to building a model of flavour, a crucial
ingredient is the justification of the assumed flavon alignments, see
Subsection~\ref{sec:align-ind}. Therefore, 
in practice, only ``simple'' alignments are adopted in explicit models. 
With the parameters $m^0_i$ of Eq.~(\ref{eq:lightnuind}) being completely
independent of the vectors $\Phi_i$ (which arise from some flavon alignment
mechanism), it is clear that the mixing matrix does not depend on the
masses. In indirect models, FD thus implies form
diagonalisability~\cite{Low:2003dz,Chen:2009um}.\footnote{However, this is not 
necessarily the case in direct models, see~\cite{King:2011zj}, where the
columns of the Dirac neutrino Yukawa matrix -- in the basis of diagonal
right-handed neutrinos -- are not related to the flavon alignments
in a simple way.} 

A special case of FD is obtained if the three contributions to the
neutrino mass matrix of Eq.~(\ref{eq:lightnuind}) feature a hierarchy $m^0_1\ll
m^0_2\ll m^0_3$. In such a scenario, which is called sequential
dominance~\cite{King:1998jw,King:2002nf,Antusch:2010tf}, see
Subsection~\ref{sec:seesaw-3},
the first term, and with it the 
vector $\Phi_1$, can be ignored to good approximation. In fact, one can even
remove the flavon $\phi^\nu_1$ from the theory altogether. This would  set $m^0_1$
automatically to zero, without affecting the pattern of the $3\times 3$
mixing. The latter can be understood by realising that the first column of the
mixing matrix is uniquely determined by requiring orthogonality to the other
two columns $\Phi_2$ and $\Phi_3$. As above, SD is a general
concept applicable to arbitrary two (or three) orthogonal flavon
alignments. Choosing $\Phi_2$ and $\Phi_3$ as given in Eq.~(\ref{Phi-1}) leads to
constrained sequential dominance (CSD)~\cite{sumrule-1}, and predicts
tri-bimaximal neutrino mixing.

\subsection{\label{sec:align-ind}Vacuum alignment mechanism in indirect
  models}

We have discussed in Subsection~\ref{sec:vacuumdir} how flavons of direct
models can be aligned using the $F$-term alignment mechanism. In indirect
models, the same mechanism is available, however, if a triplet representation
of the family symmetry is real, it is mandatory to work in a basis where this is
explicitly realised, i.e. where all group generators are real. Applications of
the $F$-term alignment mechanism in indirect models can be found
e.g. in~\cite{deMedeirosVarzielas:2005qg,Antusch:2011sx,Antusch:2011ic}. In
addition to the usual $F$-term alignment mechanism, indirect models offer an
elegant alternative possibility for achieving particular flavon vacuum
configurations. This so-called $D$-term alignment mechanism, as the name
suggests, was first implemented in supersymmetric
models~\cite{deMedeirosVarzielas:2006fc,King:2006np}, however it is also
possible to apply it in a non-supersymmetric context.

The starting point is a flavon scalar potential field which may or may not arise in
a supersymmetric model from $D$-terms,
\begin{equation}
V~=~
-m^2 \sum_i {\phi^i}^\dagger \phi^{i}_{}
~+~ \lambda \left(\sum_i {\phi^i}^\dagger \phi^{i}_{} \right)^2
~+~ \Delta V \ ,
\label{eq:ind-pot}
\end{equation}
where the index $i$ labels the components of a particular flavon triplet
$\phi$ and 
\begin{equation}
\Delta V ~=~\kappa \sum_i {\phi^i}^\dagger \phi^{i} {\phi^i}^\dagger \phi^{i}\ .
\label{nice-inv}
\end{equation}
Ignoring the term $\Delta V$ in Eq.~(\ref{eq:ind-pot}), the potential 
features an $SU(3)$ symmetry and, as a consequence, no direction of the
flavon alignment would be preferred. Inclusion of the term $\Delta V$ breaks
the $SU(3)$ symmetry of the potential and leads to minima which single out
particular vacuum alignments. With the scale of the flavon VEV depending on
$m^2$, $\lambda$ and $\kappa$, it is sufficient to consider the extrema of the
quartic term in Eq.~(\ref{nice-inv}) for a unit vector~$\Phi$. If $\kappa>0$,
it is necessary to {\it minimise} the sum $\sum_i |\Phi^i|^4$, leading to the
solution 
\be
\kappa >0 \qquad \longrightarrow \qquad 
\Phi_+ = \frac{1}{\sqrt{3}}
\begin{pmatrix}
e^{i \vartheta_1}\\e^{i \vartheta_2}\\e^{i \vartheta_3} 
\end{pmatrix} \ ,
\label{eq:111ind} 
\ee
where $\vartheta_i$ are arbitrary phases.\footnote{In principle these
phases could be rotated away by an $SU(3)$ transformation, however, this
would generally change the basis of the assumed discrete symmetry. Yet, in
specific models, these phases can typically be absorbed into a redefinition of
the physical fields that accompany these flavons.}
Such an alignment is of the form of
$\Phi_2$ in Eq.~(\ref{Phi-1}). In fact, in indirect models, where the alignment
of Eq.~(\ref{eq:111ind}) appears as a column of the Dirac neutrino Yukawa matrix,
the phases $\vartheta_i$ can be removed by a field redefinition of the 
charged leptons. In the case where $\kappa<0$, the sum $\sum_i |\Phi^i|^4$ has
to be {\it maximised}. This gives rise to the alignment
\be
\kappa <0 \qquad \longrightarrow \qquad 
\Phi_- = 
\begin{pmatrix}
e^{i \vartheta_1}\\0\\0
\end{pmatrix} \ ,
\label{eq:100ind} 
\ee
and permutations thereof. Such alignments are typically useful for constructing a
diagonal charged lepton sector. They are furthermore necessary to obtain the
alignments $\Phi_3$ in Eq.~(\ref{Phi-1}) via $SU(3)$ invariant orthogonality
conditions. Introducing a new flavon field $\phi$ which couples to the flavons
$\phi_+$ (with alignment $\Phi_+ = \Phi_2$) and $\phi_-$ 
(with alignment $\Phi_-$) as
\be
\kappa' ~\bigg| \sum_{i} {\phi^i_-}^\dagger \phi^{i}_{}~\bigg|^2
+
\kappa'' ~\bigg| \sum_{i }{\phi^i_+}^\dagger \phi^{i}_{} ~ \bigg|^2
 \ ,
\ee
we generate the alignment $\vev{\phi} \propto \Phi_3$ if $\kappa'$ and
$\kappa''$ are taken to be positive. An alignment proportional to $\Phi_1$ of
Eq.~(\ref{Phi-1}) can be obtained subsequently from orthogonality conditions
involving flavons with alignments along the directions $\Phi_2$ and $\Phi_3$.

The preceding discussion illustrates the importance of the $SU(3)$ breaking
term in Eq.~(\ref{eq:ind-pot}). It is therefore natural to identify finite
groups $G$ which have the operator in Eq.~(\ref{nice-inv}) as an invariant.
Obviously, the family symmetry $G$ must admit at least one triplet
representation, with generators which are symmetry transformations of
Eq.~(\ref{nice-inv}). As was shown in~\cite{King:2009ap}, possible candidate
symmetries include the groups $\Delta(3n^2)$~\cite{Luhn:2007uq,Fairbairn:1964},
$\Delta(6n^2)$~\cite{Fairbairn:1964,Escobar:2008vc} and 
$T_n$~\cite{Bovier:1980ga}, cf. also
Eqs.~(\ref{eq:6n2irrep},\ref{eq:tnirrep}).

All these symmetries allow for at least two quartic invariants of type 
${\bf{3\ol 3 3 \ol 3}}$, namely the $SU(3)$ invariant and the operator of
Eq.~(\ref{nice-inv}). However, four of them have additional independent quartic
invariants. These are $\Delta(24)=S_4$ with one extra invariant, as well as
$\Delta(12)=A_4$, $\Delta(27)$ and $\Delta(54)$ with two additional invariants
each~\cite{King:2009ap}. These new invariants may spoil the structure of the
vacuum derived from $\Delta V$ of Eq.~(\ref{nice-inv}) unless they are
sufficiently suppressed. From this perspective, the groups $\Delta(3n^2)$ and
$\Delta(6n^2)$ with $n>3$, as well as the groups $T_n$ are preferred
candidates for the underlying discrete family symmetry of indirect models.

We conclude the discussion of the alignments in indirect models with a possible
alternative to the invariant of Eq.~(\ref{nice-inv}), which has not received any
attention yet. A cubic term of the form 
$\phi^1 \phi^2 \phi^3$
is left invariant under the groups $\Delta(3n^2)$ and $T_n$. As such a term is
generally not real, the new term in the flavon potential reads
\be
\Delta V = \kappa (\phi^1 \phi^2 \phi^3 + H.c.) \ ,
\ee
replacing the operator in Eq.~(\ref{nice-inv}). One can easily show that, with
$\kappa<0$, such a term in the flavon potential would generate an alignment of
type $\Phi_+$, see Eq.~(\ref{eq:111ind}), with $\vartheta_3 =
-(\vartheta_1+\vartheta_2)$.

\subsection{\label{sec:indirect-flavon}Indirect models after Daya Bay and RENO}
So far, we have only discussed the flavon alignments of indirect models leading to 
tri-bimaximal mixing. The measurement of a reactor angle $\theta_{13}$ of
around $8.5^\circ$ by the Daya Bay and RENO collaborations has now ruled out models
of accurate tri-bimaximal mixing. This fact demands a modification or
extension of the common strategies for constructing indirect models.

\begin{figure}[t]
\begin{center}
\includegraphics[clip=true,trim=60mm 5mm 87mm 65mm , height=80mm]{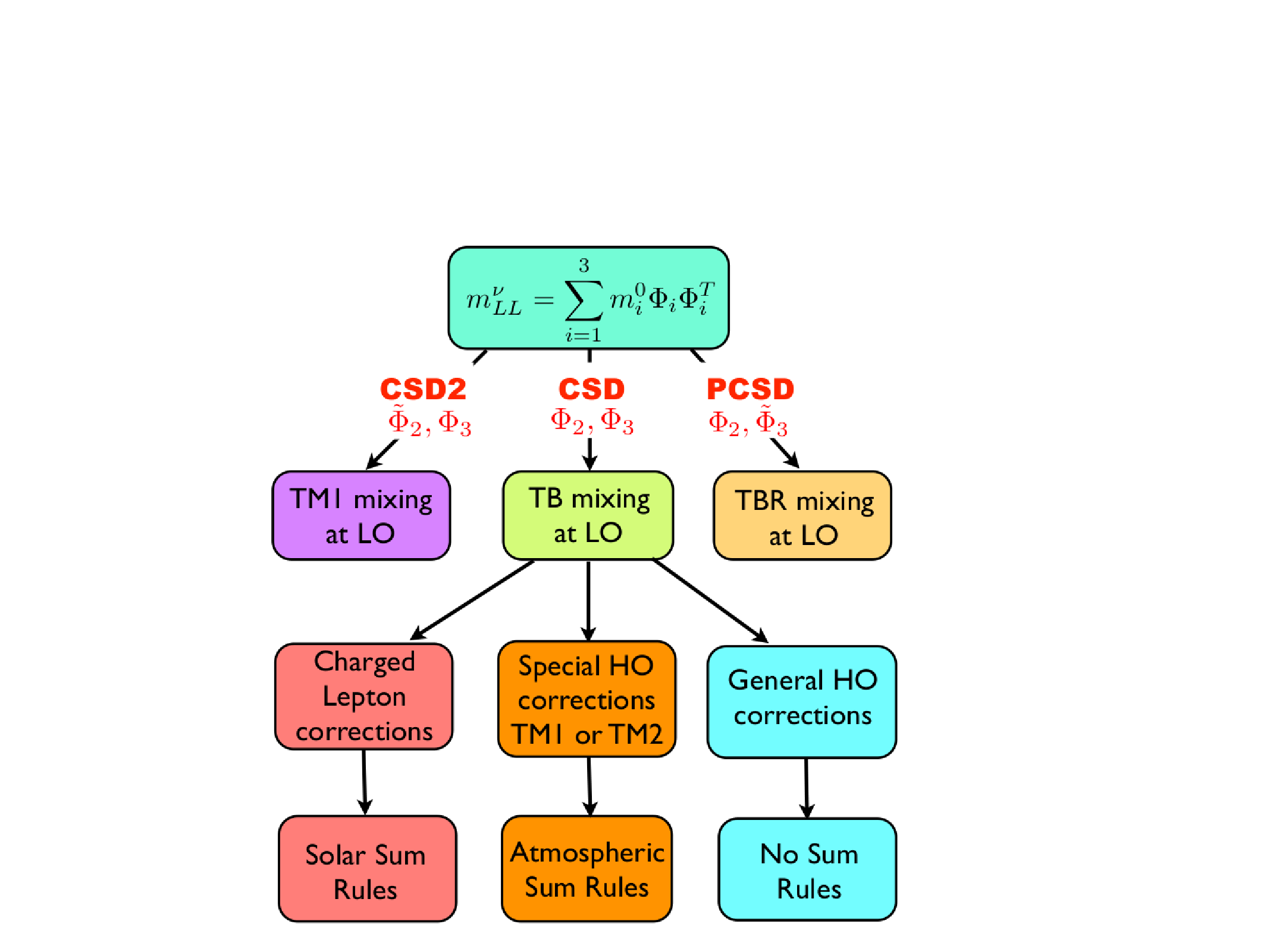}
\end{center}
\caption{\label{fig:indirect_map}\small{Possible strategies for constructing
    indirect models after Daya Bay and RENO. Starting from constrained
    sequential dominance (CSD) it is possible to add higher order (HO)
    corrections. Alternatively, it is possible to modify the flavon 
    alignments of CSD at leading order, thus giving rise to constrained
    sequential dominance 2 (CSD2) and partially constrained sequential
    dominance (PCSD).
In this diagram, we have used the acronyms 
TM=trimaximal,
TB=tri-bimaximal,
TBR=tri-bimaximal-reactor,
LO=leading order.
}}
\end{figure}

As for direct models, there are two principle paths one can pursue. The first
builds on existing indirect models of tri-bimaximal mixing which arise in the
framework of constrained sequential dominance, see the central branch of
Fig.~\ref{fig:indirect_map}. Such a leading order structure must 
be broken by higher order corrections, which can stem from either the charged lepton
or the neutrino sector. The former case requires a breaking of the accidental $T$
symmetry and leads to solar mixing sum rules as discussed in
Subsection~\ref{sec:patterns-5}. Alternatively, the higher order corrections
can break the accidental tri-bimaximal $U$ symmetry in the neutrino sector, entailing
atmospheric mixing sum rules, see Subsection~\ref{sec:patterns-6}. Breaking
the tri-bimaximal Klein symmetry of the neutrino sector completely gives rise
to arbitrary and therefore unpredictive corrections to the mixing angles. 
An example for higher order corrections which break the $U$ symmetry can be
constructed using the flavon alignments of constrained 
sequential dominance, proportional to $\Phi_2$ and $\Phi_3$, see
Eq.~(\ref{Phi-1}), and add a small perturbation along the direction $\Phi_1$ to
$\Phi_3$, see also~\cite{King:2010bk}. The form of the two flavon alignments
can then be written as 
\be
\frac{\vev{\phi_2}}{\Lambda} ~\propto ~ \Phi_2 
\ , \qquad
\frac{\vev{\phi'_3}}{\Lambda} ~\propto ~
\Phi'_3 \:=\: \Phi_3 + \epsilon \Phi_1
\ ,
\ee
with $\epsilon \ll 1$. Note that these two vectors are still orthogonal to each other,
hence the conditions of form dominance are satisfied. With these alignments,
it is straightforward to show that the resulting light neutrino mass matrix
\be
m^\nu_{LL} ~=~ m^0_2 \Phi^{}_2 \Phi_2^T + m^0_3 \Phi'_3 {\Phi'_3}^T \ ,
\label{eq:indUbreak}
\ee
breaks the original $U$ symmetry while continuing to respect the
$S$ symmetry. The latter can be easily verified by noticing that $(1,1,1)^T$
is still an eigenvector of $m^\nu_{LL}$ in Eq.~(\ref{eq:indUbreak}), meaning that
the second column of the tri-bimaximal PMNS mixing matrix, i.e. $\Phi_2$,
remains unchanged. Hence the trimaximal mixing structure TM2, see
Eq.~(\ref{TM2}), is achieved, which stabilises the solar angle. On the other
hand, the breaking of the $U$ symmetry of the neutrino mass matrix by higher
order effects allows to accommodate non-zero $\theta_{13}$.

The second strategy of constructing indirect models with sizable $\theta_{13}$
is based on new alignments at leading order. In the following we present two
examples (see the right and the left branch of Fig.~\ref{fig:indirect_map}):
partially constrained sequential dominance (PCSD)~\cite{King:2009qt} 
and constrained sequential dominance~2 (CSD2)~\cite{Antusch:2011ic}. Both
scenarios make use of two flavon triplets whose alignments are {\it not} orthogonal to
each other, in contrast to the previously discussed cases.

\subsection*{PCSD}
Partially constrained sequential dominance was first proposed
in~\cite{King:2009qt} as a simple modification of CSD, where one flavon is
aligned along the original $\Phi_2$ direction, while the alignment of the
other flavon is assumed to deviate slightly from the $\Phi_3$ direction by
filling the zero of the first component with $\epsilon\ll1$, i.e.
\be
\Phi_2
\,=\,   \frac{1}{\sqrt{3}} \begin{pmatrix}1\\1\\1\end{pmatrix}  , ~~ \qquad
\wt \Phi_3 
\,=\, \frac{1}{\sqrt{2}} \begin{pmatrix}\epsilon\\1\\-1\end{pmatrix}
  \ .
\label{eq:phi3tilde}
\ee
Note that $\Phi_2$ and $\wt \Phi_3$ are not orthogonal to each other, hence
PCSD violates form dominance at linear order in $\epsilon$.
Inserting these two alignments into Eq.~(\ref{eq:lightnuind}) yields the effective
neutrino mass matrix 
\be
m^\nu_{LL} ~=~ \frac{m^0_2 }{3}
\begin{pmatrix}
1&1&1\\
1&1&1\\
1&1&1
\end{pmatrix} ~+~
 \frac{m^0_3 }{2}
\begin{pmatrix}
\epsilon^2 &\epsilon&-\epsilon\\
\epsilon&1&-1\\
-\epsilon&-1&1
\end{pmatrix}  \ .
\ee
This matrix -- with non-zero $\epsilon$ -- is no longer diagonalised by a the
tri-bimaximal mixing matrix. Assuming $|\epsilon|\approx 0.2$ as well as a
normal neutrino mass hierarchy, i.e. $|m^0_2| \approx |\epsilon m^0_3|$,
analytic expressions for the mixing parameters valid to second order in
$\epsilon$ were derived in~\cite{King:2011ab}. These results show that to
first order in $\epsilon$, the tri-bimaximal solar and atmospheric mixing
angle predictions are maintained while the reactor angle takes a value of order
$\epsilon$. Therefore, PCSD gives rise to tri-bimaximal-reactor mixing, see
Eq.~(\ref{TBR}), at leading order. A special case of TBR mixing is obtained if the
parameter $\epsilon$ can be identified with the Wolfenstein parameter
$\lambda = 0.2253 \pm 0.0007$. As discussed in~\cite{King:2012vj}, such a situation
results in a reactor angle which satisfies $\sin \theta_{13} =
\frac{\lambda}{\sqrt{2}}$, leading to $\theta_{13}\approx 9.2^\circ$, a value
remarkably close to the one measured by Daya Bay and RENO, see Subsection~\ref{sec:patterns-3}.

The alignment of the flavon $\wt \phi_3$ in Eq.~(\ref{eq:phi3tilde}) can be
achieved through both the $F$-term and the $D$-term alignment mechanism. The
starting point are the simple alignments proportional to 
\be
\Phi_3
\,=\,   \frac{1}{\sqrt{2}} \begin{pmatrix}0\\1\\-1\end{pmatrix}  , ~~ \qquad
\Phi_x 
\,=\, \begin{pmatrix}1\\0\\0\end{pmatrix}
  \ ,
\ee
as obtained for instance in Eq.~(\ref{Phi-1}) and Eq.~(\ref{eq:100ind}), respectively.
The alignment in the direction of $\wt \Phi_3$ then arises
from successive orthogonality conditions as follows. Imposing orthogonality of
the VEV of an auxiliary flavon~${\phi_a}$ with $\vev{\phi_3}$ and $\vev{\phi_x}$ yields
\be
\vev{\phi_a} \perp \vev{\phi_3}\quad \mathrm{and} \quad
\vev{\phi_a} \perp \vev{\phi_x} \qquad \rightarrow \qquad
\frac{\vev{\phi_a}}{\Lambda} \,\propto\,
\frac{1}{\sqrt{2}} \begin{pmatrix}0\\1\\1\end{pmatrix} \ . 
\ee
Requiring the alignment of the flavon $\wt \phi_3$  to be orthogonal to the
alignment of this auxiliary flavon, we find the general structure
\be
 \vev{\wt \phi_3} \perp \vev{\phi_a}  \qquad \rightarrow \qquad
\frac{\vev{\wt \phi_3}}{\Lambda} \,\propto\,
\begin{pmatrix} 
n_1 \\ n_2\\-n_2
\end{pmatrix} \ ,
\ee
where $n_1$ and $n_2$ can take arbitrary values. For $n_1 \ll n_2$, this is
nothing but the alignment in the direction of $\wt\Phi_3$. The hierarchy
between $n_1 $ and 
$n_2$ may either be a consequence of mild tuning or, under certain
assumptions, result from a combination of a renormalisable and a
non-renormalisable term, where, after contracting the family indices, the
former is proportional to $n_1$ while the latter is proportional to $n_2$. The
necessary mass suppression of the non-renormalisable term then naturally
suppresses $\epsilon= \frac{n_1}{n_2} \ll 1$~\cite{King:2011ab}.
In order to establish a connection of $\epsilon$ and $\lambda$, one can
envisage scenarios in which the flavon $\wt\phi_3$ appears in both the
neutrino as well as in the quark sector. In the latter,  $\wt\phi_3$ has to be
responsible for generating the Cabibbo mixing~\cite{King:2012vj}.

\subsection*{CSD2}
Constrained sequential dominance 2, proposed in~\cite{Antusch:2011ic}, assumes
two flavon fields in the neutrino sector. One is aligned along the direction of $\Phi_3$ of
Eq.~(\ref{Phi-1}), while the alignment of the other flavon is a relatively simple
variation of $\Phi_2$ of Eq.~(\ref{Phi-1}), explicitly
\be
\wt \Phi_2  ~=~ \frac{1}{\sqrt{5}} \begin{pmatrix}1\\2\\0\end{pmatrix} 
\quad \mathrm{or} \quad 
 \frac{1}{\sqrt{5}} \begin{pmatrix}1\\0\\2\end{pmatrix} \ , ~~ \qquad
\Phi_3
\,=\,   \frac{1}{\sqrt{2}} \begin{pmatrix}0\\1\\-1\end{pmatrix}  
\ .\label{eq:120align}
\ee
Analogous to the case of PCSD, the alignments of the two flavons, pointing in the
directions of $\wt \Phi_2$ and $\Phi_3$, are not orthogonal to each other,
implying that CSD2 violates form dominance as well. In the following, we only
present the discussion of the first $\wt\Phi_2$ alignment of Eq.~(\ref{eq:120align}); the
alternative case of the second alignment can be treated analogously, leading
to almost identical results. Inserting the alignments of $\wt\phi_2$ and
$\phi_3$ into Eq.~(\ref{eq:lightnuind}) generates the neutrino mass matrix 
\be
m^\nu_{LL} ~=~ \frac{m^0_2 }{5}
\begin{pmatrix}
1&2&0\\
2&4&0\\
0&0&0
\end{pmatrix} ~+~
 \frac{m^0_3 }{2}
\begin{pmatrix}
0&0&0\\
0&1&-1\\
0&-1&1
\end{pmatrix}  \ ,
\ee
which is not of tri-bimaximal structure. Yet, one can immediately verify that
$(-2,1,1)^T$, i.e. the first column of the tri-bimaximal mixing matrix, is
still an eigenvector of $m^\nu_{LL}$. This shows that CSD2 necessarily leads
to the trimaximal mixing pattern TM1, see Eq.~(\ref{TM1}). With the assumption of
a normal neutrino mass hierarchy, i.e. with $|m^0_2| \approx|\epsilon m^0_3|$,
analytic expressions for the mixing parameters valid to second order
in~$\epsilon$ were derived in~\cite{Antusch:2011ic}. These results explicitly
confirm that the solar angle maintains its tri-bimaximal value at linear order
in $\epsilon$, while the deviations of the reactor and the atmospheric mixing
angles from their tri-bimaximal values are proportional to $\epsilon$, leading
to the linear mixing sum rule of Eq.~(\ref{TM1r}).

The alignment of $\wt\phi_2$ in Eq.~(\ref{eq:120align}) can be derived from a
set of orthogonality conditions similar to the situation in PCSD. In CSD2
we start from the simple alignments proportional to 
\be
\Phi_1= \frac{1}{\sqrt{6}} \begin{pmatrix} -2\\1\\1 \end{pmatrix}  , \qquad
\Phi_y= \begin{pmatrix} 0\\1\\0 \end{pmatrix}  , \qquad
\Phi_z= \begin{pmatrix} 0\\0\\1 \end{pmatrix}  \ ,
\ee
see Eqs.~(\ref{Phi-1},\ref{eq:100ind}). Demanding orthogonality of the flavon
VEV $\vev{\wt\phi_2}$ with $\vev{\phi_1}$ and $\vev{\phi_z}$ generates an
alignment in the direction of the first $\wt\Phi_2$ vector in Eq.~(\ref{eq:120align}), 
\be
\vev{\wt\phi_2} \perp \vev{\phi_1}\quad \mathrm{and} \quad
\vev{\wt\phi_2} \perp \vev{\phi_z} \qquad \rightarrow \qquad
\frac{\vev{\wt\phi_2}}{\Lambda} \,\propto\,
\frac{1}{\sqrt{5}} \begin{pmatrix}1\\2\\0\end{pmatrix} \ . 
\ee
The second $\wt\Phi_2$ vector in Eq.~(\ref{eq:120align}) can be obtained
similarly by using $\Phi_y$ instead of $\Phi_z$.

\section{\label{sec:gut}Grand unified theories of flavour}

\subsection{\label{sec:GUTs}Grand unified theories}
\begin{figure}[t]
\begin{center}
\includegraphics[width=0.96\textwidth]{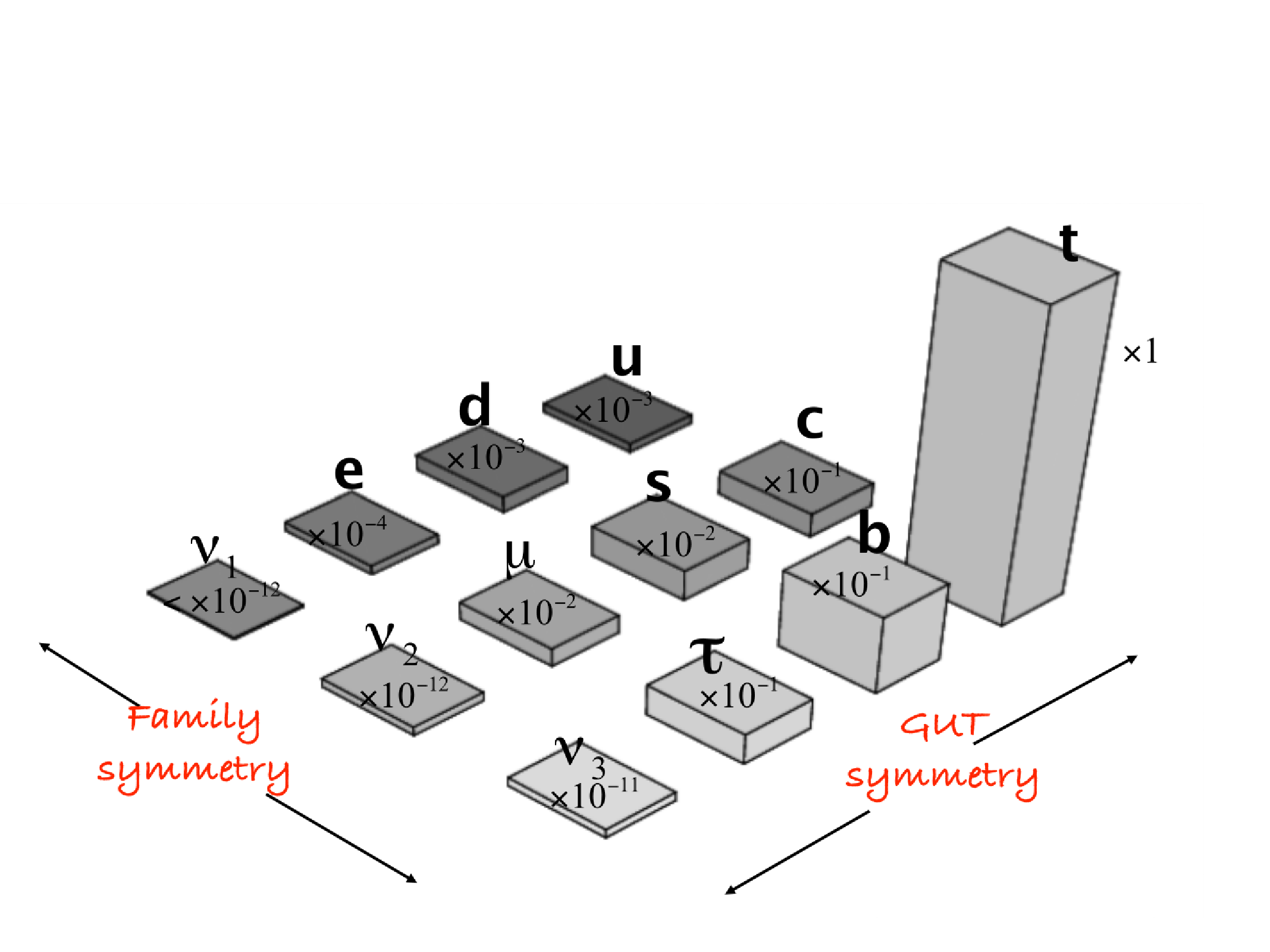}
\end{center}
\vspace*{-4mm}
    \caption{\label{masses}\small{The fermion masses are here represented by a
        lego plot. We have multiplied the masses of the bottom, charm and tau
        by $10$, the strange and muon by $10^2$, the up and down by $10^3$,
        the electron by $10^4$ to make the lego blocks visible. Assuming 
        a normal neutrino mass hierarchy, we have multiplied the third
        neutrino mass by $10^{11}$ and the second neutrino mass by $10^{12}$
        to make the lego blocks visible. This underlines how incredibly light
        the neutrinos are. The symmetry groups $G_{{\rm GUT}}$ and $G_{{\rm
            FAM}}$ act in the indicated directions.}}
\vspace*{-2mm}
\end{figure}

\begin{figure}[t]
\begin{center}
\includegraphics[width=0.93\textwidth]{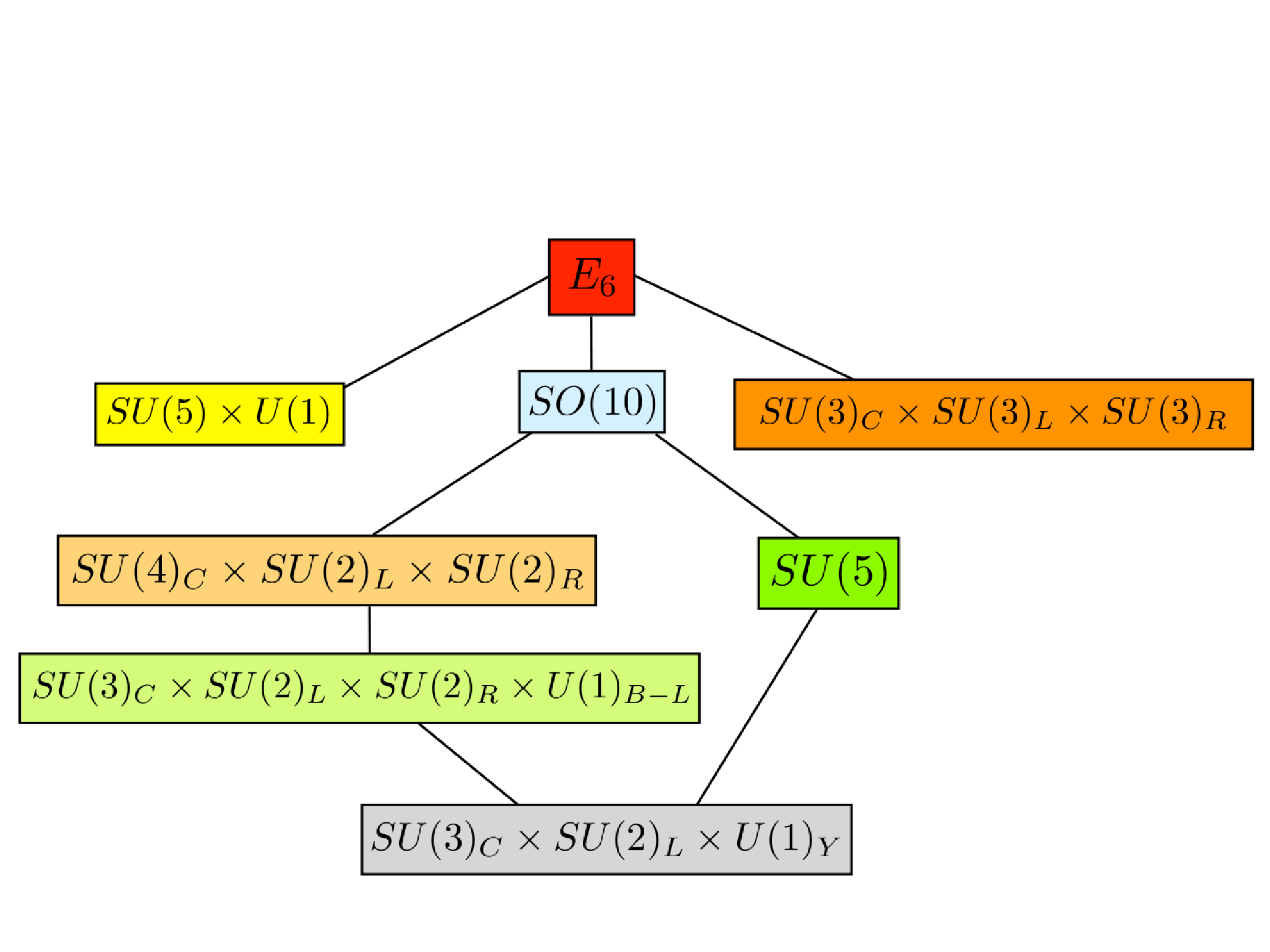}
\end{center}
\vspace*{-4mm}
    \caption{\label{GUTs}\small{Some possible candidate unified gauge groups.}}
\vspace*{-2mm}
\end{figure}

One of the exciting things about the discovery of neutrino masses
and mixing angles is that this provides additional information
about the flavour problem - the problem of understanding the origin
of three families of quarks and leptons and their masses and mixing
angles. In the framework of the see-saw mechanism, new physics beyond the
Standard Model is required to violate lepton number and generate
right-handed neutrino masses which may be as large as the
GUT scale. This is also exciting since it implies that
the origin of neutrino masses is also related to some
GUT symmetry group $G_{{\rm GUT}}$, which unifies the
fermions within each family as shown in Fig.~\ref{masses}.
Some possible candidate unified gauge groups are shown in Fig.~\ref{GUTs}.

Let us take $G_{{\rm GUT}}=SU(5)$ as an example. Each family of 
quarks (with colour $r,b,g$) and leptons fits nicely into $SU(5)$ representations 
of left-handed ($L$) fermions, $F=\overline{\bf 5}$ and $T={\bf 10}$
\be
F= \begin{pmatrix} d_r^c\\d_b^c\\d_g^c\\e^-\\-\nu_e \end{pmatrix}_L  , \qquad
T= \begin{pmatrix} 0&u_g^c&-u_b^c&u_r&d_r\\
.&0&u_r^c&u_b&d_b\\ 
.&.&0&u_g&d_g\\ 
.&.&.&0&e^c\\
.&.&.&.&0
\end{pmatrix}_L  \ ,
\ee
where $c$ denotes CP conjugated fermions. The $SU(5)$ representations
$F=\overline{\bf 5}$ and $T={\bf 10}$ decompose into multiplets of the SM
gauge group $SU(3)_C\times SU(2)_L\times U(1)_Y$ as $F=(d^c,L)$,
corresponding to, 
\be
\overline{\bf 5}=(\overline{\bf 3},{\bf 1},1/3)\oplus  ({\bf 1},\overline{\bf 2},-1/2),
\ee
and
$T=(u^c,Q,e^c)$, corresponding to,
\be
{\bf 10} =(\overline{\bf 3},{\bf 1},-2/3)\oplus  
({\bf 3},{\bf 2},1/6)\oplus ({\bf 1},{\bf 1},1).
\ee
Thus a complete quark and lepton SM family $(Q,u^c,d^c,L,e^c)$ is accommodated 
in the $F=\overline{\bf 5}$ and $T={\bf 10}$ representations, with
right-handed neutrinos, whose CP conjugates are denoted as $\nu^c$, being singlets of $SU(5)$, $\nu^c={\bf 1}$. The Higgs doublets $H_u$ and $H_d$
which break electroweak symmetry in a two Higgs doublet model
are contained in the $SU(5)$ multiplets $H_{\bf 5}$ and $H_{\overline{\bf 5}}$.

The Yukawa couplings for one family of quarks and leptons are given by,
\be
y_u H_{{\bf 5}i}T_{jk}T_{lm}\epsilon^{ijklm}+ y_{\nu}H_{{\bf 5}i}F^i\nu^c+
y_d H_{\overline{\bf 5}}^iT_{ij}F^j,
\ee
where $\epsilon^{ijklm}$ is the totally antisymmetric tensor of $SU(5)$ with
$i,j,j,k,l=1,\ldots, 5$,
which decompose into the SM Yukawa couplings
\be
y_u H_uQu^c+  y_{\nu}H_uL\nu^c+
y_d (H_dQd^c+H_de^cL).
\ee
Notice that the Yukawa couplings for down quarks and charged leptons are equal
at the GUT scale. Generalising this relation to all three families we find the $SU(5)$ prediction for Yukawa matrices,
\be
Y_d=Y_e^T,
\ee
which is successful for the third family, but fails badly for the first and
second families. Georgi and Jarlskog~\cite{Georgi:1979df} suggested to
include a higher Higgs representation 
$H_{\overline{\bf 45}}$ which is responsible for the 2-2 entry of the down and charged lepton Yukawa matrices. Dropping $SU(5)$ indices for clarity,
\be
(Y_{d})_{22} H_{\overline{\bf 45}}T_2F_2 ,
\ee
decomposes into the second family SM Yukawa couplings
\be
(Y_{d})_{22}(H_dQ_2d_2^c-3H_de_2^cL_2),
\ee
where the factor of $-3$ is an $SU(5)$ Clebsch-Gordan
coefficient.\footnote{In this setup, $H_d$ is the light linear combination of the
electroweak doublets contained in $H_{\bf{\ol 5}}$~and~$H_{\bf{\ol {45}}}$.}
Assuming a hierarchical Yukawa matrix with a zero Yukawa element (texture) in
the 1-1 position, results in the GUT scale Yukawa relations, 
\be
y_b = y_{\tau}, \quad  y_s = \frac{y_{\mu}}{3}, \quad y_d = 3y_e, 
\ee
which, after renormalisation group running effects are taken into account, are consistent with the low energy masses.
The precise viability of these relations has been widely discussed in the light of recent progress in 
lattice theory which enable more precise values of quark masses to be determined, especially the strange quark mass
(see, e.g.,~\cite{Ross:2007az}). In supersymmetric (SUSY) theories with low values of the ratio of Higgs vacuum expectation values, the relation for 
the third generation $y_b = y_{\tau}$ at the GUT scale remains viable,
but a viable GUT scale ratio of $y_\mu/y_s$ is more accurately achieved within
SUSY $SU(5)$ GUTs using a Clebsch factor of $9/2$, as proposed
in~\cite{Antusch:2009gu}, which is 50\% higher than the Georgi-Jarlskog
prediction of $3$.

\subsection{\label{sec:GUTxFam}Combining GUTs and family symmetry}
As already remarked in Subsection~\ref{sec:patterns-3}, it is a remarkable
fact that the smallest leptonic mixing angle, 
the reactor angle, is of a similar magnitude to the largest quark mixing
angle, the Cabibbo angle, indeed they  
may even be equal to each other up to a factor of $\sqrt{2}$. Such
relationships may be a hint of a connection between leptonic mixing and quark
mixing, where such a connection might be achieved using
GUTs~\cite{Antusch:2011qg,Marzocca:2011dh}. 
For example, the Georgi-Jarlskog relations discussed above already lead to the 
left-handed charged lepton mixing angle having a simple relation with the
right-handed down-type quark mixing angle 
$\theta_{12}^{e_L} \approx \theta_{12}^{d_R}/3$ where the approximation assumes hierarchical Yukawa matrices,
with the 1-1 elements being approximately zero. If the upper $2\times 2$
Yukawa matrices are symmetric (as motivated by the successful
Gatto-Sartori-Tonin (GST) relation~\cite{Gatto:1968ss} which relates the 12
mixing $\theta_{12}^{d_{L,R}}$ to the down and strange mass by
$\theta_{12}^{d_{L,R}}  \approx\sqrt{m_d/m_s}$) then we may drop the $L,R$
subscripts and this relation simply becomes  $\theta_{12}^{e} =
\theta_{12}^{d}/3$. In large classes of models, including  those discussed
later, the quark mixing originates predominantly from the down-type quark
sector, in which case this relation becomes $\theta_{12}^{e} = \theta_C/3$.
If one starts from TB mixing in the neutrino sector, resulting from some discrete family symmetry,
then, using the results in Subsection~\ref{sec:patterns-4} such a charged lepton correction results in a reactor angle in the lepton sector of $\theta_{13}\approx \theta_C/(3\sqrt{2})$ as discussed for example in~\cite{sumrule-1}. 
This is a factor of 3 too small to account for the
observed reactor angle, but it illustrates how the reactor angle could possibly be related to the Cabibbo angle using GUTs. 
Indeed it has been suggested that perhaps the charged lepton mixing angle is exactly equal to the Cabibbo angle
in some GUT model, leading to $\theta_{13}\approx
\theta_C/\sqrt{2}$~\cite{King:2012vj,Antusch:2012fb,Zhang:2012mn}. However it
is non-trivial to reconcile such large charged lepton mixing with 
the successful relationships between charged lepton and down-type quark masses, and it seems more likely that
charged lepton mixing is not entirely responsible for the reactor angle.

The above discussion provides an additional motivation for
combining GUTs with discrete family symmetry in order to account for the reactor angle.
Putting these two ideas together we are suggestively led to a framework of
new physics beyond the Standard Model based on commuting
GUT and family symmetry groups,
\beq
G_{{\rm GUT}}\times G_{{\rm FAM}} .
\label{symmetry}
\eeq
There are many possible candidate GUT and family symmetry groups
some of which are listed in Table~\ref{table3}. Unfortunately the
model dependence does not end there, since the details of the symmetry
breaking vacuum plays a crucial role in specifying the model and
determining the masses and mixing angles, resulting in many models.
These models may be classified according to the particular
GUT and family symmetry they assume as shown in Table~\ref{table3}.

\begin{table}[t]
\begin{center}
\begin{tabular}{|l|c|c|c||}
\hline
\hline
$\begin{array}{cc}
 & G_{{\rm GUT}} \\
G_{{\rm FAM}} &
\end{array}$
&
$\begin{array}{cc}
SU(2)_L\times U(1)_Y \\ ~
\end{array}$
&
$\begin{array}{cc}
SU(5) \\ ~
\end{array}$
&
$\begin{array}{cc}
SO(10) \\ ~
\end{array}$
\\\hline\hline

~$S_3$& \cite{survey-SM-S3}  & & \cite{survey-SO10-S3}  \\\hline

~$A_4$ &
\cite{survey-SM-A4,Brahmachari:2008fn,King:2011zj,Antusch:2011ic,King:2011ab,Hernandez:2012ra,Ma:2012ez,Altarelli:2012bn}& 
\cite{survey-SU5-A4,Cooper:2012wf}
&  \\\hline

~$T'$ & &\cite{survey-SU5-Tprime} &  \\\hline

~$S_4$ &
\cite{survey-SM-S4,Altarelli:2009gn,King:2011zj,Hernandez:2012ra,Altarelli:2012bn}
& \cite{Meloni:2011fx,Hagedorn:2012ut}& \cite{survey-SO10-S4} \\\hline

~$A_5$ &\cite{Hernandez:2012ra,Cooper:2012bd}  & &   \\\hline

~$T_7$ & \cite{survey-SM-T7,Luhn:2012bc}& &   \\\hline

~$\Delta(27)$  & \cite{survey-SM-Delta27}& & \cite{survey-PS-Delta6n2} \\\hline

~$\Delta(96)$   & \cite{Toorop:2011jn,Ding:2012xx}& \cite{King:2012in}&   \\\hline

~$D_N$ &\cite{survey-SM-DN} & &   \\\hline

~$Q_N$ &  \cite{survey-SM-QN}& &   \\\hline

~other  & \cite{survey-SM-other} & &   \\
\hline
\hline
\end{tabular}
\end{center}
\caption{\label{table3}\small{Some candidate GUT and discrete family symmetry
    groups, and the papers that use these symmetries to successfully describe
    the solar, atmospheric and reactor neutrino data.}}
\end{table}

Another complication is that the masses and mixing angles determined
in some high energy theory must be run down to low energies using
the renormalisation group equations.
Large radiative corrections are seen
when the see-saw parameters are tuned,
since the spectrum is sensitive to small changes in the parameters,
and this effect is sometimes used to magnify small mixing angles
into large ones.

In natural models with a normal mass hierarchy based on SD the parameters are
not tuned, since the hierarchy and large atmospheric and
solar angles arise naturally as discussed in the previous section.
Therefore in SD models the radiative
corrections to neutrino masses and mixing angles are only expected
to be a few per cent, and this has been verified numerically.

\section{\label{sec:examp}Model examples}
In this section we give three examples of SUSY GUTs of flavour based on
the (semi-)direct 
approach, which can account for all quark and lepton masses and
mixing, including the observed reactor angle 
which only gets a small correction from charged lepton mixing.
The first example is based on the minimal family symmetry 
$A_4$ combined with the minimal GUT $SU(5)$. This is
actually a semi-direct model, since only half the Klein symmetry is contained
in $A_4$, resulting in TM2 mixing, but, like all semi-direct models, it cannot
explain the relative smallness of the reactor angle. 
The second example does explain the smallness of the reactor angle
compared to the atmospheric or solar angles by embedding the $A_4$ into an $S_4$
family symmetry. This allows a direct model where the Klein symmetry resulting
from $S_4$ is half broken by a rather special higher order correction,
resulting again in TM2 mixing as in the $A_4$ case. The third model based on
$\Delta(96)$ is an example of a direct model with a larger family symmetry
where the Klein symmetry corresponds  to bi-trimaximal mixing in which the
reactor angle is already non-zero at the leading order, and where the 
small charged lepton correction with an assumed zero phase brings it into
agreement with the Daya Bay and RENO measurements of $\theta_{13}\approx 8.5^\circ$. 

It is worth mentioning that all three models require a $U(1)_R$
symmetry in order to achieve the desired flavon vacuum alignment, see
Subsection~\ref{sec:vacuumdir}. The $U(1)_R$ charges of the different
superfields are assigned in a standard way, i.e. the quark and lepton
superfields carry charge $+1$,  while the Higgs and the flavon fields are
neutral. After supersymmetry breaking the models therefore 
feature a residual discrete symmetry, called $R$-parity, forbidding the
dangerous lepton and baryon number violating terms $LLe^c$, $LQd^c$ and
$u^cd^cd^c$ which, if simultaneously present, would mediate rapid proton decay
in conflict with experimental bounds on the proton lifetime. $R$-parity does,
however, not forbid the effective operators $QQQL$ and
$u^cu^cd^ce^c$. In principle, these could endanger the stability of the
proton in the presented models. However, whether or not this leads to a
conflict with existing constraints, is a subtle quantitative question which depends
on the details of the underlying renormalisable theory 
(set of messenger fields and their masses). For an extensive review on proton
stability in grand unified theories we refer the reader to~\cite{Nath:2006ut}.

\subsection[${A_4 \times SU(5)}$: a semi-direct model of trimaximal mixing]{\label{sec:a4tmgut}$\bs{A_4 \times SU(5)}$: a semi-direct model of trimaximal mixing}

Our first example of a $G_{\mathrm{FAM}}\times G_{\mathrm{GUT}}$ model with
large $\theta_{13}$ is based on the Altarelli-Feruglio~$A_4$
model of leptons~\cite{Altarelli:2005yp,Altarelli:2005yx}. Working in a
supersymmetric $SU(5)$ setting, the three matter families of $F={\bf \ol 5}$
and $T={\bf {10}}$, see Subsection~\ref{sec:GUTs}, transform under $A_4$ as
${\bf 3}$ and ${\bf 1'',1',1}$, respectively.  The see-saw mechanism is
implemented in the model by introducing right-handed neutrinos~$\nu^c$ living in
the ${\bf 3}$ of $A_4$. The Higgs fields $H_{\bf 5}^{}$, $H_{\bf \ol 5}$ and
$H_{\bf \ol {45}}$ furnish the one-dimensional $A_4$ representations ${\bf
  1}$, ${\bf 1'}$ and ${\bf 1''}$.  The latter gives rise to the intriguing
Georgi-Jarlskog (GJ) relations~\cite{Georgi:1979df}. 
The family symmetry breaking flavon fields are $SU(5)$ singlets and can be
divided into fields which appear in the neutrino sector $\phi^\nu_{\bf r}$ and fields which
appear in the quark sector $\phi^q_{\bf r}$. The $A_4$ family symmetry is
enriched by the shaping symmetry $U(1)\times Z_2\times Z_3\times Z_5$ in order
to control the coupling of the flavons to the different matter sectors. The complete
charge assignments of the matter, Higgs and flavon superfields is presented in
Table~\ref{tab:A4-assignments}. 

\begin{table}[t]
\begin{center}
$$
\begin{array}{|c|c|c|c|c|c|c|c|c|c|c|c|c||c|c|c|c|c|c|}\hline
\!\!\phantom{\Big|}&\multicolumn{5}{c|}{\mathrm{matter~fields}} &
 \multicolumn{3}{c|}{\mathrm{Higgs~fields}}  & 
 \multicolumn{10}{c|}{\mathrm{flavon~fields}} \\ \cline{2-19}
\!\!\phantom{\Big|} & \nu^c &  F &T_1&T_2 &T_3& 
H^{}_{\bf 5} & H_{\bf{\ol{5}}} & H_{\bf{\ol{45}}} &
\phi^\nu_{\bf 3} & \phi^\nu_{\bf 1} & \phi^\nu_{\bf 1'} & \phi^\nu_{\bf 1''} &
\phi^q_{\bf 3}  & \phi^q_{\bf{1}} & \phi^q_{\bf 1'} & \phi^q_{\bf 1''} &
\wt \phi^q_{\bf 1' } & \wt \phi^q_{\bf 1} \\\hline
~\!SU(5)\!\!\!\phantom{\Big|}~ & \bf 1 & \bf \ol 5 & \bf 10  & \bf 10 & \bf 10& 
\bf  5 &\bf \ol 5 &\bf \ol{45} &
\bf 1&\bf 1&\bf 1&\bf 1&\bf 1&\bf 1&\bf 1&\bf 1& \bf 1  & \bf 1\\\hline
A_4\!\!\phantom{\Big|} & \bf 3&\bf 3&\bf 1''&\bf 1'&\bf 1&
\bf 1&\bf 1'&\bf 1''&
\bf 3&\bf 1&\bf 1'&\bf 1''&
\bf 3&\bf 1&\bf 1' & \bf 1'' & \bf 1' &\bf 1\\ \hline
U(1)\!\!\phantom{\Big|} & 1&-1&3&3&0&0&-1&-2&
-2&-2&-2&-2&2&-1&-1&-1 &-5&2 \\\hline
Z_2\!\!\phantom{\Big|} & 0&0&0&0&0&0&0&1&0&0&0&0&0&1&0&0&1&0 \\\hline
Z_3\!\!\phantom{\Big|} & 1&2&2&0&0&0&1&1&1&1&1&1&0&1&2&2&2&0 \\\hline
Z_5\!\!\phantom{\Big|} & 1&4&0&0&0&0&1&1&3&3&3&3&0&0&0&0&0&0 \\\hline
\end{array}
$$
\end{center}
\caption{\label{tab:A4-assignments}\small{The charge assignments of the matter,
Higgs and flavon superfields in the $A_4 \times SU(5)$ model
of~\cite{Cooper:2012wf}. The shaping symmetry $U(1)\times Z_2\times Z_3\times
Z_5$ constrains the set of operators allowed in the superpotential.}}  
\end{table}

A discussion of all the different aspects of the model, including the vacuum
alignment,  can be found in~\cite{Cooper:2012wf}. Here we mainly focus our
attention on the neutrino sector. The leading order renormalisable operators
of the neutrino superpotential which are invariant under all imposed
symmetries of the model, see Table~\ref{tab:A4-assignments}, are
\be
{W}_{\nu}~\sim~ F\nu^c H^{}_{\bf 5} +
\left(\phi^\nu_{\bf 3}+ \phi^\nu_{\bf 1} 
+\phi^\nu_{\bf 1'} + \phi^\nu_{\bf 1''} \right)\nu^c\nu^c \ ,
\label{eqn:neutrino}
\ee
where dimensionless order one coupling coefficients are suppressed. 
Note that the Dirac Yukawa term does not involve any flavon field, hence, the Dirac
neutrino Yukawa matrix does not break the $A_4$ symmetry. Inserting the Higgs
VEV $v_u$ (the VEV of the electroweak doublet contained in $H^{}_{\bf 5}$)
then generates an $A_4$ invariant Dirac neutrino mass matrix $m_{LR}$. As a
consequence, the mixing in the neutrino sector originates solely from the
right-handed neutrino mass matrix whose form depends on the chosen basis of
$A_4$  as well as the vacuum configuration of the flavon fields. 
Working in the $A_4$ basis of Appendix~\ref{app:CGs}, it has been shown
in~\cite{Cooper:2012wf} that the two triplet flavon alignments
\be
\vev{\phi^\nu_{\bf 3}} ~=~\varphi^\nu_{\bf 3} \begin{pmatrix}1\\1\\1\end{pmatrix} \ ,\qquad
\vev{\phi^q_{\bf 3}} ~=~\varphi^q_{\bf 3} \begin{pmatrix}1\\0\\0\end{pmatrix} \ ,
\ee
can be obtained from the $F$-term alignment mechanism along the lines
of~\cite{Altarelli:2005yx}. This necessitates the introduction of a $U(1)_R$
symmetry, a set of driving fields, see Subsection~\ref{sec:vacuumdir}, as well
as the auxiliary flavon field $\wt \phi^q_{\bf {1}}$ which has the same
charges as the triplet flavon $\phi^q_{\bf {3}}$, except for $A_4$.
With these alignments as well as VEVs for the one-dimensional flavon fields,
using the Clebsch-Gordan coefficients in Appendix~\ref{app:CGs},
the Dirac neutrino mass matrix and the right-handed neutrino mass matrix take
the form\footnote{As the mass term of the right-handed neutrinos in the
superpotential of Eq.~(\ref{eqn:neutrino}) involves the CP conjugate fields
$\nu^c$, the mass matrix in the conventions of Subsection~\ref{sec:seesaw-1}
is obtained using the complex conjugate flavons VEVs ${\varphi^\nu_{\bf
    r}}^\ast$, cf. also Eq.~(\ref{eq:Mast-MRR}). However, for notational
clarity, we drop the $^\ast$ here and in the following.} 
\be
m_{LR} ~\approx ~\begin{pmatrix}
                     1 & 0 & 0\\
		     0 & 0 & 1\\
		     0 & 1 & 0
\end{pmatrix} v_u  \ ,  
\ee

\be
M_{RR} ~\approx ~
 \varphi^\nu_{\bf 3} \begin{pmatrix}
      2  &  -  1       & -1    \\
-1&2&-1 \\
-1&-1&2  
\end{pmatrix}  
+ 
\varphi^\nu_{\bf 1}
\begin{pmatrix}
      1&0&0\\
0&0&1 \\
0&1&0
\end{pmatrix}  
+
 \varphi^\nu_{\bf 1'} 
\begin{pmatrix}
      0&0&1\\0&1&0\\1&0&0
\end{pmatrix}  
+
 \varphi^\nu_{\bf 1''} 
\begin{pmatrix}
    0&1&0\\1&0&0\\0&0&1 
\end{pmatrix}  
.\label{matrixneutrA4}
\ee
The effective light neutrino mass matrix emerges from these by applying the type~I
see-saw formula of Eq.~(\ref{seesaw}), i.e.
\be
m^\nu_{LL} = - m_{LR} M_{RR}^{-1} m_{LR}^T \ .
\ee
Due to the trivial structure of the Dirac neutrino mass matrix $m_{LR}$, the
neutrino mixing matrix is identical to the unitary matrix which
diagonalises $M_{RR}$ (and automatically also $M_{RR}^{-1}$), except for a
permutation of the second and the third row~\cite{King:2011zj}. We remark that
the light neutrino masses are not related by a mass sum rule since the
right-handed neutrino mass term involves four independent flavon
fields. The inverse mass sum rule for $A_4$ quoted in
Eq.~(\ref{inversemasssum}) with $\gamma = 1$ and $\delta = -2$
can only be recovered by removing the flavons
$\phi^\nu_{\bf 1'}$ and $\phi^\nu_{\bf 1''}$, which in turn corresponds to the
well-known case of tri-bimaximal neutrino mixing. 

Rather than diagonalising $M_{RR}$ explicitly, let us discuss its symmetries. The
first two terms in Eq.~(\ref{matrixneutrA4}) are symmetric under the
tri-bimaximal Klein generators $S$ and $U$ of Eq.~(\ref{eq:TBsu}). The third and
the fourth terms break the tri-bimaximal structure, however, in a special way.
It is straightforward to prove explicitly that $S^TM_{RR} S =M_{RR}$ is still
respected.  A simple way of seeing this is by noticing that all neutrino
flavon VEVs remain unchanged under the $A_4$ transformation~$S$, see
Appendix~\ref{app:CGs}. On the other hand, the $U$ matrix of 
Eq.~(\ref{eq:TBsu}) does not form part of $A_4$. In order for $M_{RR}$ to be also
symmetric under $U$, hence entailing tri-bimaximal neutrino mixing, one would
have to require $\varphi^\nu_{\bf 1' } = \varphi^\nu_{\bf 1''}$. 
In~\cite{Altarelli:2005yp,Altarelli:2005yx} this condition among the a priori
unrelated VEVs is realised by not including flavons in the ${\bf 1'}$ and
${\bf 1''}$  representations of $A_4$ in the first place. Alternatively, the two non-trivial
one-dimensional representations can be unified into a doublet of $S_4$, see
Appendix~\ref{app:CGs}. A suitable VEV alignment of such a doublet can relate
the two components such as to generate a right-handed neutrino mass matrix of
tri-bimaximal form, see e.g.~\cite{King:2011zj}. In general, however,
$\varphi^\nu_{\bf 1' } \neq \varphi^\nu_{\bf 1''}$ and there is no accidental
$U$ symmetry in an $A_4$ model with neutrino  flavons in all possible
representations of the family symmetry. 

With $m_{LR}$ being invariant under the full $A_4$ family symmetry and $M_{RR}$
being symmetric under $S$, the type~I see-saw mechanism generates a light
effective neutrino mass matrix which also respects the $S$ symmetry. This
symmetry can be translated to a particular mixing pattern by considering an
eigenvector $\vec v$ of $S$ with eigenvalue $+1$. One can easily check that
the only solution to $S \vec v = \vec v $
is of the form $\vec v \propto (1,1,1)^T$. 
Using this and the invariance of the light neutrino mass matrix $m^\nu_{LL}$
under $S$ as well as $S^T=S$, we obtain
\be
m^\nu_{LL} \vec v  ~=~ S m^\nu_{LL} S \vec v   ~=~ S m^\nu_{LL}  \vec v \ ,
\ee
which shows that $m^\nu_{LL}  \vec v$ is an eigenvector of $S$ with
eigenvalue $+1$, and thus 
\be
m^\nu_{LL}  \vec v ~\propto ~\vec v \ .
\ee
As $\vec v\propto (1,1,1)^T$ is an eigenvector of the neutrino mass matrix,
the normalised vector $\frac{\vec v}{|\vec v|}$ corresponds to a column of the
neutrino mixing matrix. Except for being orthogonal to $\vec v$, the other two
columns are not specified by the $S$ symmetry. In order to be meaningful for
physics, the vector $\frac{\vec v}{|\vec v|}$ has to be identified with the
second column of the neutrino mixing matrix, so that the trimaximal pattern
TM2 of Eq.~(\ref{TM2}) ensues. As 
mentioned earlier, this special mixing pattern allows for a large reactor
angle while keeping the solar angle at its tri-bimaximal value at leading order.

It is now important to notice that the TM2 mixing sum rule of Eq.~(\ref{TM2r}) applies
only to the neutrino sector. In other words, the TB deviation parameters as
well as the CP phase should carry a neutrino index, i.e. $s^\nu,a^\nu,r^\nu$
and $\delta^\nu$. In order to find the deviation parameters of the physical
PMNS matrix, it is necessary to add the effect of the charged lepton
corrections. As is demonstrated in~\cite{Cooper:2012wf}, the mass matrices of
the charged fermions in the $A_4\times SU(5)$ model are given by
\be
M_u\sim\begin{pmatrix}
      \overline{\epsilon}^6 & \overline{\epsilon}^6 & \overline{\epsilon}^3 \\
      \overline{\epsilon}^6 & \overline{\epsilon}^3 & \overline{\epsilon}^3 \\
      \overline{\epsilon}^3 & \overline{\epsilon}^3 & 1
     \end{pmatrix}v_u \ , 
\quad
 M_d\sim\begin{pmatrix}
         \epsilon^6 & \epsilon^4 & \epsilon^4 \\
	 \epsilon^4 & \epsilon^3 & \epsilon^3 \\
	 \epsilon^7 & \epsilon^6 & \epsilon
        \end{pmatrix} v_d \ , 
\quad
M_e\sim\begin{pmatrix}
         -3\epsilon^6 & \epsilon^4 & \epsilon^7 \\
	 \epsilon^4 & -3\epsilon^3 & -3\epsilon^6 \\
	 \epsilon^4 & -3\epsilon^3 & \epsilon
        \end{pmatrix} v_d\ ,
\label{eq:mudeA4}
\ee
where the scales of the flavon VEVs were assumed to be
\be
\varphi^q_{\bf r} ~\sim ~ 
\wt \varphi^q_{\bf 1} ~\sim ~ 
\epsilon M \ ,\qquad
\wt \varphi^q_{\bf 1'} ~\sim ~ 
\epsilon^2 M \ .
\ee
$M$ denotes a messenger mass which we allow to vary for the up-type and the
down-type quarks, thus justifying a different expansion parameter in $M_u$
($\ol \epsilon$) and $M_{d,e}$ ($\epsilon$). The factors of $-3$ in $M_e$
correspond to an $SU(5)$ Clebsch-Gordan coefficient which originates from
Georgi-Jarlskog terms of the form $ F T H_{\bf \ol{45}}$~\cite{Georgi:1979df},
multiplied by appropriate products of flavon fields. Finally, $v_d$ denotes the VEV of
the light combination of the electroweak doublets contained in $H_{\bf
  \ol{5}}$  and $H_{\bf \ol{45}}$.  

With the structure of $M_e$ given in Eq.~(\ref{eq:mudeA4}), the only significant
left-handed charged lepton mixing $V_{e_L}$ is the 12 mixing $\theta^e_{12}
\approx \epsilon / 3$.  The parameter $\epsilon$ can be approximated by the
Wolfenstein parameter $\lambda$ as the effect of the left-handed up-type quark
mixing on the CKM matrix is negligible. Combining the TM2 mixing of the
neutrino sector and charged lepton corrections with $\theta^e_{12}\approx
\lambda/3$, see Subsection~\ref{sec:patterns-4}, leads to the sum rule
bounds~\cite{Cooper:2012wf} 
\be
|a|\lesssim \frac{1}{2}\left(r+\frac{\lambda}{3}\right) |\cos \delta| \ ,
\qquad
|s| \lesssim \frac{\lambda}{3} \ ,
\label{eq:TM2rulebounds}
\ee
where $s,a,r$ are the physical TB deviation parameters of the PMNS matrix and
$\delta$ denotes the physical CP phase.

We conclude the discussion of the $A_4\times SU(5)$ model of trimaximal
neutrino mixing by pointing out that this framework does not provide any
explanation for the suppression of the reactor angle compared to the solar or
atmospheric angles. Therefore, this model relies on mild tuning of
parameters. In the next subsection we show how to obtain such a suppression in
the context of an $S_4$ model of tri-bimaximal mixing in which the $U$
symmetry gets broken by higher order corrections.

\subsection[${S_4 \times SU(5)}$: a direct model of tri-bimaximal mixing plus
  corrections]{\label{sec:s4su5}$\bs{S_4 \times SU(5)}$: a direct model of
  tri-bimaximal mixing with corrections}

In this subsection we present the main ingredients of the supersymmetric
$S_4\times SU(5)$ model of~\cite{Hagedorn:2012ut}. It is based on an earlier direct
model~\cite{Hagedorn:2010th} which has been ruled out by the measurement of
$\theta_{13} \approx 8.5^\circ$. In order to accommodate this experimental
result, the model of~\cite{Hagedorn:2010th} has simply been augmented  with an
extra $S_4$ singlet flavon field $\eta$. The three families of $SU(5)$ matter multiplets
$F={\bf \ol 5}$ and $T={\bf 10}$ transform under $S_4$ as ${\bf 3}$ and ${\bf
  2+1}$, respectively. We furthermore introduce three right-handed neutrinos
$\nu^c$ which are unified in the ${\bf 3}$ of~$S_4$ and allow for the type~I
see-saw mechanism. The Higgs sector is $S_4$ blind and comprises
the standard $SU(5)$ Higgses in the ${\bf 5}$ and ${\bf \ol 5}$, plus an
additional Georgi-Jarlskog Higgs in the ${\bf \ol{45}}$. The family symmetry
is broken by a set of flavon fields transforming in various representations
of $S_4$. In order to control the coupling of the flavon fields to different
matter sectors, we impose a global $U(1)$ shaping symmetry. The complete
charge assignments of matter, Higgs and flavon fields are listed in
Table~\ref{tab:S4-assignments}. 

\begin{table}[t]
\begin{center}
$$
\begin{array}{|c|c|c|c|c|c|c|c|c|c|c|c|c|c|c|c||c|}\hline
\!\!\phantom{\Big|}&\multicolumn{4}{c|}{\mathrm{matter~fields}} &
 \multicolumn{3}{c|}{\mathrm{Higgs~fields}}  & 
 \multicolumn{9}{c|}{\mathrm{flavon~fields}} \\ \cline{2-17}
\!\!\phantom{\Big|} & T_3 & T & F & \nu^c & H^{}_{\bf 5} & H_{\bf{\ol{5}}} & H_{\bf{\ol{45}}} &
\phi^u_{\bf 2} & \wt\phi^u_{\bf 2} & \phi^d_{\bf 3} & \wt\phi^d_{\bf 3} &
\phi^d_{\bf 2}  & \phi^\nu_{\bf{ 3'}} & \phi^\nu_{\bf 2} & \phi^\nu_{\bf 1} & \eta\\\hline
~\!SU(5)\!\!\!\phantom{\Big|}~ & \bf 10 & \bf 10 & \bf \ol 5 & \bf 1 &\bf  5 &\bf \ol 5 &\bf \ol{45}
&\bf 1&\bf 1&\bf 1&\bf 1&\bf 1&\bf 1&\bf 1&\bf 1& \bf 1\\\hline
S_4\!\!\phantom{\Big|} & \bf 1&\bf 2&\bf 3&\bf 3&\bf 1&\bf 1&\bf 1&\bf 2&\bf 2&\bf 3&\bf 3&\bf
2&\bf 3'&\bf 2&\bf 1 & \bf 1\\ \hline
U(1)\!\!\phantom{\Big|} & 0&5&4&-4&0&0&1&-10&0&-4&-11 &1 &8&8&8& 7 \\\hline
\end{array}
$$
\end{center}
\caption{\label{tab:S4-assignments}\small{The charge assignments of the matter,
Higgs and flavon superfields in the $S_4 \times SU(5)$ model
of~\cite{Hagedorn:2012ut}. The $U(1)$ shaping symmetry constrains the set of
operators allowed in the superpotential.}}  
\end{table}

With the model formulated at the effective level, it is straightforward to
derive the leading operators of the matter superpotential which are invariant
under all imposed symmetries. Assuming a generic messenger mass $M$ of order
the GUT scale, and  suppressing all dimensionless order one coupling
coefficients, we find 
\bea
W &~\sim~  & T_3T_3H^{}_{\bf 5} + \frac{1}{M} T T  \phi^u_{\bf 2} H^{}_{\bf 5}  +
\frac{1}{M^2} TT \phi^u_{\bf 2} \wt\phi^u_{\bf 2} H^{}_{\bf 5} 
\label{eq:s4-up}
\\ 
&&+ \frac{1}{M} F T_3 \phi^d_{\bf 3} H_{\bf \ol{5}} + \frac{1}{M^2} (F
\wt\phi^d_{\bf 3})^{}_{\bf{1}} ( T \phi^d_{\bf 2} )^{}_{\bf{1}} H_{\bf \ol{45}}
+ \frac{1}{M^3} (F \phi^d_{\bf 2} \phi^d_{\bf 2})^{}_{\bf{3}} ( T \wt\phi^d_{\bf
  3} )^{}_{\bf{3}} H_{\bf \ol{5}}
\label{eq:s4-down}
\\
&&+ F \nu^c H^{}_{\bf 5} + \nu^c \nu^c \phi^\nu_{\bf 1} +  \nu^c \nu^c \phi^\nu_{\bf 2} +  \nu^c \nu^c
\phi^\nu_{\bf {3'}}
 \Red{+ \frac{1}{M} \nu^c\nu^c \phi^d_{\bf 2} \eta }
 \ .
\label{eq:s4-nu}
\eea 
The terms in Eq.~(\ref{eq:s4-nu}), may be compared to the
neutrino sector of the $A_4$ model in Eq.~(\ref{eqn:neutrino}).
In the $S_4$ model it is the last term highlighted in \Red{red} colour which provides the
source of the higher order correction to the right-handed neutrino mass matrix
which is essential in generating a large reactor angle.
In principle, all independent invariant products of the $S_4$ representations
have to be considered for each of these terms; in practice, there is
often only one possible choice. In our example, the second and the third term
of Eq.~(\ref{eq:s4-down}) would give rise to several independent terms. However,
the contractions specified by the subscripts ${\bf 1}$ and ${\bf 3}$ single
out a unique choice. Within a given UV completion, the existence and
non-existence of certain messenger fields can justify such a construction.

The Yukawa matrices are generated when the flavon fields acquire their
VEVs. The explicit form of these matrices depends on the $S_4$ basis which we
choose as given in Appendix~\ref{app:CGs}. Adopting the $F$-term alignment
mechanism which requires to introduce a $U(1)_R$ symmetry as well as new
driving fields, see Subsection~\ref{sec:vacuumdir}, is has been shown
in~\cite{Hagedorn:2010th,Hagedorn:2012ut} that the following alignments can be
obtained,
\bea
\label{VEVup}
&& \langle \phi^u_{\bf 2} \rangle ~ = ~ \varphi^u_{\bf 2} \begin{pmatrix} 0 \\ 1 \end{pmatrix} \; , \;\;
\langle \wt\phi^u_{\bf 2} \rangle ~=~
\wt\varphi^u_{\bf 2} \begin{pmatrix}0\\1\end{pmatrix} \; , \;\;
\\ \label{VEVdown}
&& \langle \phi^d_{\bf 3} \rangle ~ = ~ \varphi^d_{\bf 3} \begin{pmatrix}0\\1\\0\end{pmatrix} \; , \;\;
\langle \wt\phi^d_{\bf 3} \rangle ~ =~
\wt\varphi^d_{\bf 3} \begin{pmatrix}0\\-1\\1\end{pmatrix} \; , \;\;
\langle \phi^d_{\bf 2} \rangle ~=~ \varphi^d_{\bf 2} \begin{pmatrix}1\\0\end{pmatrix} \; , \;\;
\\ \label{VEVnu}
&& \langle\phi^\nu_{\bf{3'}} \rangle~=~ \varphi^\nu_{\bf{3'}} \begin{pmatrix}1\\1\\1 \end{pmatrix} \; , \;\;
\langle\phi^\nu_{\bf 2}\rangle~=~ \varphi^\nu_{\bf 2} \begin{pmatrix}1\\1 \end{pmatrix} \; , \;\;
\langle \phi^\nu_{\bf 1} \rangle~=~ \varphi^\nu_{\bf 1} \ .
\eea
Inserting these vacuum alignments and the Higgs VEVs $v_u$  and $v_d$ yields a
diagonal up-type quark mass matrix $M_u\approx \mathrm{diag}\,(\varphi^u_{\bf
  2}\wt \varphi^u_{\bf 2} /M^2 ,  \varphi^u_{\bf 2} /M \,,  1 ) \,v_u$ as well
as down-type quark and charged  lepton mass matrices 
\bea
\label{matrixMd}
M_d& \approx&
\begin{pmatrix}
0 & {(\varphi^d_{\bf 2})^2 \wt\varphi^d_{\bf 3}}/{M^3}  
& -{(\varphi^d_{\bf 2})^2\wt\varphi^d_{\bf 3}}/{M^3}   \\[0.5mm]
-{(\varphi^d_{\bf 2})^2 \wt\varphi^d_{\bf 3}}/{M^3}   
& {\varphi^d_{\bf 2}\wt\varphi^d_{\bf 3}}/{M^2} 
&   -{\varphi^d_{\bf 2} \wt\varphi^d_{\bf 3}}/{M^2}  +
{(\varphi^d_{\bf 2})^2 \wt\varphi^d_{\bf 3}}/{M^3}   \\[0.5mm]
0  & 0 & {\varphi^d_{\bf 3}}/{M}
\end{pmatrix} v_d  \ ,\\
\label{matrixMe}
M_e &\approx &
\begin{pmatrix}
0 & -{(\varphi^d_{\bf 2})^2 \wt\varphi^d_{\bf 3}}/{M^3}  & 0\\[0.5mm]
{(\varphi^d_{\bf 2})^2 \wt\varphi^d_{\bf 3}}/{M^3}   
& -3 \, {\varphi^d_{\bf 2}\wt\varphi^d_{\bf 3}}/{M^2}  & 0\\[0.5mm]
-{(\varphi^d_{\bf 2})^2 \wt\varphi^d_{\bf 3}}/{M^3}   
& 3 \, {\varphi^d_{\bf 2}\wt\varphi^d_{\bf 3}}/{M^2}  +
{(\varphi^d_{\bf 2})^2 \wt\varphi^d_{\bf 3}}/{M^3}  
&{ \varphi^d_{\bf 3}}/{M}
\end{pmatrix} v_d  \ .
\eea
The factors of $-3$ in $M_e$ originate from the second term of
Eq.~(\ref{eq:s4-down}) involving the Georgi-Jarlskog Higgs field
$H_{\bf{\ol{45}}}$~\cite{Georgi:1979df}. 
Note that the 1-2 and 2-1 entries, which originate from
the same superpotential term, have identical absolute values; together with the
zero texture in the 1-1 entry, this allows for a simple realisation of the GST relation in
the $S_4\times SU(5)$ model. In the neutrino sector we find the
Dirac neutrino mass matrix and the right-handed neutrino mass matrix
\be
m_{LR} \approx \begin{pmatrix}
                     1 & 0 & 0\\
		     0 & 0 & 1\\
		     0 & 1 & 0
\end{pmatrix} v_u  \, , \quad
M_{RR} \approx \begin{pmatrix}
                     \varphi^\nu_{\bf 1} + 2 \varphi^\nu_{\bf 3'} &
                     \varphi^\nu_{\bf 2} -  \varphi^\nu_{\bf 3'} \Red{+
                     \frac{\varphi^d_{\bf 2}\vev{\eta}}{M}}
		              & \varphi^\nu_{\bf 2} -\varphi^\nu_{\bf 3'}\\
		       \varphi^\nu_{\bf 2} -  \varphi^\nu_{\bf 3'} \Red{+
                     \frac{\varphi^d_{\bf 2}\vev{\eta}}{M}}&
                       \varphi^\nu_{\bf 2} + 2  \varphi^\nu_{\bf 3'}
		              &  \varphi^\nu_{\bf 1} -  \varphi^\nu_{\bf 3'}\\
		      \varphi^\nu_{\bf 2} - \varphi^\nu_{\bf 3'} &
                      \varphi^\nu_{\bf 1} - \varphi^\nu_{\bf 3'}
		              &  \varphi^\nu_{\bf 2} + 2  \varphi^\nu_{\bf 3'}
                        \Red{+\frac{\varphi^d_{\bf 2}\vev{\eta}}{M}}
\end{pmatrix}  .\label{matrixneutr}
\ee
It is clear from Eqs.~(\ref{matrixMd}-\ref{matrixneutr}) that the fermion
masses and mixings are solely determined by the scales of the flavon VEVs. In
order to achieve viable GUT scale hierarchies of the quark masses and
mixing angles~\cite{Ross:2007az}, we have to assume
\be
\varphi_{\bf 2}^u \sim \wt \varphi_{\bf 2}^u\sim \lambda^4 M \ , \qquad
\varphi_{\bf 3}^d \sim \lambda^2 M \ , \quad
\wt\varphi_{\bf 3}^d \sim \lambda^3 M \ , \quad 
\varphi_{\bf 2}^d \sim \lambda M \ ,
\ee
where $\lambda$ denotes the Wolfenstein parameter. With these magnitudes, the
charged fermion mass matrices are fixed completely,
\be
M_u \sim
\begin{pmatrix}
\lambda^8 & 0&0   \\
0 & \lambda^4 & 0 \\
0  & 0 & 1
\end{pmatrix} v_u  \ , ~\quad
M_d \sim
\begin{pmatrix}
0 & \lambda^5 & \lambda^5   \\
\lambda^5 & \lambda^4 &  \lambda^4 \\
0  & 0 & \lambda^2
\end{pmatrix} v_d  \ , ~\quad
M_e \sim
\begin{pmatrix}
0 & \lambda^5 &0     \\
\lambda^5 & 3\lambda^4 &  0 \\
\lambda^5  & 3\lambda^4 & \lambda^2
\end{pmatrix} v_d  \ .
\label{eq:s4masslam}
\ee
Due to the GJ factor of $-3$ and the texture zero in the 1-1 entry, we
obtain viable charged lepton masses. With the vanishing off-diagonals in the third
column of $M_e$, there is only a non-trivial 12 mixing in the left-handed
charged lepton mixing $V_{e_L}$, see Subsection~\ref{sec:patterns-4}. This
mixing, $\theta^e_{12} \approx \lambda/3$, will contribute to the total
PMNS mixing as a charged lepton correction.

Turning to the neutrino sector, we first observe that the Dirac neutrino
Yukawa term does not involve any flavon field. As the family symmetry
$S_4$ remains unbroken by $m_{LR}$, the mixing pattern of the effective light
neutrino mass matrix $m^\nu_{LL}$ (obtained from the type~I see-saw mechanism)
is exclusively determined by the structure of $M_{RR}$. Dropping the higher order
terms which are written in \Red{red}, we note that the leading order structure of 
$M_{RR}$, and with it $m^\nu_{LL}$,  is of tri-bimaximal form.\footnote{Similar to the
$A_4\times SU(5)$ model of Subsection~\ref{sec:a4tmgut}, the masses of the light
  neutrinos are not related by any mass sum rule as the right-handed neutrino mass
  matrix $M_{RR}$ is generated from the VEVs of three independent flavon fields.} This can be easily
seen by verifying that the flavon alignments of Eq.~(\ref{VEVnu}) are left
invariant under the $S$ and $U$ transformations of Appendix~\ref{app:CGs}.
This leading order tri-bimaximal structure yields light neutrino masses
$m^\nu\sim 0.1\,\mathrm{eV}$ if we set $\varphi^\nu_{\bf{1,2,3'}} \sim
\lambda^4 M$. As we want to break the TB Klein symmetry by
means of the flavon~$\eta$ at higher order, we set $\vev{\eta} \sim
\lambda^4M$. Then the TB breaking effect is suppressed by one power of
$\lambda$ compared to the leading order. The effective flavon 
$\phi^d_{\bf 2}\eta$ transforms as an $S_4$ doublet with an alignment proportional
to $(1,0)^T$. As can be seen from  the $S_4$ generators of the doublet
representation, see Appendix~\ref{app:CGs}, this alignment breaks the $U$
symmetry but respects $S$. This directly proves that $M_{RR}$ as well as
$m^\nu_{LL}$ are both invariant under~$S$, which in turn entails the TM2 neutrino
mixing pattern, where the second column of the mixing matrix is proportional
to $(1,1,1)^T$, cf.~Eq.~(\ref{TM2}). 
The physical PMNS matrix is obtained from multiplying the TM2 neutrino
mixing with the left-handed charged lepton mixing, see
Eq.~(\ref{Eq:PMNS_Definition}). As a result, we find the same sum rule bound as
in the $A_4\times SU(5)$ model of Subsection~\ref{sec:a4tmgut}, given
explicitly in Eq.~(\ref{eq:TM2rulebounds})~\cite{Hagedorn:2012ut}.

In summary, the measurement of large $\theta_{13}$ has ruled out 
the original $S_4\times SU(5)$ model~\cite{Hagedorn:2010th} which predicted
accurate tri-bimaximal neutrino mixing plus small charged lepton
corrections. A modest extension of the particle content can induce a breaking of the
$U$ symmetry of the TB Klein symmetry at relative order $\lambda$. The
required new flavon field allows for large $\theta_{13}$ and does not destroy
the successful predictions of the original model, i.e. it does not have any
significant effects on the quark or flavon sectors of the model.

\subsection[${\Delta(96) \times SU(5)}$: a direct model of bi-trimaximal mixing]{\label{sec:delta96}$\bs{\Delta(96) \times SU(5)}$: a direct model of bi-trimaximal mixing} 

The direct model discussed in this subsection is based on the observation that larger
family symmetry groups can contain physically interesting $Z_2\times Z_2$
subgroups which, in the basis of diagonal charged leptons, differ from the
well-known TB Klein symmetry~\cite{Toorop:2011jn,deAdelhartToorop:2011re}.
The first model of leptons adopting the family symmetry $\Delta(96)$ was
constructed in~\cite{Ding:2012xx}. Here we present the first (supersymmetric) 
grand unified model based on $\Delta(96)\times SU(5)$~\cite{King:2012in}. 

The group $\Delta(96)$ is a member of the series of groups
$\Delta(6n^2)$~\cite{Escobar:2008vc} with $n=4$. Like its subgroup
$S_4=\Delta(6\cdot 2^2)$, it can be obtained from three generators $S$, $T$,~$U$. 
The group has ten irreducible representations: ${\bf 1}$, ${\bf 1'}$,
${\bf 2}$,  ${\bf 3}$, ${\bf \ol 3}$, ${\bf \wt 3}$, ${\bf 3'}$, ${\bf \ol
  3'}$, ${\bf \wt 3'}$,  ${\bf 6}$. The generators of the one- and
two-dimensional representations are identical to the corresponding $S_4$
representations, see Appendix~\ref{app:CGs}. The triplet ${\bf 3}$ is
generated by
\be
S= \frac{1}{3}
\begin{pmatrix}
 -1 & 2 & 2 \\
 2 & -1 & 2 \\
 2 & 2 & -1
\end{pmatrix} \! , 
~~\,
U=\frac{1}{3} \begin{pmatrix}
 -1+\sqrt{3} & -1-\sqrt{3} & -1 \\
 -1-\sqrt{3} & -1 & -1+\sqrt{3} \\
 -1 & -1+\sqrt{3} & -1-\sqrt{3}
\end{pmatrix}  \! ,
~~\,
\tg=\begin{pmatrix}
\omega^2 & 0&0\\
0&1&0\\
0&0&\omega
\end{pmatrix} \! ,\,
 \label{eq:96-SUT3}
\ee
where $\omega=e^{\frac{2\pi i}{3}}$. Notice that $S$ is identical to the
tri-bimaximal $S$ symmetry of Eq.~(\ref{eq:TBsu}). Invariance of the neutrino
sector of a $\Delta(96)$ model under $S$ therefore implies TM2 neutrino
mixing. The complex conjugate representation ${\bf \ol 3}$ is generated by the
matrices of Eq.~(\ref{eq:96-SUT3}) with $T\rightarrow T^\ast$. The third triplet
${\bf \wt 3}$ is obtained from 
\be
S=
\begin{pmatrix}
 1 & 0 & 0 \\
 0 & 1 & 0 \\
 0 & 0 & 1
\end{pmatrix}  , 
\qquad
U=\frac{1}{3} \begin{pmatrix}
 -2&-2&1\\
-2&1&-2\\
1&-2&-2 
\end{pmatrix}   ,
\qquad
\tg=\begin{pmatrix}
\omega^2 & 0&0\\
0&1&0\\
0&0&\omega
\end{pmatrix}  .
 \label{eq:96-SUT3tilde}
\ee
The generators of the three representations ${\bf 3}$, ${\bf \ol 3}$, ${\bf
  \wt 3}$ all have determinant $+1$. This is not so for the other three
representations ${\bf 3'}$, ${\bf \ol 3'}$, ${\bf \wt 3'}$ which can be
obtained from the unprimed triplets by simply changing the overall sign of the
corresponding $U$ generator. Concerning the sextet representation and for more
details on the group theory of $\Delta(96)$ such as Kronecker products and
Clebsch-Gordan coefficients, we refer to the extensive appendix of~\cite{King:2012in}.

The construction of the $\Delta(96)\times SU(5)$ model follows closely the
logic of the $S_4 \times SU(5)$ model in~\cite{Hagedorn:2010th}. In particular,
the flavons of the up-type and the down-type quark sector are almost
identical, and the GJ mechanism is also 
implemented along with the GST relation. 
The complete charge assignments of the matter, Higgs and flavon fields of the
$\Delta(96)$ model~\cite{King:2012in} are listed in  Table~\ref{trans}. In
addition to a $U(1)$ shaping symmetry -- defined in terms of suitably chosen
integers  $x,y,z,w$ -- a $Z_3$ factor has been introduced to forbid dangerous
terms in the superpotential which, otherwise, would be allowed.

\begin{table}[t]
\centering{{
$\begin{array}{|c|c|c|c|c|c|c|c|c|c|c|c|c|c|c|c|}\hline
\phantom{\Big|}& \multicolumn{4}{c|}{\text{matter~fields}} &  \multicolumn{3}{c|}{\text{Higgs~fields}} &
 \multicolumn{8}{c|}{\text{flavon~fields}} \\\cline{2-16}
\phantom{\Big|}&T_3& T & F & \nu^c & H_{\bf{5}}&H_{\bf{\overline{5}}}&
 H_{\bf{\overline{45}}}& \phi^u_{\bf 2}& \wt{\phi}^u_{\bf 2} &
 \phi_{\bf \overline{3}}^d & \wt{\phi}_{\bf \overline{3}}^d & \phi_{\bf 2}^d & \phi^{\nu}_{\bf{\overline{3}^{\prime}}} & \phi^{\nu}_{\bf{\widetilde{3}^{\prime}}}&\phi^{\nu}_{\bf{\widetilde{3}}}\\\hline
\phantom{\Big|}   SU(5)\phantom{\Big|}      & \bf{10} & \bf{10}&
\bf{\overline{5}} & \bf{1}& \bf{5} & \bf{\overline{5}}&  \bf{\overline{45}} & \bf{1} &\bf{1}&\bf{1}&\bf{1}&\bf{1}&\bf{1}&\bf{1}&\bf{1}\\\hline
\phantom{\Big|} \Delta(96)\phantom{\Big|}    & \bf{1}  &\bf{2} & \bf{3} & \bf{\overline{3}} & \bf{1}& \bf{1}& \bf{1}  & \bf{2} & \bf{2}&  \bf{\overline{3}} & \bf{\overline{3}}& \bf{2} & \bf{\overline{3}^{\prime}}  & \bf{\widetilde{3}^{\prime}}&\bf{\widetilde{3}}\\\hline
\phantom{\Big|}U(1)\phantom{\Big|}  & 0 & x & y & -y& 0 &0 & z &-2x & 0 &-y& -x-y-2z &z&2y&2y&w\\\hline
\phantom{\Big|}Z_3\phantom{\Big|}    & 0 & 0   &2 & 1 & 0&1 & 1&0&0&0&0&0&1&1&1\\\hline
\end{array}$}}
\caption{\label{trans}\small{The charge assignments of the matter, Higgs and
    flavon superfields in the $\Delta(96)\times SU(5)$ model
    of~\cite{King:2012in}.  The $U(1)$ shaping symmetry is defined by four
    independent integers $x$, $y$, $z$, and $w$.}}
\end{table}

With the particle content and the symmetries of Table~\ref{trans}, the leading
order operators of the matter superpotential take the form
\bea
W &~\sim~  & T_3T_3H^{}_{\bf 5} + \frac{1}{M} T T  \phi^u_{\bf 2} H^{}_{\bf 5}  +
\frac{1}{M^2} TT \phi^u_{\bf 2} \wt\phi^u_{\bf 2} H^{}_{\bf 5} 
\label{eq:96-up}
\\ 
&&+ \frac{1}{M} F T_3 \phi^d_{\bf \ol 3} H_{\bf \ol{5}} + \frac{1}{M^2} (F
\wt\phi^d_{\bf \ol 3})^{}_{\bf{1}} ( T \phi^d_{\bf 2} )^{}_{\bf{1}} H_{\bf \ol{45}}
+ \frac{1}{M^3} (F \phi^d_{\bf 2} \phi^d_{\bf 2})^{}_{\bf{3}} ( T \wt\phi^d_{\bf
  \ol 3} )^{}_{\bf{\ol 3}} H_{\bf \ol{5}}~~~~
\label{eq:96-down}
\\
&&+ F \nu^c H^{}_{\bf 5} + \nu^c \nu^c \phi^\nu_{\bf {\ol 3'}} +  \nu^c \nu^c \phi^\nu_{\bf {\wt 3'}}  \ .
 \phantom{\frac{1}{M}} 
\label{eq:96-nu}
\eea 
Dimensionless order one coupling coefficients are suppressed, and $M$ is a
generic messenger mass scale. The subscripts on the parentheses denote the
specific contractions being taken from the $\Delta(96)$ tensor product
contained inside the parentheses.

As has been elaborated in~\cite{King:2012in}, the $F$-term alignment mechanism
can be used to derive the vacuum alignment of the flavon fields as\footnote{The alignments presented in Eq.~(\ref{VEVnu96}) do not break the
neutrino Klein symmetry generated by $S$ and~$U$. While
$\vev{\phi^\nu_{\bf{\ol 3'}}}$ corresponds to the most general such alignment
of a flavon in the ${\bf \ol 3'}$, this is not the case 
for~$\vev{\phi^\nu_{\bf{\wt 3'}}}$. It is straightforward to verify that the most
general alignment of a  flavon in the ${\bf \wt 3'}$ of $\Delta(96)$ 
which satisfies $S \vev{\phi^\nu_{\bf{\wt 3'}}} = U \vev{\phi^\nu_{\bf{\wt 3'}}} =
\vev{\phi^\nu_{\bf{\wt 3'}}}$, with $U$ being the negative of the matrix shown
in Eq.~(\ref{eq:96-SUT3tilde}), takes the form $\langle\phi^\nu_{\bf \wt 3'}\rangle 
\propto \left(v_1, \frac{v_1+v_3}{2},v_3\right)^T$.} 
\bea
\label{VEVup96}
&& \langle \phi^u_{\bf 2} \rangle ~ = ~ \varphi^u_{\bf 2} \begin{pmatrix} 0 \\ 1 \end{pmatrix} \; , \;\;
\langle \wt\phi^u_{\bf 2} \rangle ~=~
\wt\varphi^u_{\bf 2} \begin{pmatrix}0\\1\end{pmatrix} \; , \;\;
\\ \label{VEVdown96}
&& \langle \phi^d_{\bf \ol 3} \rangle ~ = ~ \varphi^d_{\bf \ol 3} \begin{pmatrix}0\\0\\1\end{pmatrix} \; , \;\;
\langle \wt\phi^d_{\bf \ol 3} \rangle ~ =~
\wt\varphi^d_{\bf \ol 3} \begin{pmatrix}0\\1\\-1\end{pmatrix} \; , \;\;
\langle \phi^d_{\bf 2} \rangle ~=~ \varphi^d_{\bf 2} \begin{pmatrix}1\\0\end{pmatrix} \; , \;\;
\\ \label{VEVnu96}
&& \langle\phi^\nu_{\bf{\ol 3'}} \rangle~=~ \varphi^\nu_{\bf{\ol 3'}} \begin{pmatrix}1\\1\\1 \end{pmatrix} \; , \;\;
\langle\phi^\nu_{\bf \wt 3'}\rangle~=~ \varphi^\nu_{\bf \wt
  3'} \begin{pmatrix}\omega^2 \\ -\frac{1}{2} \\ \omega  \end{pmatrix}  \ .
\eea
Inserting the flavon and the Higgs VEVs yields the charged fermion mass matrices
$M_u\approx
\mathrm{diag}\,(\varphi^u_{\bf 2}\wt \varphi^u_{\bf 2} /M^2 , 
 \varphi^u_{\bf  2} /M \,,  1 ) \,v_u$ 
and
\bea
\label{matrixMd96}
M_d& \approx&
\begin{pmatrix}
0 & {(\varphi^d_{\bf 2})^2 \wt\varphi^d_{\bf \ol 3}}/{M^3}  
& -{(\varphi^d_{\bf 2})^2\wt\varphi^d_{\bf \ol 3}}/{M^3}   \\[0.5mm]
{(\varphi^d_{\bf 2})^2 \wt\varphi^d_{\bf \ol 3}}/{M^3}   
& {\varphi^d_{\bf 2}\wt\varphi^d_{\bf \ol 3}}/{M^2} -
{(\varphi^d_{\bf 2})^2 \wt\varphi^d_{\bf \ol 3}}/{M^3}
&   -{\varphi^d_{\bf 2} \wt\varphi^d_{\bf \ol 3}}/{M^2}     \\[0.5mm]
0  & 0 & {\varphi^d_{\bf \ol 3}}/{M}
\end{pmatrix} v_d  \ ,\\
\label{matrixMe96}
M_e &\approx &
\begin{pmatrix}
0 & {(\varphi^d_{\bf 2})^2 \wt\varphi^d_{\bf \ol 3}}/{M^3}  & 0\\[0.5mm]
{(\varphi^d_{\bf 2})^2 \wt\varphi^d_{\bf \ol 3}}/{M^3}   
& -3 \, {\varphi^d_{\bf 2}\wt\varphi^d_{\bf \ol 3}}/{M^2}  -
{(\varphi^d_{\bf 2})^2 \wt\varphi^d_{\bf \ol 3}}/{M^3}  & 0\\[0.5mm]
-{(\varphi^d_{\bf 2})^2 \wt\varphi^d_{\bf \ol 3}}/{M^3}   
& 3 \, {\varphi^d_{\bf 2}\wt\varphi^d_{\bf \ol 3}}/{M^2}  
&{ \varphi^d_{\bf \ol 3}}/{M}
\end{pmatrix} v_d  \ .
\eea
In the neutrino sector we obtain the Dirac neutrino mass matrix and the
right-handed neutrino mass matrix
\be
m_{LR} \approx \begin{pmatrix}
                     1 & 0 & 0\\
		     0 & 1 & 0\\
		     0 & 0 & 1
\end{pmatrix} v_u  \, , \qquad
M_{RR} \approx \begin{pmatrix} 
-2 \varphi^\nu_{\bf \ol 3'}  +\omega\varphi^\nu_{\bf \wt 3'} &
\varphi^\nu_{\bf \ol 3'}  +\omega^2\varphi^\nu_{\bf \wt 3'} &
   \varphi^\nu_{\bf \ol 3'}    -\frac{1}{2}\varphi^\nu_{\bf \wt 3'} 
\\
\varphi^\nu_{\bf \ol 3'}  +\omega^2\varphi^\nu_{\bf \wt 3'} &
-2\varphi^\nu_{\bf \ol 3'}  -\frac{1}{2}\varphi^\nu_{\bf \wt 3'} &
    \varphi^\nu_{\bf \ol 3'}  +\omega\varphi^\nu_{\bf \wt 3'} 
\\
\varphi^\nu_{\bf \ol 3'}  -\frac{1}{2}\varphi^\nu_{\bf \wt 3'} &
\varphi^\nu_{\bf \ol 3'}  +\omega\varphi^\nu_{\bf \wt 3'} &
-2\varphi^\nu_{\bf \ol 3'}  +\omega^2\varphi^\nu_{\bf \wt 3'} 
\end{pmatrix}  .
\label{matrixneutr96}
\ee
The mass matrices $M_u$, $M_d$ and $M_e$ of the $\Delta(96)$ model are almost
identical to those of the $S_4$ model presented in
Subsection~\ref{sec:s4su5}. Indeed, with a similar hierarchy of flavon VEVs,
\be
\varphi_{\bf 2}^u \sim \wt \varphi_{\bf 2}^u\sim \lambda^4 M \ , \qquad
\varphi_{\bf \ol 3}^d \sim \lambda^2 M \ , \quad
\wt\varphi_{\bf \ol 3}^d \sim \lambda^3 M \ , \quad 
\varphi_{\bf 2}^d \sim \lambda M \ ,
\ee
one obtains the charged fermion mass matrices of Eq.~(\ref{eq:s4masslam}). As in
Subsection~\ref{sec:s4su5}, the left-handed charged lepton mixing is
described by a unitary matrix with angles $\theta^e_{13}\approx
\theta^e_{23}\approx~0$ and $\theta^e_{12} \approx \lambda/3$ as well as 
a general phase $\delta^e_{12}$, see e.g. Eq.~(\ref{U12}). 

In the neutrino sector, the right-handed neutrino mass matrix $M_{RR}$ involves
the VEVs of only two flavon fields. The three eigenvalues of $M_{RR}$ are
therefore related, leading to the mass sum rule for the light neutrinos
reported in Eq.~(\ref{inversemasssum}) with $\gamma = 1$ and $ \delta = \pm 2i$.
$M_{RR}$ is furthermore symmetric
under the Klein symmetry generated by $S$ and $U$ of Eq.~(\ref{eq:96-SUT3}). This
can be shown either explicitly by calculating $S^T M_{RR} S = U^T M_{RR} U = M_{RR}$,
or by realising that the alignments of the two neutrino flavons $\phi^\nu_{\bf
  \ol 3'}$ and $\phi^\nu_{\bf \wt 3'}$ of Eq.~(\ref{VEVnu96}) are left invariant
under both $S$ and $U$ (in the appropriate representations). 
With $m_{LR}$ being proportional to the identity matrix, the effective light
neutrino mass matrix $m^\nu_{LL}$, obtained from the type~I see-saw formula,
is symmetric under $S$ and $U$ as well. This particular Klein symmetry which
originates from the family symmetry $\Delta(96)$ corresponds to the so-called
bi-trimaximal mixing pattern~\cite{King:2012in,Toorop:2011jn} in  
the neutrino sector,
\be
\label{Unu96}
U^{\nu}_{\mathrm{BT}}=
\begin{pmatrix}
 a_+ & \frac{1}{\sqrt{3}} & a_- \\
 -\frac{1}{\sqrt{3}} & \frac{1}{\sqrt{3}} & \frac{1}{\sqrt{3}} \\
a_- & -\frac{1}{\sqrt{3}} & a_+
\end{pmatrix} , ~~~ \mathrm{with} ~~
a_\pm=\frac{1}{2}\left(1\pm\frac{1}{\sqrt{3}}\right) \ .
\ee
The bi-trimaximal mixing structure, which is a special form of TM2 mixing,
does not involve any CP phases and translates to the following values of the
neutrino mixing angles\footnote{Notice that this leading order result is
a  realisation of bi-large mixing as defined in~\cite{Boucenna:2012xb-1}.} 
\be
\label{neutangles96}
\theta^{\nu}_{13}=\sin ^{-1}(a_-) \approx 12.2^{\circ} \ ,
\qquad  
\theta^{\nu}_{12}=\theta^{\nu}_{23}=\tan ^{-1}\left(\sqrt{3}-1\right)
\approx 36.2^{\circ} \ .
\ee
It is a remarkable fact that $\theta^\nu_{13}$ deviates from the
experimentally measured value of the reactor angle $\theta_{13}\approx
8.5^\circ$ by about $3^\circ$, a deviation which is typical for charged lepton
corrections to a vanishing 13 neutrino mixing angle 
in $SU(5)$ GUTs with implemented Georgi-Jarlskog mechanism.  
Indeed, taking into account the left-handed charged lepton 12 mixing with
$\theta^e_{12} \approx \lambda/3$ and phase $\delta^e_{12}$, see
Subsection~\ref{sec:patterns-4}, leads to the relation
\be
\sin\theta_{13}
\approx a_- - \mbox{$\frac{1}{\sqrt{3}}$}  \theta^e_{12}\cos \delta^e_{12} \ . 
\ee
As the charged lepton correction to $\theta_{13}$ has to be negative in order to
hit the measured value, one must assume $\delta^e_{12}\approx 0$. This entails
a vanishing Dirac CP phase 
\be 
\delta\approx 0 \ .
\ee 
The final PMNS mixing matrix is then real (up to Majorana phases) and has the form,
\begin{eqnarray}\label{Eq:PMNS}
U_{\mathrm{PMNS}} \approx 
\left(\begin{array}{ccc}
a_+c^e_{12} + \frac{1}{\sqrt{3}}  s^e_{12}
&
 \frac{1}{\sqrt{3}}c^e_{12} - \frac{1}{\sqrt{3}}  s^e_{12}
&
a_-c^e_{12} - \frac{1}{\sqrt{3}}  s^e_{12}
\!\\
a_+s^e_{12} - \frac{1}{\sqrt{3}} c^e_{12}
&
\frac{1}{\sqrt{3}}s^e_{12} + \frac{1}{\sqrt{3}} c^e_{12} 
&
a_-s^e_{12} + \frac{1}{\sqrt{3}} c^e_{12}
\!\\
a_-
&
-  \frac{1}{\sqrt{3}} 
&
a_+
\end{array}
\right) ,
\end{eqnarray}
leading to the following phenomenologically viable lepton mixing angles,
\be 
\theta_{13}\approx 9.6^{\circ} \ ,
\qquad
\theta_{12}\approx 32.7^{\circ} \ , 
\qquad
\theta_{23}\approx 36.9^{\circ} \ .
\ee
The charged lepton corrections are thus crucial for direct models based on a
$\Delta(96)$ family symmetry. Furthermore, a specific choice of the phase
$\delta^e_{12}$ is required, which must eventually be explained in a more complete
model, for example along the lines of the models proposed
in~\cite{Antusch:2011sx}, see also end of Subsection~\ref{sec:vacuumdir}.

\section{\label{sec:conclude}Conclusion}
Neutrino physics has progressed at a breathtaking rate over the last decade
and a half, as discussed in Section~\ref{sec:intro}, with a major discovery almost every
other year, as can be seen by glancing at the milestones in
Subsection~\ref{milestones}. The big experimental result of 2012 is the
measurement by Daya Bay and RENO of the reactor angle, which has had a major
impact on neutrino physics, making the early measurement of CP violation
possible. It has also ruled out a large number of models which predicted the
reactor angle to be zero, including simple patterns of lepton mixing
known as BM, TB or GR which are presented in Subsection~\ref{sec:patterns-1}. As
discussed in Section~\ref{sec:FS}, these simple patterns can result from
discrete symmetries, reviewed in  Section~\ref{sec:nut}, such as $A_4$, $S_4$
or $A_5$. On the other hand, anarchy  expected the reactor angle to be rather
large and this is what was observed. Does this mean that we should give up on
the symmetry approach in favour of anarchy?  

In this review the idea of giving up on the symmetry approach in favour of
anarchy is given short shrift for good reason, namely some simple mixing
patterns remain a good approximate characterisation of lepton mixing. 
For example, we have seen that it is possible to describe the current pattern
of lepton mixing by simply perturbing around TB mixing.
This perturbation can be parameterised in terms of the reactor ($r$), solar
($s$) and atmospheric ($a$) deviations from TB mixing introduced in
Subsection~\ref{sec:patterns-2}, where such perturbations are no larger than
the Wolfenstein parameter $\lambda$, which apparently coincides with the
reactor parameter $r$. Indeed TBC mixing provides a good approximation to the
observed lepton mixing.  

From the symmetry perspective, the measurement of the reactor angle has caused
theorists to work harder to explain the observed deviations from TB mixing,
which in the near future could also include the atmospheric deviation
parameter $a$ and the solar deviation parameter $s$. Indeed there are already
hints from the global fits to oscillation data that both these parameters
could be non-zero, as discussed in Subsection~\ref{sec:glob}. The models based on  
small discrete family symmetries such as $A_4$, $S_4$ or $A_5$ may be viewed
as predicting BM, TB or GR lepton mixing only at the LO in the absence of HO
operator corrections or charged lepton corrections, not to mention
renormalisation group running or canonical normalisation corrections. Indeed
there are many sources of corrections that can modify the naive simple mixing patterns apparently predicted by discrete family symmetry.

If we are very lucky then it is possible that only a special subset of these corrections are important, as discussed in Section~\ref{sec:patterns}. For example, if only Cabibbo-like charged lepton corrections are important then this leads to solar mixing sum rules. Although, in the framework of GUT models, it might seem natural to equate charged lepton corrections to the Cabibbo angle, this assumption is at odds with the simplest type of relations between charged lepton masses and down-type quark masses, where such relations would prefer the charged lepton mixing angle to be about a third of the Cabibbo angle. Alternatively, in the absence of charged lepton corrections,
if only certain kinds of HO corrections are important then TB mixing could be reduced down to TM1 or TM2 mixing, where only half of the original Klein symmetry of the neutrino mass matrix is broken, and this leads to atmospheric sum rules. In either case, solar and atmospheric sum rules provide interesting relations involving the deviation parameters $r,s,a$ together with $\cos \delta$ which can be tested in future neutrino oscillation experiments.

We have distinguished between two general approaches of using discrete family
symmetry to build realistic models, 
referred to as direct or indirect.
In the direct approach the Klein symmetry of the neutrino mass matrix,
discussed in Subsection~\ref{sec:kleinse}, is identified as a subgroup of the 
discrete family symmetry (with a different subgroup preserved in the charged lepton sector), while in the indirect approach the Klein symmetry emerges accidentally after the discrete family symmetry is completely broken. 
Both direct and indirect approaches rely on spontaneous breaking of the
discrete family symmetry via new scalar fields which develop VEVs, referred to
here as flavons, as discussed generally in Subsection~\ref{sec:FNs}. 

A common feature of both the simplest direct or indirect models is the use of
the type~I see-saw mechanism, reviewed in Section~\ref{sec:seesaw}, with form dominance. In fact the different types of see-saw mechanism are generally reviewed along with the sequential dominance mechanism. 
As discussed in Subsection~\ref{sec:ImB-1}, form dominance corresponds to the columns of the Dirac mass matrix 
being proportional to the columns of the PMNS matrix in the basis where the right-handed neutrino mass matrix is diagonal. The virtue is that this usually leads to a form diagonalisable physical neutrino mass matrix, with mixing angles being independent of neutrino masses. In the case of the simplest indirect models, form dominance is usually identified with constrained sequential dominance featuring a normal mass hierarchy with the lightest neutrino mass being approximately zero. The downside of form dominance is that, since the columns of the Dirac mass matrix are orthogonal, leptogenesis is identically zero so corrections to form dominance may be required.

The direct approach is discussed in detail in Section~\ref{sec:DmB}, including
flavon vacuum alignment and model building strategies following Daya Bay and
RENO. In the direct approach it is possible to achieve the simple patterns of
lepton mixing, namely BM, TB or GR as an approximation to lepton mixing using
small discrete family symmetries such as $A_4$, $S_4$ or $A_5$, and then
consider the effect of corrections as discussed above. If we are lucky then
such  
corrections could either preserve the Klein symmetry in the neutrino sector
and break the symmetry in the charged lepton sector, leading to solar sum
rules, or preserve half the original Klein symmetry in the neutrino sector,
leading to TM mixing and atmospheric sum rules. Indeed it is possible to start
with only half the original Klein symmetry in the neutrino sector arising as a
subgroup of the family symmetry, as in $A_4$ for example, which we refer to as
the semi-direct approach. More generally, however, the whole original Klein
symmetry will be broken by HO corrections, and also charged lepton,
renormalisation group and canonical normalisation corrections will also be present. Alternatively it is possible to use larger discrete family symmetries in which the Klein symmetry already gives a non-zero reactor angle at the LO. For example 
$\Delta(96)$ can give bi-trimaximal mixing, where the reactor angle is non-zero and the solar and atmospheric angles start out equal, with all angles receiving modest charged lepton corrections. 

An analogous discussion for the indirect approach is provided in
Section~\ref{sec:ImB}. An important difference is that TB mixing at the LO can
be achieved not only from groups in which the Klein symmetry $Z_2\times Z_2$
can be embedded 
but also in groups of odd order such as $\Delta(27)$ or $T_7$, or more generally infinite classes of groups.
The types of correction to TB mixing discussed above are also possible for the
indirect case, perhaps leading to solar and atmospheric sum rules if we are
lucky. However, the indirect approach offers new alternatives to achieving a
reactor angle already at the LO, without resorting to large discrete family
symmetry groups, by using new vacuum alignments to construct the neutrino mass
matrix. Such new alignments are more easily achieved with smaller groups since
the discrete family symmetry is completely broken and we are freed from the
requirement of identifying the Klein symmetry as a subgroup. Examples of new
vacuum alignments, which can be achieved even in $A_4$, include PCSD and CSD2,
where PCSD can give the successful TBC mixing if the misalignment parameter is
identified with the Wolfenstein parameter. 

We have already explained that we prefer the symmetry approach to, say, anarchy, since simple symmetric mixing patterns still provide a good approximation to reality. However there is a deeper reason why we prefer to use symmetry, namely to address the flavour problem. In our view, to abandon the symmetry approach would completely miss the opportunity provided by neutrino physics of elucidating the flavour problem, the problem of all quark and lepton masses and mixing parameters, including CP violating phases. The history of physics,
if it tells us anything at all, teaches us that symmetry and unification have always provided a guiding light in the path to understanding deep problems in physics. Therefore in this review we have considered models based not only on discrete family symmetry,  but also using GUTs, in order to address the flavour problem. 
Motivated by such considerations, in Section~\ref{sec:gut} we have briefly reviewed grand unified theories and how they may be combined with discrete family symmetry to describe all quark and lepton masses and mixing, tabulating some recent examples of this kind. 

Finally in Section~\ref{sec:examp} we have discussed three model examples which
combine an $SU(5)$ GUT with the discrete family symmetries $A_4$, $S_4$ and
$\Delta(96)$. These models are presented in sufficient detail to illustrate the complexity of the current state of the art of GUTs of flavour that is required to account for all quark and lepton masses and mixing. Critics will use these examples as evidence that the effort of constructing such models is not worth the trouble and question what has been achieved by all this complexity. They will also point out that the number of input parameters in these models exceeds the number of mass and mixing parameters that we are trying to explain. However this last observation misses the point. What is relevant is the number of predictions (or postdictions) not the number of parameters. Typically each of these models contains half a dozen relationships such as the GST and GJ relations which agree with experiment, and neutrino mass and mixing sum rules giving predictions for future neutrino experiments. Thus one feels that something has been understood by constructing these models, including the mass hierarchy and origin of the three families as well as the milder neutrino hierarchy with bi-large lepton mixing.

The flavour problem is not going away, in fact since 1998 it has got significantly worse with at least seven new parameters arising from the neutrino sector on top of the three charged lepton masses and 
the ten flavour parameters from the quark sector. However the neutrino parameters also provide some clues such as small neutrino masses, bi-large mixing and a Cabibbo-like reactor angle. There is a ghost of a chance that these clues may be just enough to allow us to unlock the whole flavour puzzle.  
We are not there yet, but the hope is that the details of the admittedly rather complicated models given in this review may inform and inspire new young researchers to do better. 

It is still not too late for theorists to redeem their past failures to successfully predict anything in the neutrino sector by 
making a genuine prediction which can be tested by experiment.
Indeed there is still room to make predictions for the solar and atmospheric deviation parameters $s,a$ as well 
as $\cos \delta$, or to relate them via sum rules to the reactor parameter $r$ since 
all these parameters can and will be measured in future high precision
neutrino experiments. Also one can make predictions for 
the pattern of neutrino masses including their ordering and their scale.
A crucial question is whether neutrinos are Dirac or Majorana particles.
In the former case neutrino mass could have the same origin as that of 
the quarks and charged leptons, while in the latter
case something qualitatively different may be involved such as the see-saw mechanism for example.
In the absence of experimental information about these questions, we must admit that we do not yet understand the origin of neutrino mass, which remains one of the biggest unsolved mysteries of the Standard Model.

\begin{center}
{\bf Acknowledgements}
\end{center}
We thank A. Merle and A. Stuart for providing Fig.~\ref{doublebeta2}, and 
S. Morisi for the permission to use his code to generate Fig.~\ref{fig:mixingfits}.
SFK acknowledges partial support from the STFC Consolidated ST/J000396/1
grant. SFK and CL acknowledge partial support from the EU ITN grants UNILHC
237920 and INVISIBLES 289442.

\appendix

\section{\label{equivalence}Equivalence of different parametrisations}

In this appendix we exhibit the equivalence of different
parametrisations of the lepton mixing matrix.
A $3\times 3$ unitary matrix may be parametrised by 3 angles and 6
phases. We shall find it convenient to parametrise a unitary
matrix $V^{\dagger}$ by:\footnote{It is convenient to define the
parametrisation of $V^{\dagger}$ rather than $V$
because the lepton mixing matrix involves $V_{\nu_L}^{\dagger}$
and the neutrino mixing angles will play a central r\^{o}le.}
\beq
V^{\dagger}=P_2R_{23}R_{13}P_1R_{12}P_3 , 
\label{V1}
\eeq
where $R_{ij}$ are a sequence of real rotations corresponding to the
Euler angles $\theta_{ij}$, and $P_i$ are diagonal phase matrices.
The Euler matrices are given by
\begin{equation}
R_{23}=
\left(\begin{array}{ccc}
1 & 0 & 0 \\
0 & c_{23} & s_{23} \\
0 & -s_{23} & c_{23} \\
\end{array}\right) , \quad
R_{13}=
\left(\begin{array}{ccc}
c_{13} & 0 & s_{13} \\
0 & 1 & 0 \\
-s_{13} & 0 & c_{13} \\
\end{array}\right), \quad
R_{12}=
\left(\begin{array}{ccc}
c_{12} & s_{12} & 0 \\
-s_{12} & c_{12} & 0\\
0 & 0 & 1 \\
\end{array}\right) ,
\end{equation}
where $c_{ij} = \cos\theta_{ij}$ and $s_{ij} = \sin\theta_{ij}$.
The phase matrices are given by
\beq
P_1=
\left( \begin{array}{ccc}
1 & 0 & 0    \\
0 & e^{i\chi} & 0 \\
0 & 0 & 1
\end{array}
\right),\quad
P_2=
\left( \begin{array}{ccc}
1 & 0 & 0    \\
0 & e^{i\phi_2} & 0 \\
0 & 0 & e^{i\phi_3}
\end{array}
\right),\quad
P_3=
\left( \begin{array}{ccc}
e^{i\omega_1} & 0 & 0    \\
0 & e^{i\omega_2} & 0 \\
0 & 0 & e^{i\omega_3}
\end{array}
\right).
\eeq

By commuting the phase matrices to the left, it is not difficult to
show that the parametrisation in Eq.~(\ref{V1}) is equivalent to
\beq
V^{\dagger}=PU_{23}U_{13}U_{12},
\label{V2}
\eeq
where $P=P_1P_2P_3$ and
\begin{equation}
U_{23}=
\left(\begin{array}{ccc}
1 & 0 & 0 \\
0 & c_{23} & s_{23}e^{-i\delta_{23}} \\
0 & -s_{23}e^{i\delta_{23}} & c_{23} \\
\end{array}\right),
\label{U23}
\end{equation}
\begin{equation}
U_{13}=
\left(\begin{array}{ccc}
c_{13} & 0 & s_{13}e^{-i\delta_{13}} \\
0 & 1 & 0 \\
-s_{13}e^{i\delta_{13}} & 0 & c_{13} \\
\end{array}\right),
\label{U13}
\end{equation}
\begin{equation}
U_{12}=
\left(\begin{array}{ccc}
c_{12} & s_{12}e^{-i\delta_{12}} & 0 \\
-s_{12}e^{i\delta_{12}} & c_{12} & 0\\
0 & 0 & 1 \\
\end{array}\right),
\label{U12}
\end{equation}
where
\bea
\delta_{23}&=&\chi+\omega_2-\omega_3,
\\
\delta_{13}&=&\omega_1-\omega_3,
\\
\delta_{12}&=&\omega_1-\omega_2.
\eea
The matrix $U$ is an example of a unitary matrix,
and as such
it may be parametrised by either of the equivalent forms
in Eqs.~(\ref{V1}) or (\ref{V2}).
If we use the form in Eq.~(\ref{V2}) then the phase matrix $P$ on the
left may always be removed by an additional charged lepton phase rotation
$\Delta V_{e_L}=P^\dagger$,
which is always possible since
right-handed charged lepton phase rotations can always make the charged
lepton masses real. Therefore $U$ can always be parametrised by
\beq
U=U_{23}U_{13}U_{12},
\label{MNS2A}
\eeq
which involves just three irremovable physical phases $\delta_{ij}$.
In this parametrisation the Dirac phase $\delta$
which enters the CP odd part of
neutrino oscillation probabilities is given by
\beq
\delta = \delta_{13}-\delta_{23}-\delta_{12}.
\label{DiracA}
\eeq

Another common parametrisation of the lepton mixing matrix is
\beq
U=R_{23}U_{13}R_{12}P_0,
\label{MNS3}
\eeq
where
\beq
P_0=
\left( \begin{array}{ccc}
e^{i\beta_1} & 0 & 0    \\
0 & e^{i\beta_2} & 0 \\
0 & 0 & 1
\end{array}
\right),
\eeq
and in Eq.~(\ref{MNS3}) $U_{13}$ is of the form in
Eq.~(\ref{U13}) but with $\delta_{13}$ replaced by the Dirac phase $\delta$.
The parametrisation in Eq.~(\ref{MNS3}) can be transformed into
the parametrisation in Eq.~(\ref{MNS2A}) by commuting the phase matrix $P_0$
in Eq.~(\ref{MNS3}) to the left, and then removing the phases
on the left-hand side by charged lepton phase rotations.
The two parametrisations are then related by the phase relations
\bea
\delta_{23}&=&\beta_2,
\\
\delta_{13}&=&\delta + \beta_1,
\\
\delta_{12}&=&\beta_1-\beta_2.
\eea
The use of the parametrisation in Eq.~(\ref{MNS3}) is widespread in the
literature, however it is often more convenient to use the parametrisation in
Eq.~(\ref{MNS2A}) which is trivially related to Eq.~(\ref{MNS3}) by the above phase
relations.

\section{\label{deviations}Deviations from TB mixing to second order }

\subsection{PMNS matrix expansion}
The PMNS matrix when expanded to first order in the three real parameters $r,s,a$ 
defined in Eq.~(\ref{rsadef}) reads ~\cite{King:2007pr}
\begin{eqnarray}
U_{\mathrm{PMNS}} \approx
\left( \begin{array}{ccc}
\sqrt{\frac{2}{3}}(1-\frac{1}{2}s)  & \frac{1}{\sqrt{3}}(1+s) & \frac{1}{\sqrt{2}}re^{-i\delta } \\
-\frac{1}{\sqrt{6}}(1+s-a + re^{i\delta })  & \frac{1}{\sqrt{3}}(1-\frac{1}{2}s-a- \frac{1}{2}re^{i\delta })
& \frac{1}{\sqrt{2}}(1+a) \\
\frac{1}{\sqrt{6}}(1+s+a- re^{i\delta })  & -\frac{1}{\sqrt{3}}(1-\frac{1}{2}s+a+ \frac{1}{2}re^{i\delta })
 & \frac{1}{\sqrt{2}}(1-a)
\end{array}
\right)P_{12}.~~~
\label{MNS1}
\end{eqnarray}
The second order corrections to the PMNS matrix elements are~\cite{King:2007pr},
\begin{eqnarray}
\Delta U_{e1} & \approx & \sqrt{\frac{2}{3}}(-\frac{1}{4}r^2 -\frac{3}{8}s^2), \nonumber \\
\Delta U_{e2} & \approx & \frac{1}{\sqrt{3}}( -\frac{1}{4}r^2) ,\nonumber \\
\Delta U_{e3} & \approx & 0,\nonumber \\
\Delta U_{\mu 1} & \approx & -\frac{1}{\sqrt{6}}(\frac{1}{2}rse^{i\delta }  -rae^{i\delta }  +sa +a^2) ,\nonumber \\
\Delta U_{\mu 2} & \approx & \frac{1}{\sqrt{3}}(-\frac{1}{2}rse^{i\delta }  -\frac{1}{2}rae^{i\delta }  +\frac{1}{2}sa - \frac{3}{8}s^2-a^2),
\nonumber \\
\Delta U_{\mu 3} & \approx & \frac{1}{\sqrt{2}}(-\frac{1}{4}r^2),
\nonumber \\
\Delta U_{\tau 1} & \approx &
\frac{1}{\sqrt{6}}(\frac{1}{2}rse^{i\delta }  +rae^{i\delta }  +sa),
\nonumber \\
\Delta U_{\tau 2} & \approx &
-\frac{1}{\sqrt{3}}(\frac{1}{2}rse^{i\delta }  -\frac{1}{2}rae^{i\delta }  -\frac{1}{2}sa - \frac{3}{8}s^2),
\nonumber \\
\Delta U_{\tau 3} & \approx &
\frac{1}{\sqrt{2}}(-\frac{1}{4}r^2 -a^2).
\label{MNS2}
\end{eqnarray}
The Jarlskog CP invariant to second order is then given by~\cite{King:2007pr}
\be
J\approx (\frac{r}{6} + \frac{rs}{12} ) \sin \delta .
\ee

\subsection{Neutrino oscillations in matter}\label{matter}
In this appendix we present the complete formulae for neutrino oscillations in the
presence of matter of constant density to second order in the quantities
$r,s,a$ and $\Delta_{21}$, where it is assumed that
$\Delta_{21} \ll 1$.
Here $\Delta_{ij}=1.27\Delta m_{ij}^2L/E$ with $L$ the oscillation length in km,
$E$ the beam energy in GeV, and $\Delta m_{ij}^2=m_i^2-m_j^2$ in eV$^2$.
We write
$\Delta = \Delta_{31}$, $\alpha = \frac{\Delta m_{21}^2}{\Delta m_{31}^2}$
and $A=\frac{VL}{2\Delta}$ where $V$ is the potential expressed in units of eV as
\be
V\approx 7.56\times 10^{-14}\rho \ N_e ,
\ee
where $\rho$ is the matter density of the Earth in units of g/cm$^3$ and
$N_e\approx 0.5$ is the number of electrons per nucleon in the Earth.
The constant density approximation is good when the neutrino beam only passes
through the Earth's crust where $\rho \approx 3$ g/cm$^3$ or the
Earth's mantle where $\rho \approx 4.5$ g/cm$^3$.

The complete set of neutrino oscillation probabilities for electron neutrino
or muon neutrino beams in the
presence of matter of constant density to second order in the parameters
$r,s,a$ and $\alpha$ are~\cite{King:2007pr}:
\beq
P_{ee}  =  1 - \frac{8}{9}\alpha^2 \frac{\sin ^2 A\Delta }{A^2}
-2r^2\frac{\sin ^2 (A-1)\Delta }{(A-1)^2},
\ee
\bea
P_{e\mu} & = & \frac{4}{9}\alpha^2 \frac{\sin ^2 A\Delta }{A^2}
+ r^2\frac{\sin ^2 (A-1)\Delta }{(A-1)^2} \nonumber \\
& + &
\frac{4}{3}r\alpha  \cos (\Delta - \delta ) \frac{\sin A\Delta }{A} \frac{\sin (A-1)\Delta }{(A-1)},
\eea
\bea
P_{e\tau} & = &
\frac{4}{9}\alpha^2 \frac{\sin ^2 A\Delta }{A^2}
+ r^2\frac{\sin ^2 (A-1)\Delta }{(A-1)^2} \nonumber \\
& - &
\frac{4}{3}r\alpha  \cos (\Delta - \delta ) \frac{\sin A\Delta }{A} \frac{\sin (A-1)\Delta }{(A-1)},
\eea
\bea
P_{\mu e} & = & \frac{4}{9}\alpha^2 \frac{\sin ^2 A\Delta }{A^2}
+ r^2\frac{\sin ^2 (A-1)\Delta }{(A-1)^2} \nonumber \\
& + &
\frac{4}{3}r\alpha  \cos (\Delta + \delta ) \frac{\sin A\Delta }{A} \frac{\sin (A-1)\Delta }{(A-1)},
\eea
\bea
P_{\mu \mu} & = & 1
- (1-4a^2)\sin ^2 \Delta
+ \frac{2}{3}(1-s)\alpha \Delta \sin 2 \Delta \nonumber \\
& - & \frac{4}{9} \alpha^2 \frac{\sin ^2 A\Delta }{A^2}
-\frac{4}{9}\alpha^2 \Delta^2 \cos 2\Delta \nonumber \\
& + &
\frac{4}{9}\alpha^2 \frac{1}{A}
\left( \sin \Delta \frac{\sin A\Delta }{A} \cos (A-1)\Delta - \frac{\Delta }{2}\sin 2\Delta  \right)
\nonumber \\
& - & r^2\frac{\sin ^2 (A-1)\Delta }{(A-1)^2} \nonumber \\
& - & \frac{1}{A-1}r^2
\left( \sin \Delta \cos A\Delta \frac{\sin (A-1)\Delta }{(A-1)} - \frac{A}{2}\Delta \sin 2\Delta  \right)
           \nonumber \\
& - &
\frac{4}{3}r\alpha  \cos \delta \cos \Delta \frac{\sin A\Delta }{A} \frac{\sin (A-1)\Delta }{(A-1)},
\eea
\bea
P_{\mu \tau} & = &
(1-4a^2)\sin ^2 \Delta
- \frac{2}{3}(1-s)\alpha \Delta \sin 2 \Delta
+ \frac{4}{9}\alpha^2 \Delta^2 \cos 2\Delta   \nonumber \\
& - &
\frac{4}{9}\alpha^2 \frac{1}{A}
\left( \sin \Delta \frac{\sin A\Delta }{A} \cos (A-1)\Delta - \frac{\Delta }{2}\sin 2\Delta  \right)
\nonumber \\
& + &
\frac{1}{A-1}r^2
\left( \sin \Delta \cos A\Delta \frac{\sin (A-1)\Delta }{(A-1)} - \frac{A}{2}\Delta \sin 2\Delta  \right)
           \nonumber \\
& + &
\frac{4}{3}r\alpha  \sin \delta \sin \Delta \frac{\sin A\Delta }{A}
\frac{\sin (A-1)\Delta }{(A-1)}.
\eea

\section{\label{app:CGs}Generators and Clebsch-Gordan coefficients of 
${\boldsymbol{S_4}}$, ${\boldsymbol{A_4}}$ and ${\boldsymbol{T_7}}$}
In this section we list the generators of the groups $S_4$, $A_4$ and $T_7$
in the basis where the order three generator is diagonal. As this basis is particularly
convenient for model building purposes, we state the corresponding 
(basis dependent) Clebsch-Gordan coefficients for all non-trivial Kronecker products.
We first consider the two intimately linked groups $S_4$ and $A_4$, before
discussing the group~$T_7$. 

\subsection[The groups ${{S_4}}$ and ${{A_4}}$]{The groups
  ${\boldsymbol{S_4}}$ and ${\boldsymbol{A_4}}$}
The permutation group $S_4$ can be defined in terms of three generators
  $S$, $T$ and $U$ satisfying the presentation rules~\cite{Hagedorn:2010th}
\be
S^2=T^3=U^2=(ST)^3 = (SU)^2 = (TU)^2=(STU)^4 = 1 \ .
\ee
Dropping the generator $U$ and with it all relations involving $U$, we obtain
the presentation of the alternating group $A_4$.

The triplet matrix representations of the three $S_4$ generators 
in the $T$-diagonal basis can be obtained from
Eq.~(\ref{eq:s4irrep}). Noticing that the $b$ generator (corresponding to $U$) in
Eq.~(\ref{eq:delta6n2}) occurs only quadratically, we immediately find another
triplet representation by changing the overall sign of $U$. The obtained
irreducible representations are called ${\bf 3}$ and ${\bf 3'}$,
respectively. Likewise we find the two singlet representations ${\bf 1}$ and
${\bf 1'}$. Summing up the square of the dimensions of these representations,
$1^2+1^2+3^2+3^2=20$, shows that there exists only one more irreducible
representation, namely the doublet ${\bf 2}$. Its matrix representation is
presented, together with the other irreducible representations in the
following table. 
$$
\begin{array}{c|ccc|c}\toprule
S_4 & A_4 & S & T & U \\\midrule
{\bf 1,1'} & {\bf 1} & 1&1 &\pm 1 \\[2mm]
{\bf 2} & \begin{pmatrix}{\bf 1''}\\{\bf 1'}\end{pmatrix} & 
\begin{pmatrix} 1 & 0 \\ 0&1 \end{pmatrix} 
&\begin{pmatrix} \omega & 0 \\ 0&\omega^2 \end{pmatrix}  
& \begin{pmatrix} 0& 1 \\ 1&0 \end{pmatrix} \\[4mm]
{\bf 3,3'} & {\bf 3}& 
\frac{1}{3}\begin{pmatrix} -1 & 2&2 \\ 2&-1&2 \\2&2&-1 \end{pmatrix} 
&\begin{pmatrix} 1&0&0\\ 0&\omega^2 &0 \\ 0&0&\omega \end{pmatrix}  
& \mp \begin{pmatrix} 1&0&0\\0&0& 1 \\ 0&1&0 \end{pmatrix} 
\\\bottomrule
\end{array}
$$
The same table also shows the representations of the $S_4$ subgroup $A_4$,
generated by $S$ and $T$ only. Dropping the $U$ generator, it is clear that
both triplets of $S_4$ coincide with the single $A_4$ triplet. Likewise, the
two $S_4$ singlets correspond to the trivial singlet of $A_4$. The $S_4$
doublet, on the other hand, becomes reducible once the $U$ generator is
removed. Hence, it decomposes into two separate non-trivial
irreducible representations of $A_4$, ${\bf 1''}$ and ${\bf 1'}$.

The non-trivial $S_4$ product rules in the $T$-diagonal basis are listed
below, where we use  the number of primes within the expression
\be
{\bs{\alpha}}^{(\prime)}  \otimes {\bs {\beta}}^{(\prime)} ~\rightarrow
~{\bs{\gamma}}^{(\prime)} \ , \label{CGnotation} 
\ee
to classify the results. We denote this number by $n$, e.g. in ${\bf 3}\otimes
{\bf 3}^\prime \rightarrow {\bf 3}^\prime$ we get $n=2$.
%
%
%%%%%%%%%%%%%%%%%%%%%%%%%%%
%%%%    S4 Clebsch Gordan coefficients    %%%%  
%%%%%%%%%%%%%%%%%%%%%%%%%%%
%
%
$$
\begin{array}{lll}
{\bf 1}^{(\prime)} \otimes {\bf 1}^{(\prime)} ~\rightarrow ~{\bf
  1}^{(\prime)} ~~
\left\{ \begin{array}{c} 
~\\n=\mathrm{even}\\~
\end{array}\right.
&%
\left.
\begin{array}{c} 
{\bf 1}^{\phantom{\prime}} \otimes {\bf 1}^{\phantom{\prime}} ~\rightarrow ~{\bf 1}^{\phantom{\prime}}\\
{\bf 1}^{{\prime}} \otimes {\bf 1}^{{\prime}} ~\rightarrow ~{\bf 1}^{\phantom{\prime}}\\
{\bf 1}^{\phantom{\prime}} \otimes {\bf 1}^{{\prime}} ~\rightarrow ~{\bf 1}^{{\prime}}
\end{array}
\right\}
&
\alpha \beta \ ,
\\[10mm]
{\bf 1}^{(\prime)} \otimes \;{\bf 2} \;~\rightarrow \;~{\bf 2}^{\phantom{(\prime)}}~~ \left\{
\begin{array}{c}
n=\mathrm{even} \\
n=\mathrm{odd}
\end{array} \right.
&%
\left.
\begin{array}{c}
{\bf 1}^{\phantom{\prime}} \otimes {\bf 2} ~\rightarrow ~{\bf 2} \\
{\bf 1}^{\prime} \otimes {\bf 2} ~\rightarrow ~{\bf 2}\\
\end{array}\;~
\right\}
&
 \alpha \begin{pmatrix} \beta_1 \\ (-1)^n \beta_2\end{pmatrix}  ,
\\[7mm]
{\bf 1}^{(\prime)} \otimes {\bf 3}^{(\prime)} ~\rightarrow ~{\bf 3}^{(\prime)}
~~ \left\{ \begin{array}{c}
~\\[3mm]n=\mathrm{even} \\[3mm]~
\end{array}\right.
&%
\left. 
\begin{array}{c}
{\bf 1}^{\phantom{\prime}} \otimes {\bf 3}^{\phantom{\prime}} ~\rightarrow ~{\bf 3}^{\phantom{\prime}}
\\
{\bf 1}^{{\prime}} \otimes {\bf 3}^{{\prime}} ~\rightarrow ~{\bf 3}^{\phantom{\prime}}
\\
{\bf 1}^{\phantom{\prime}} \otimes {\bf 3}^{{\prime}} ~\rightarrow ~{\bf 3}^{{\prime}}
\\
{\bf 1}^{{\prime}} \otimes {\bf 3}^{\phantom{\prime}} ~\rightarrow ~{\bf 3}^{{\prime}}
\end{array}
\right\}
&
 \alpha   \begin{pmatrix} \beta_1 \\  \beta_2\\\beta_3 \end{pmatrix}  ,
\\[12.2mm]
{\bf 2} \;\; \otimes \;\;{\bf 2} \;~\rightarrow \;~{\bf 1}^{(\prime)} ~~ \left\{\begin{array}{c}
n=\mathrm{even}\\
n=\mathrm{odd}
\end{array}\right.
&%
\left.
\begin{array}{c}
{\bf 2} \otimes {\bf 2} ~\rightarrow ~{\bf 1}^{\phantom{\prime}} \\
{\bf 2} \otimes {\bf 2} ~\rightarrow ~{\bf 1}^{{\prime}} 
\end{array}~\;
\right\}
&
 \alpha_1 \beta_2 + (-1)^n \alpha_2 \beta_1 \ , 
\\[7mm]
{\bf 2} \;\;\otimes \;\; {\bf 2} ~\;\rightarrow \;~{\bf 2}^{\phantom{(\prime)}} ~~ \left\{ \begin{array}{c}
~\\[-3mm] n=\mathrm{even}\\[-3mm]~
\end{array}\right.
&%
\left.
\begin{array}{c}
~\\[-3mm]
{\bf 2} \otimes {\bf 2} ~\rightarrow ~{\bf 2} \\[-3mm]~
\end{array}~~\,
\right\}
&
   \begin{pmatrix} \alpha_2 \beta_2 \\  \alpha_1\beta_1 \end{pmatrix} , 
%\\[6mm]
%
%
\end{array}
$$
$$
\begin{array}{lll}
{\bf 2}\;\; \otimes \; {\bf 3}^{{(\prime)}} ~\rightarrow ~{\bf 3}^{{(\prime)}} ~~ \left\{\begin{array}{c}
~\\[-2mm] n=\mathrm{even}\\ \\[2mm]
n=\mathrm{odd}\\[-2mm]~
\end{array}\right.
&%
\left.
\begin{array}{c}
{\bf 2} \otimes {\bf 3}^{\phantom{\prime}} ~\rightarrow ~{\bf 3}^{\phantom{\prime}} \\
{\bf 2} \otimes {\bf 3}^{{\prime}} ~\rightarrow ~{\bf 3}^{{\prime}} \\[3mm]
{\bf 2} \otimes {\bf 3}^{\phantom{\prime}} ~\rightarrow ~{\bf 3}^{{\prime}} \\
{\bf 2} \otimes {\bf 3}^{{\prime}} ~\rightarrow ~{\bf 3}^{\phantom{\prime}} 
\end{array}\;
\right\}
&
 \alpha_1 \begin{pmatrix} \beta_2 \\  \beta_3\\\beta_1 \end{pmatrix} + (-1)^n
\alpha_2 \begin{pmatrix} \beta_3 \\  \beta_1\\\beta_2 \end{pmatrix}  ,
\\[13.5mm]
{\bf 3}^{(\prime)} \otimes {\bf 3}^{(\prime)} ~\rightarrow ~{\bf 1}^{(\prime)}
~~ \left\{ \begin{array}{c}
~\\n=\mathrm{even}\\~
\end{array}\right.
&%
\left.\begin{array}{c}
{\bf 3}^{\phantom{\prime}} \otimes {\bf 3}^{\phantom{\prime}} ~\rightarrow ~{\bf 1}^{\phantom{\prime}}
\\
{\bf 3}^{{\prime}} \otimes {\bf 3}^{{\prime}} ~\rightarrow ~{\bf 1}^{\phantom{\prime}}
\\
{\bf 3}^{\phantom{\prime}} \otimes {\bf 3}^{{\prime}} ~\rightarrow ~{\bf 1}^{{\prime}}
\end{array}\right\}
&
 \alpha_1 \beta_1 +\alpha_2\beta_3+\alpha_3\beta_2 \ ,
\\[9mm]
{\bf 3}^{(\prime)} \otimes {\bf 3}^{(\prime)} ~\rightarrow ~{\bf 2}^{\phantom{(\prime)}} ~~
\left\{ \begin{array}{c}
~\\[-3mm]
n=\mathrm{even}\\ \\[1mm]
n=\mathrm{odd}\\[-4.5mm]~
\end{array}\right.
&%
\left.\begin{array}{c}
{\bf 3}^{\phantom{\prime}} \otimes {\bf 3}^{\phantom{\prime}} ~\rightarrow ~{\bf 2} \\
{\bf 3}^{{\prime}} \otimes {\bf 3}^{{\prime}} ~\rightarrow ~{\bf 2} \\[3mm]
{\bf 3}^{\phantom{\prime}} \otimes {\bf 3}^{{\prime}} ~\rightarrow ~{\bf 2} \\
\end{array}\;
\right\}
&
\begin{pmatrix} \alpha_2 \beta_2 +\alpha_3 \beta_1+\alpha_1\beta_3\\ 
(-1)^n(\alpha_3 \beta_3+\alpha_1\beta_2+\alpha_2\beta_1) \end{pmatrix} ,
\\[10.5mm]
{\bf 3}^{(\prime)} \otimes {\bf 3}^{(\prime)} ~\rightarrow ~{\bf 3}^{(\prime)}
~~ \left\{\begin{array}{c}
~\\n=\mathrm{odd}\\~
\end{array}\right.
&%
\left.\begin{array}{c}
{\bf 3}^{\phantom{\prime}} \otimes {\bf 3}^{\phantom{\prime}} ~\rightarrow ~{\bf 3}^{{\prime}}
\\
{\bf 3}^{\phantom{\prime}} \otimes {\bf 3}^{{\prime}} ~\rightarrow ~{\bf 3}^{\phantom{\prime}}
\\
{\bf 3}^{{\prime}} \otimes {\bf 3}^{{\prime}} ~\rightarrow ~{\bf 3}^{{\prime}}
\end{array}\right\}
& 
\begin{pmatrix} 
2 \alpha_1 \beta_1-\alpha_2\beta_3-\alpha_3\beta_2 \\  
2 \alpha_3 \beta_3-\alpha_1\beta_2-\alpha_2\beta_1 \\  
2 \alpha_2 \beta_2-\alpha_3\beta_1-\alpha_1\beta_3 
 \end{pmatrix} ,
\\[9mm]
{\bf 3}^{(\prime)} \otimes {\bf 3}^{(\prime)} ~\rightarrow ~{\bf 3}^{(\prime)}~~\left\{\begin{array}{c}
~\\n=\mathrm{even}\\~
\end{array}\right.
&%
\left.\begin{array}{c}
{\bf 3}^{\phantom{\prime}} \otimes {\bf 3}^{\phantom{\prime}} ~\rightarrow ~{\bf 3}^{\phantom{\prime}}
\\
{\bf 3}^{{\prime}} \otimes {\bf 3}^{{\prime}} ~\rightarrow ~{\bf 3}^{\phantom{\prime}}
\\
{\bf 3}^{\phantom{\prime}} \otimes {\bf 3}^{{\prime}} ~\rightarrow ~{\bf 3}^{{\prime}}
\end{array}\right\}
&
\begin{pmatrix} 
\alpha_2\beta_3-\alpha_3\beta_2 \\  
\alpha_1\beta_2-\alpha_2\beta_1 \\  
\alpha_3\beta_1-\alpha_1\beta_3 
 \end{pmatrix}  .
\end{array}\\[3mm]
$$
%
%
%%%%%%%%%%%%%%%%%%%%%%%%%%%
%%%%    S4 Clebsch Gordan coefficients    %%%% 
%%%%%%%%%%%%%%%%%%%%%%%%%%%
%
%
The $A_4$ Clebsch-Gordan coefficients can be obtained from these
expressions by simply dropping all primes and identifying the components of
the $S_4$ doublet ${\bf 2}$ as the ${\bf 1''}$ and ${\bf 1'}$ representations
of $A_4$. We thus find the non-trivial $A_4$ products, explicitly,

%
%
%%%%%%%%%%%%%%%%%%%%%%%%%%%
%%%%    A4 Clebsch Gordan coefficients    %%%%  
%%%%%%%%%%%%%%%%%%%%%%%%%%%
%
%
$$
\begin{array}{lcl}
{\bf 1'} \otimes {\bf 1''} ~\rightarrow ~{\bf 1} 
&&
\alpha \beta \ ,\\[3mm]
{\bf 1'} \otimes {\bf 3} ~\rightarrow ~{\bf 3}  
&&
\alpha \begin{pmatrix} 
\beta_3 \\  
\beta_1 \\  
\beta_2 
 \end{pmatrix}  , \\[8mm]
{\bf 1''} \otimes {\bf 3} ~\rightarrow ~{\bf 3}  
&&
\alpha \begin{pmatrix} 
\beta_2 \\  
\beta_3 \\  
\beta_1 
 \end{pmatrix}  , \\[8mm]
{\bf 3} \otimes {\bf 3} ~\rightarrow ~{\bf 1} 
&&
\alpha_1\beta_1 +\alpha_2\beta_3+\alpha_3\beta_2 \ ,\\[2mm]
{\bf 3} \otimes {\bf 3} ~\rightarrow ~{\bf 1'} 
&&
\alpha_3\beta_3 +\alpha_1\beta_2+\alpha_2\beta_1 \ ,\\[2mm]
{\bf 3} \otimes {\bf 3} ~\rightarrow ~{\bf 1''} 
&&
\alpha_2\beta_2 +\alpha_3\beta_1+\alpha_1\beta_3 \ ,\\[3mm]
{\bf 3} \otimes {\bf 3} ~\rightarrow ~{\bf 3}  + {\bf 3} 
&&
\begin{pmatrix} 
2 \alpha_1 \beta_1-\alpha_2\beta_3-\alpha_3\beta_2 \\  
2 \alpha_3 \beta_3-\alpha_1\beta_2-\alpha_2\beta_1 \\  
2 \alpha_2 \beta_2-\alpha_3\beta_1-\alpha_1\beta_3 
 \end{pmatrix}
+
\begin{pmatrix} 
\alpha_2\beta_3-\alpha_3\beta_2 \\  
\alpha_1\beta_2-\alpha_2\beta_1 \\  
\alpha_3\beta_1-\alpha_1\beta_3 
 \end{pmatrix}   . 
\end{array}
$$
%
%%%%%%%%%%%%%%%%%%%%%%%%%%%
%%%%    A4 Clebsch Gordan coefficients    %%%% 
%%%%%%%%%%%%%%%%%%%%%%%%%%%

\subsection[The group ${T_7}$]{The group $\bs{T_7}$}
The Frobenius group $T_7=Z_7 \rtimes Z_3$ is obtained from two generators
$a$ and $c$ obeying the presentation, see e.g.~\cite{Luhn:2007yr},
\be
<a,c\,|\,a^3=c^7=1 \,,\, aca^{-1} = c^2> \ . \label{presentation-t7}
\ee
Notice that with $k=2$ in Eq.~(\ref{eq:pres-tn}), the condition $1+2+2^2 =
0~\mathrm{mod}~7$ holds. A triplet representation with non-diagonal order
three generator is given in Eq.~(\ref{eq:tnirrep}), where $\eta=e^{\frac{2\pi
i}{7}}$. To diagonalise~$a$, we apply the basis transformation of
Eq.~(\ref{eq:diagonalize-a}) (followed by the phase transformation $T^2$ in
order to bring the $c$ generator into a more appealing form), 
resulting in the matrix representation of the triplet ${\bf 3}$ as shown in
the below table. Clearly, $T_7$ also 
contains another triplet representation given by the complex conjugate ${\bf
  \ol 3}$ of the ${\bf 3}$. Furthermore there are three singlet representations.
$$
\begin{array}{ccc}
& a & c \\[1mm] \hline \\[-3mm]
{\bf 1} & 1& 1 \\[2mm]
{\bf 1'} & \omega  & 1 \\[2mm]
{\bf 1''} & \omega^2& 1 \\[2mm]
{\bf 3} & \begin{pmatrix} 1&0&0\\0&\omega^2&0\\ 0&0&\omega  \end{pmatrix} & 
\frac{\eta}{3} \begin{pmatrix} 
1+\eta+\eta^3 &
\omega^2+\omega \eta   + \eta^3&
\omega+\omega^2 \eta+\eta^3 \\
\omega+\omega^2 \eta+\eta^3 &
1+\eta+\eta^3&
\omega^2+\omega \eta   + \eta^3\\
\omega^2+\omega \eta   + \eta^3 & 
\omega+\omega^2 \eta+\eta^3 &
1+\eta+\eta^3 
\end{pmatrix}
 \\[9mm]
{\bf \ol 3} & \begin{pmatrix} 1&0&0\\0&\omega&0\\ 0&0&\omega^2  \end{pmatrix} & 
\frac{\eta^6}{3} \begin{pmatrix} 
1+\eta^6+\eta^4 &
\omega+\omega^2 \eta^6   + \eta^4&
\omega^2+\omega \eta^6+\eta^4 \\
\omega^2+\omega \eta^6+\eta^4 &
1+\eta^6+\eta^4&
\omega+\omega^2 \eta^6   + \eta^4\\
\omega+\omega^2 \eta^6   + \eta^4 & 
\omega^2+\omega \eta^6+\eta^4 &
1+\eta^6+\eta^4 
\end{pmatrix}
\end{array}
$$
Although the order-seven generators of the triplet representations look rather
involved, the Clebsch-Gordan coefficients take a relatively simple
form. Omitting the trivial products, i.e. those involving the singlet ${\bf 
  1}$ as well as products of only one-dimensional irreducible representations,
the product rules are reported below. Again, we use the convention that the components of
the first representation of any given product ${\bs{\alpha} \otimes \bs{\beta}}$ are denoted
by $\alpha_i$ while we use $\beta_i$ for the components of the second
representation. 

$$
\begin{array}{lcl}
\bf 1' \otimes  3 ~\rightarrow ~{\bf 3}
&~&
\alpha \begin{pmatrix} 
\beta_2\\\beta_3\\\beta_1 
 \end{pmatrix} \ , \qquad \qquad
{{\bf 1'' \otimes  3}} ~\rightarrow ~ {\bf 3} 
~~~~~~~~
\alpha \begin{pmatrix} 
\beta_3\\\beta_1\\\beta_2 
 \end{pmatrix} \ ,
\\[9mm]
\bf 1' \otimes  \ol 3 ~\rightarrow ~{\bf \ol 3}
&&
\alpha \begin{pmatrix} 
\beta_3\\\beta_1\\\beta_2
 \end{pmatrix} \ , \qquad \qquad
{{\bf 1'' \otimes  \ol 3}} ~\rightarrow ~ {\bf \ol 3} 
~~~~~~~~
\alpha \begin{pmatrix} 
\beta_2\\\beta_3\\\beta_1 
 \end{pmatrix} \ ,
%\\[9mm]
%
%
\end{array}
$$
$$
\begin{array}{lcl}
{\bf 3 \otimes \ol 3} ~\rightarrow ~ {\bf 1}
&&
\alpha_1\beta_1+\alpha_2 \beta_2+\alpha_3\beta_3 \ , \\[2mm]
{\bf 3 \otimes \ol 3} ~\rightarrow ~ {\bf 1'}
&&
\alpha_1\beta_2+\alpha_2 \beta_3+\alpha_3\beta_1 \ , \\[2mm]
{\bf 3 \otimes \ol 3} ~\rightarrow ~ {\bf 1''}
&&
\alpha_1\beta_3+\alpha_2 \beta_1+\alpha_3\beta_2 \ , \\[2mm]
{\bf 3 \otimes \ol 3} ~\rightarrow ~ {\bf 3}
&&
\begin{pmatrix} 
\alpha_1\beta_1+\omega^2 \alpha_2 \beta_2+\omega \alpha_3\beta_3  \\
\alpha_1\beta_3+\omega^2 \alpha_2 \beta_1+\omega \alpha_3\beta_2 \\
\alpha_1\beta_2+\omega^2 \alpha_2 \beta_3+\omega \alpha_3\beta_1
 \end{pmatrix} \ ,\\[9mm]
{\bf 3 \otimes \ol 3} ~\rightarrow ~ {\bf \ol 3}
&&
\begin{pmatrix} 
\alpha_1\beta_1+\omega \alpha_2 \beta_2+\omega^2 \alpha_3\beta_3  \\
\alpha_3\beta_1+\omega \alpha_1\beta_2+\omega^2 \alpha_2 \beta_3  \\
 \alpha_2 \beta_1+\omega \alpha_3\beta_2+\omega^2 \alpha_1\beta_3
 \end{pmatrix} \ ,\\[10mm]
{\bf 3 \otimes  3}~\rightarrow ~ {\bf 3}
&&
\begin{pmatrix} 
\alpha_1\beta_1+ \alpha_2 \beta_3+ \alpha_3\beta_2  \\
\omega (\alpha_1\beta_2+ \alpha_2 \beta_1+\alpha_3\beta_3) \\
\omega^2 (\alpha_1\beta_3+ \alpha_2 \beta_2+ \alpha_3\beta_1)
 \end{pmatrix} \ ,\\[9mm]
{\bf 3 \otimes  3}~\rightarrow ~ {\bf \ol 3} + {\bf \ol 3}
&&
\begin{pmatrix} 
2 \alpha_1\beta_1- \alpha_2 \beta_3- \alpha_3\beta_2  \\
2 \alpha_2\beta_2- \alpha_3 \beta_1- \alpha_1\beta_3  \\
2 \alpha_3\beta_3- \alpha_1 \beta_2- \alpha_2\beta_1
 \end{pmatrix} +
 \begin{pmatrix} 
 \alpha_2 \beta_3- \alpha_3\beta_2  \\
 \alpha_3 \beta_1- \alpha_1\beta_3  \\
 \alpha_1 \beta_2- \alpha_2\beta_1
 \end{pmatrix} \ .
\end{array}
$$

\end{document}